\documentclass[manuscript,screen]{acmart}
\usepackage{array}          
\usepackage{multirow}      
\usepackage{booktabs}            
\usepackage{tcolorbox}   
\usepackage{caption}       
\usepackage{arydshln}
\usepackage{subfigure}
\usepackage[utf8]{inputenc}
\usepackage{graphicx}
\usepackage{colortbl}
\usepackage{enumitem}
\usepackage{amsmath}    
\usepackage{mathtools}  
\usepackage{algorithm}
\usepackage{algpseudocode}
\usepackage{booktabs}
\usepackage{adjustbox}
\usepackage{pifont}
\usepackage{tabularx}
\newcolumntype{L}[1]{>{\raggedright\arraybackslash}p{#1}}
\newcolumntype{Y}{>{\raggedright\arraybackslash}X}

\renewcommand{\arraystretch}{1.15}

\theoremstyle{definition}

\theoremstyle{remark}

\newcolumntype{R}[1]{>{\small\raggedright\arraybackslash}m{#1}}      
\newcolumntype{L}[1]{>{\raggedright\arraybackslash}p{#1}} 

\begin{document}
\title{Think Like an Engineer: A Neuro-Symbolic Collaboration Agent for Generative Software Requirements Elicitation and Self-Review}

\author{Sai Zhang}
\email{zhang_sai@tju.edu.cn}\thanks{S. Zhang also with CSIRO's Data61}
\orcid{0000-0001-8972-2824}
\affiliation{
    \institution{College of Intelligence and Computing, Tianjin University}
    \country{China}
    \postcode{300072}
  }

\author{Zhenchang Xing}
\email{zhenchang.xing@data61.csiro.au}
\orcid{0000-0001-7663-1421}
\affiliation{
  \institution{CSIRO's Data61}
  \country{Australia}
  \postcode{0200-2600-2601}
}

\author{Jieshan Chen}
\email{Jieshan.Chen@data61.csiro.au}
\orcid{0000-0002-2700-7478}
\affiliation{
  \institution{CSIRO's Data61}
  \country{Australia}
  \postcode{0200-2600-2601}
}

\author{Dehai Zhao}
\email{dehai.zhao@data61.csiro.au}
\orcid{0000-0003-3637-4939}
\affiliation{
  \institution{CSIRO's Data61}
  \country{Australia}
  \postcode{0200-2600-2601}
}

\author{ZiZhong Zhu}
\email{zizhong_zhu@tju.edu.cn}
\orcid{0000-0003-2303-8741}
\affiliation{
\institution{School of New Media and Communication, Tianjin University}
\country{China}
}

\author{Xiaowang Zhang$^*$}\thanks{*Corresponding author}
\email{xiaowangzhang@tju.edu.cn}
\orcid{0000-0002-3931-3886}
\affiliation{
  \institution{College of Intelligence and Computing, Tianjin University}
  \country{China}}

\author{Zhiyong Feng}
\email{zyfeng@tju.edu.cn}
\orcid{0000-0001-8158-7453}
\affiliation{
  \institution{College of Intelligence and Computing, Tianjin University}
  \country{China}}

\author{Xiaohong Li}
\email{xiaohongli@tju.edu.cn}
\orcid{0000-0002-0752-6764}
\affiliation{
\institution{College of Intelligence and Computing, Tianjin University}
\country{China}
}

\renewcommand{\shortauthors}{Sai Zhang et al.}

\begin{abstract}
The vision of End-User Software Engineering (EUSE) is to empower non-professional users with full control over the software development lifecycle. It aims to enable users to drive generative software development using only natural language requirements. However, since end-users often lack knowledge of software engineering, their requirement descriptions are frequently ambiguous, raising significant challenges to generative software development. Although existing approaches utilize structured languages like Gherkin to clarify user narratives, they still struggle to express the causal logic between preconditions and behavior actions. This paper introduces RequireCEG, a requirement elicitation and self-review agent that embeds causal-effect graphs (CEGs) in a neuro-symbolic collaboration architecture. Large language models (LLMs) perform high-level semantic reasoning, while CEGs provide logic symbolic review, complementing each other to resolve ambiguities. RequireCEG first uses a feature tree to analyze user narratives hierarchically, clearly defining the scope of software components and their system behavior requirements. Next, it constructs the CEGs based on the elicited requirements, capturing the causal relationships between atomic preconditions and behavior actions. Additionally, RequireCEG incorporates a novel self-healing CEGs check mechanism to ensure both the format and logical semantic correctness. Finally, the constructed CEGs are used to review and optimize Gherkin scenarios, ensuring consistency between the generated Gherkin requirements and the system behavior requirements elicited from user narratives. To evaluate our method, we created the RGPair benchmark dataset and conducted extensive experiments. Results show that RequireCEG improves the quality of requirements, functional diversity, and consistency compared with existing methods. In tests with five public websites, RequireCEG achieved an 87\% coverage rate and improved diversity by 51.88\% over state-of-the-art methods.
\end{abstract}



\maketitle

\section{Introduction}
End-User Software Engineering (EUSE) \cite{burnett2004end, ko2011state} helps end-user programmers \cite{barricelli2019end, nardi1993small} develop applications more systematically and with higher quality. In EUSE, end-users control the entire software development lifecycle and can build complete applications from natural language requirements alone. Yet, because they lack formal software-engineering training, their narratives are often ambiguous, making requirement elicitation in generative software development difficult \cite{robinson2024requirements}.

\begin{figure}[h]
	\centering
	\includegraphics[width=\textwidth]{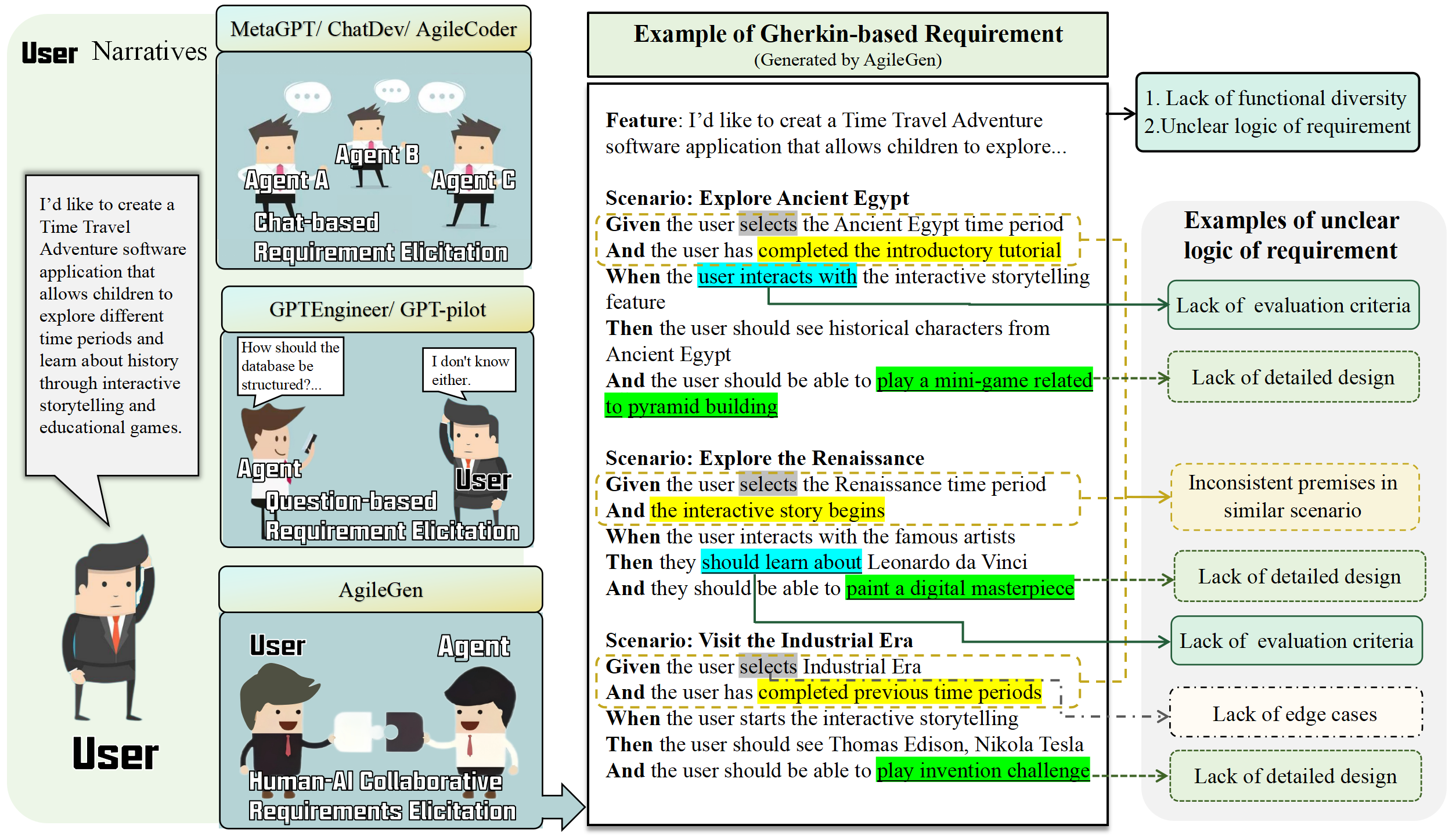}
	\caption{\label{fig:fig0} Using the "Time Travel Adventure" requirement as an example, the problem-statement diagram shows that the Gherkin requirements produced by AgileGen~\cite{zhang2024empowering} show limited functional diversity and unclear logical semantics.}
\end{figure}

In recent years, Large Language Models (LLMs) have shown remarkable planning and reasoning abilities \cite{achiam2023gpt}, making it increasingly feasible to generate complete applications directly from a user narrative \cite{robinson2024requirements}. As illustrated in Fig. \ref{fig:fig0}, existing generative Requirement Elicitation approaches fall into three categories: (1) Chat-based Requirement Elicitation (MetaGPT \cite{hong2023metagpt}, ChatDev \cite{qian2024chatdev}, AgileCoder \cite{nguyen2024agilecoder}). Multi-round role-playing among LLMs agents (e.g., product manager, architect) is used to simulate the entire development process. Because the pipeline is fully automated end-to-end, user involvement in the lifecycle is limited, and the delivered software can diverge from user expectations. (2) Question-based Requirement Elicitation (GPT-Engineer\footnote{\url{https://github.com/AntonOsika/gpt-engineer}}, GPT-pilot\footnote{\url{https://github.com/Pythagora-io/gpt-pilot}}). These agents follow a built-in workflow that proactively asks clarifying questions during requirement elicitation to refine the requirements. The approach assumes that users possess professional domain knowledge, raising the expertise required of end users. (3) Human–AI Collaborative Requirement Elicitation (AgileGen \cite{zhang2024empowering}). The user narrative is transformed into Gherkin scenarios, and users steer the generative software development lifecycle by approving or revising these scenarios, thereby lowering the professional barrier to participation.

The Human-AI collaborative requirement elicitation method introduces structured Gherkin language~\cite{parsa2023acceptance} as a means to enhance end-user engagement and clarify requirements. Still, it is lacking in causal logic analysis and reasoning. As a result, it may still lead to unclear logic of generated Gherkin when handling user narratives involving complex business scenarios.
As illustrated in Fig. \ref{fig:fig0}, the Gherkin requirements produced by AgileGen~\cite{zhang2024empowering} highlight this problem: 
From the Cross-scenario view, all three scenarios describe similar behavior actions, yet their preconditions are inconsistent. Scenarios 1 and 3 explicitly require that "the introductory tutorial (or the previous periods) is completed" at the start. While Scenario 2 states, "The interactive story begins," leaving the starting order unclear. Inside each scenario, vague phrases like "should learn about" and "User interacts with" provide no way to evaluate success. Likewise, higher-level features such as "paint a digital masterpiece" or "play a mini-game related to pyramid building" list no concrete steps, objectives, or evaluation criteria.

The ambiguity observed in the scenarios above stems from the absence of causal relationship analysis and constraint during the requirements elicitation process. Without these safeguards, the relationship between preconditions and behavior actions remains unclear in the Gherkin specifications. Traditional, enterprise-level requirements engineering methods \cite{van2000requirements} rely on multiple rounds of interviews and document reviews conducted by specialists. This process is both costly and time-consuming, and that depends heavily on sustained input from domain experts. When business contexts grow complex and diverse, purely manual approaches are prone to omissions and subjective misjudgments \cite{marques2024using}. Therefore, leveraging the strong analytical and reasoning capabilities of LLMs emerges as a promising alternative. Nonetheless, a significant challenge remains: bridging the gap between the flexibility of natural-language narratives and the formal logic required by software components so that LLMs can identify the causal relationships implicit in user narratives and translate them into implementable software artifacts.

To tackle the above challenges, we introduce RequireCEG, a Requirement Elicitation and self-review agent that supports end-user software development. RequireCEG is designed as a neuro-symbolic collaboration architecture, in which LLMs perform high-level semantic reasoning and Causal-Effect Graphs (CEGs) conduct logical symbolic review. The two mechanisms work in tandem to remove ambiguity in Gherkin scenarios.
Unlike AgileGen~\cite{zhang2024empowering}, whose approach relies on surface-level Gherkin template style matching generation and provides limited support for detecting missing or inconsistent requirements. RequireCEG injects analyst-level reasoning by pairing LLM-guided feature-tree scoping with autonomous CEG construction and review, thereby exposing implicit requirements and guaranteeing logical consistency in the resulting Gherkin.
RequireCEG is divided into three stages:
\textbf{(1) Software components of feature tree hierarchical elicit. }
A top-down hierarchical analysis of the user narrative using a feature tree, scoping the features and enhancing the diversity of software components while specifying system behavior requirements from both user operation and system response perspectives.
\textbf{(2) CEGs construction and self-healing check. }
To uncover the causal relationships within system behavior requirements, RequireCEG constructs CEGs that identify atomic conditions and actions and establish explicit logic for the causal relationships between them based on the requirement. Additionally, we introduce a self-healing formal check and causal intervention mechanism to keep the CEGs reflecting the intended causal relationships in the system behavior requirements.
\textbf{(3) Self-review of the Gherkin scenario by CEGs.}
The healing CEGs are used to review the Gherkin scenarios, guaranteeing consistency between the Gherkin scenarios and the system behavior requirements elicited from the user narrative. 

This core task involves transforming a user narrative into Gherkin-described requirement scenarios—a domain-specific language that is both end-user-readable and automatically testable, fosters human-AI collaboration~ \cite{zhang2024empowering,robinson2024requirements}. To support evaluation, we build RGPair,  derived from 40 high-quality open-source GitHub projects. RGPair contains 413 Gherkin feature files paired with 40 natural language requirements, providing a solid basis for assessing the transformation of natural language requirements into Gherkin scenarios.
Our contributions are summarized as follows:
\begin{itemize}
 \item We present the requirement elicitation and self-review neuro-symbolic collaboration architecture, RequireCEG, which empowers Large Language Models (LLMs) to review the causal relationships between preconditions and behavior actions in scenarios of Gherkin, thereby addressing the shortage of logical constraints in Requirement Elicitation.

\item To bridge the gap between the flexibility of natural-language narratives and the formal logic of software components, we use the feature tree to hierarchically elicit the diverse software components involved in the user narrative. By describing each behavior of a component from both the user operation and the system response perspectives, the method can plan behavior descriptions for complex businesses.

\item To ensure that the CEGs faithfully capture the causal relationships in the system behavior requirement, we introduce the first LLM-based self-healing mechanism for causal intervention detection and formal checking. This mechanism automatically repairs formal or semantic defects in the CEGs, increasing the reliability of its logical constraints.

\item To eliminate ambiguities between preconditions and behavior actions in Gherkin scenarios, we introduce a CEG-based review method that maps the causal logic in CEGs onto Gherkin scenarios, editing or augmenting the pre-generated scenarios to ensure consistency with system behavior requirements elicited from user narratives. To evaluate RequireCEG, we build RGPair, a benchmark comprising 413 Feature files paired with their corresponding 40 natural-language requirements, extracted from 40 high-quality public GitHub repositories. This provides an essential resource for future research on converting natural-language requirements into Gherkin scenarios. 
Our dataset and the code for the results are available for download \footnote{https://github.com/HarrisClover/RequireCEG}.
\end{itemize}

\section{Motivation and Design Rationale}
RequireCEG is designed to be equipped with the analytical skills of a requirements analyst, enabling autonomous requirement elicitation and review to support End-User Software Engineering. It combines conceptual processes from software engineering with the reasoning and analytical capabilities of Large Language Models (LLMs), achieving complementary strengths to ensure diversity and consistency in requirements.
In the following sections, we present the motivation and core design rationale of RequireCEG, as illustrated in Table~\ref{tab:summary}.

\begin{table}[htbp]
  \caption{Summary of Questions, Challenges, Strategies, Goals, and Contributions of RequireCEG}\label{tab:summary}
  \centering
  {
  \footnotesize
  \setlength{\tabcolsep}{3pt}
  \renewcommand{\arraystretch}{1.1}
  \begin{tabularx}{\textwidth}{@{}
      L{2.1cm}  
      L{2.2cm}  
      Y         
      Y        
      L{2.2cm}  
    @{}}
    \toprule
    \textbf{Q}uestions & \textbf{S}trategies &\textbf{C}hallenges & Contributions & \textbf{G}oals\\
    \midrule
    \textbf{Q1}: Shared End-User Language & \textbf{S1}: Gherkin &
    \textbf{C1}: Ambiguity between preconditions and actions &
    CEGs-driven Gherkin self-review: fix or complement Given-When-Then scenarios & \textbf{G1}: Consistency of Gherkin requirements\\
    \midrule
    \textbf{Q2}: Unclear Requirement Logic& \textbf{S2}: Causal-Effect Graph &
    \textbf{C2}: Capture implicit causal relationships in requirements &
    Self-healing CEG: logical formal check and semantic check of causal intervention &
    \textbf{G2}: Reliability of expressing logical relationships\\
    \midrule
    \textbf{Q3}: Lack of Functional Diversity & \textbf{S3}: Feature Tree &
    \textbf{C3}: Bridging flexible narratives with structured component logic &
    Feature-tree hierarchical elicitation with user/system perspectives& \textbf{G3}: Increased functional diversity in requirements\\
    \bottomrule
  \end{tabularx}
  }
\end{table}

\subsection{Motivation: Issues of LLM-based Requirements Generation}
\label{sec:LLMq}
End-user software Engineering aims enable users to build complete applications using only natural language narratives (\textbf{Q1}) \cite{robinson2024requirements}. To assess progress toward this goal, we evaluate AgileGen, a leading human-AI collaborative development agent.
We use a user narrative from the Software Requirement Description Dataset (SRDD) \cite{qian-etal-2024-chatdev}, describing a "Time Travel Adventure" app:
"I'd like to create a Time Travel Adventure software application that allows children to explore different time periods and learn about history through interactive storytelling and educational games."
Based on this narrative, we conduct two experiments with AgileGen~\cite{zhang2024empowering}: 1) Generate multiple requirements from the same user narrative. 2) Generate multiple applications from the same requirements.

\begin{figure}[t]
	\centering
	\includegraphics[width=0.93\textwidth]{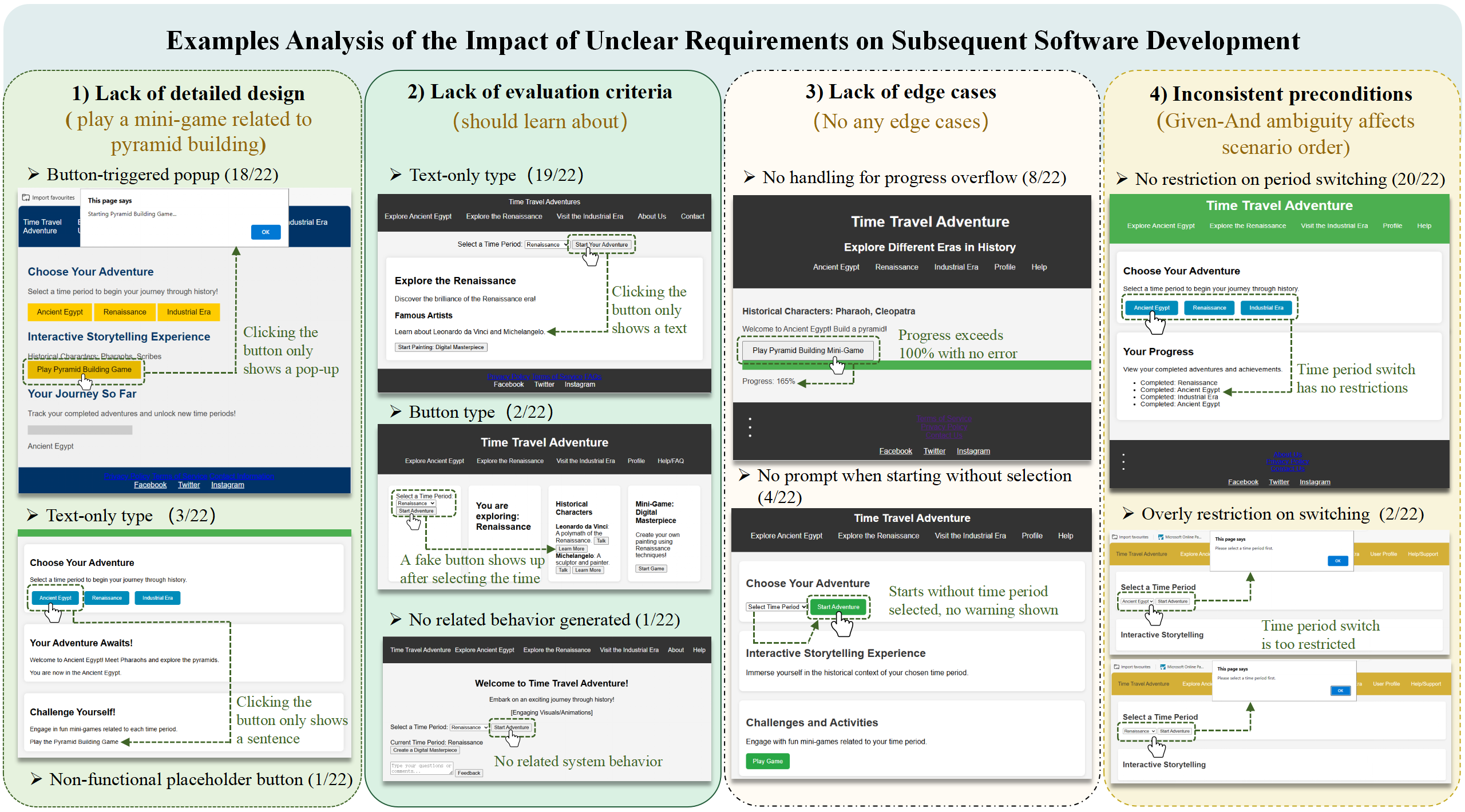}
\caption{\label{fig:fig4} Using the Time Travel Adventure application as a case study, we demonstrate how unclear requirements impact downstream software development. We used AgileGen to generate applications 22 times based on the Gherkin requirement shown in Fig.~\ref{fig:fig0}.}
\end{figure}

\textbf{1) Generate multiple requirements from the same user narrative.}
We generated 22 requirements for the "Time Travel Adventure" narrative using AgileGen to explore common issues in requirement generation.
Structurally, all outputs included exactly one Feature (100\%), with 59.1\% containing 3–4 scenarios and 40.9\% containing 5–6. This single-feature format shows a lack of functional diversity (\textbf{Q3}). In real-world educational software, additional features such as login or parental control are typically expected.
Scenario content analysis revealed four frequent problems \textbf{Q2} (each in over 50\% of cases):
\begin{itemize}
    \item (1) Lack of detailed design (77.2\%): Many scenarios omit concrete business logic. For example, vague statements like "play a mini-game related to pyramid building" lack implementation-specific detail.
    \item (2) Lack of evaluation criteria (72.8\%): Scenarios contain unverifiable statements such as "should learn about" or "should understand", which cannot be evaluated or tested.
    \item (3) Lack of edge cases (72.7\%): Most scenarios do not consider error handling or alternative flows, e.g., missing conditions like "select" vs. "not select".
    \item (4) Inconsistent preconditions in similar scenarios (63.6\%): Scenarios define inconsistent preconditions (e.g., one assumes prior stages are completed, another does not), leading to behavioral ambiguity.
\end{itemize}

\textbf{2) Generate multiple applications from the same requirements.}
 We selected one representative Gherkin requirement (frequently occurring issues, shown in Fig.\ref{fig:fig0}, the following four types of requirement flaws directly impacted the generated applications, As shown in Fig.~\ref{fig:fig4}:
\begin{itemize}
    \item (1) Missing detailed design: Not detailed behavior like "play a mini-game related to pyramid building" were poorly implemented. In 81.8\% of cases, a button opened a pop-up saying "Game Started" without real gameplay. In 13.6\%, only static text like "Play the Pyramid Building Game" appeared. In 4.5\%, a placeholder button had no function.
    
    \item (2) Missing evaluation criteria: Vague statement like "should learn about" led to weak implementations. In 86.4\% of outputs, a line of text such as "Learn about Leonardo da Vinci" was shown. In 9\%, buttons like "Learn More" or "Talk" had no response. In 4.5\%, the feature was missing entirely.
    
    \item (3) Missing edge cases: Lack of edge cases handling caused robustness issues. For example, 36.4\% of applications failed to respond when a progress bar exceeded 100\%, and 18.2\% gave no warning when users clicked "Start" without a selection.
    
    \item (4) Inconsistent preconditions: Scenario inconsistencies caused the functional disorder. In 90.9\% of outputs, users could switch historical periods freely, even when a fixed order was intended. In 9.1\%, switching was overly restricted and blocked progress unnecessarily.
\end{itemize}

In summary, this study identifies two major issues in requirements generated by LLMs. First, the requirement logic is often unclear (\textbf{Q2}), as evidenced by a lack of detailed design, lack of evaluation criteria, lack of edge cases, and inconsistent preconditions in similar scenarios. Second, there is a lack of functional diversity (\textbf{Q3}). These shortcomings can lead to the generation of software that deviates from real business objectives, ultimately hindering the effectiveness of End-User Software Engineering. Therefore, developing more suitable methods for expressing requirements in end-user contexts (\textbf{Q1}) remains a critical direction for future research.

\subsection{The Shared Language of End-User Software Engineering: Gherkin}
\label{sec:2.2}
\begin{figure}[t]
	\centering
	\subfigure[Practitioners’ views on how human factors affect performance in RE activities~\cite{laplante2022requirements[5]2}.]{%
		\centering
		\includegraphics[width=0.4\textwidth]{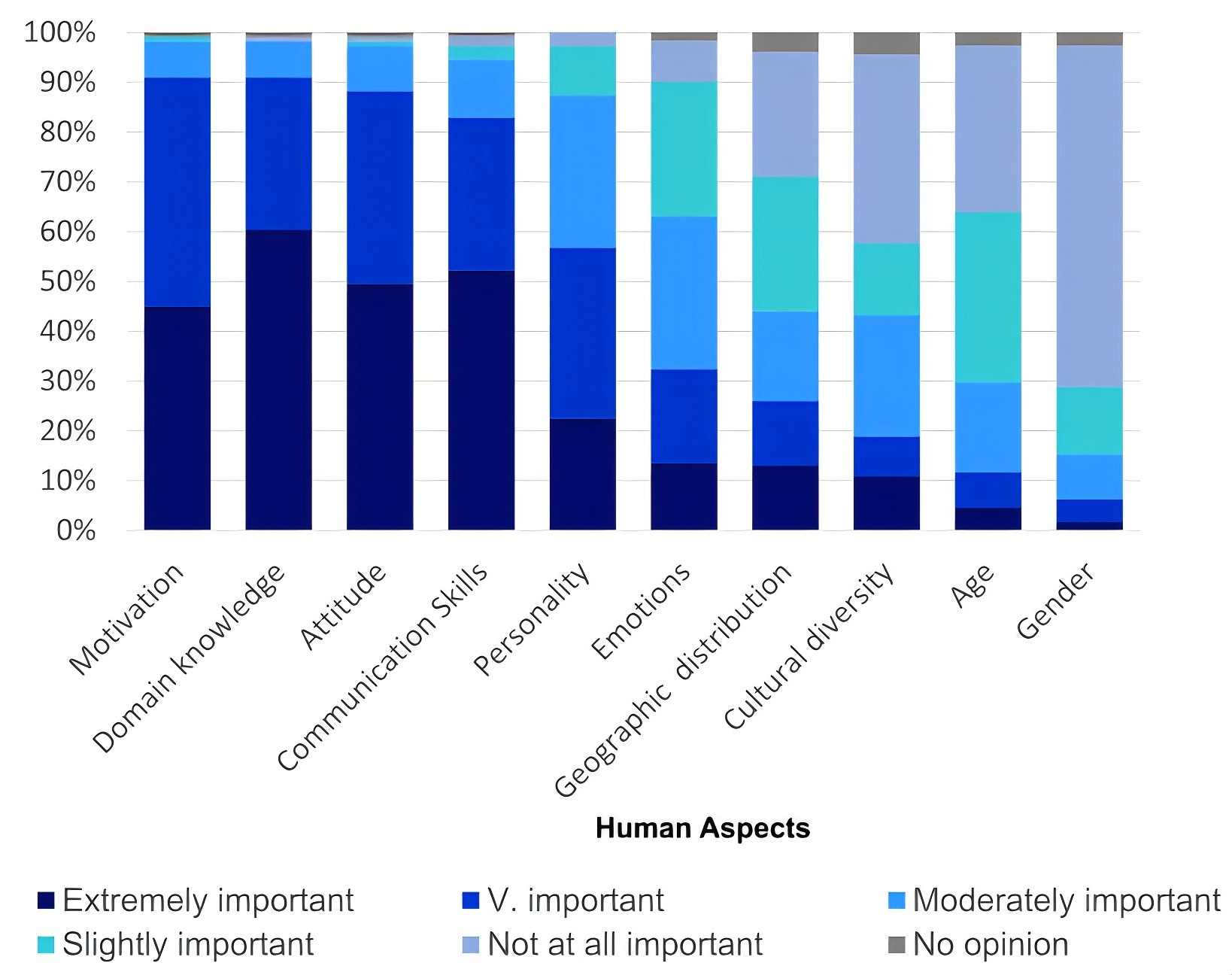}
		\label{fig:sec_2_fig1_a}
	}
    \hspace{0.05\textwidth}
	\subfigure[In 2024, Shexmo Santos et al.~\cite{santos4933832perception} conducted a survey to identify key challenges practitioners face when adopting BDD.]{%
		\centering
		\includegraphics[width=0.4\textwidth]{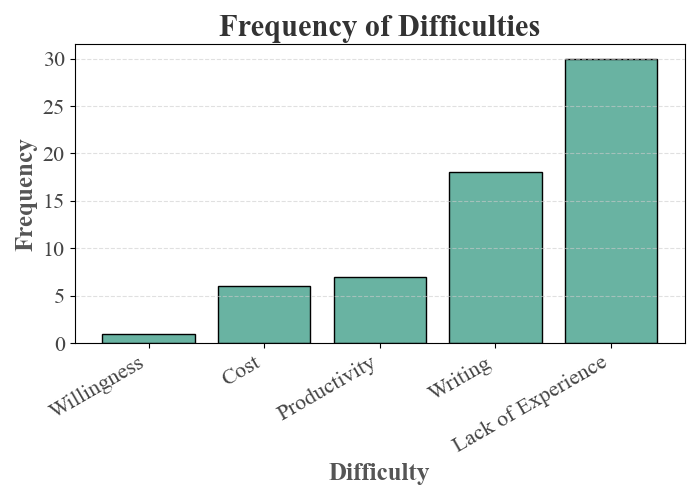}
		\label{fig:sec_2_fig1_b}
	}
        \label{fig:sec_2_fig1}
	\caption{Challenges in Applying BDD (Behavior-Driven Development) \& Human aspects on the performance of individuals in Requirements Engineering (RE)-related activities}
\end{figure}

End-user software engineering (EUSE) was initially proposed to enhance software quality through collaboration with end users \cite{burnett2004end}. Studies show that the number of end users developing software now far exceeds that of professional developers, their software remains error-prone \cite{burnett2014future}. With the rise of large language models (LLMs), EUSE is evolving toward enabling users to build applications from natural language requirements \cite{robinson2024requirements}.
As shown in Fig.~\ref{fig:sec_2_fig1_a}, 91\% of participants cited domain knowledge as critical to requirements engineering. Yet, many end users lack software expertise, making it difficult to define clear acceptance criteria, often leading to mismatches between expectations and results \cite{zhang2024empowering}.
To address this, there is growing need in a shared language between users and AI agents (\textbf{Q1}). According to \cite{robinson2024requirements}, this language should meet the following criteria:

1) It should provide essential structure.

2) It should help users understand what can and cannot be built.

3) It should surface implicit requirements that users may not have explicitly expressed.

\begin{table}[ht]
\centering
\footnotesize
\caption{Gherkin Feature Example Template~\cite{parsa2023acceptance}}
\label{tab:Gherkin_example}
\begin{tabular}{
    L{0.45\textwidth}
    R{0.4\textwidth}
}
\toprule
\textbf{Feature:} \textit{[Title (one line describing the feature or story)]} & \textbf{A Software Component} \\
\midrule
\multicolumn{2}{l}{} \\[-1.5ex]
\textbf{Narrative:} & \multirow[c]{4}{*}{\textbf{Software Component Description}} \\
\textbf{As a} [\textit{role}] & \\
\textbf{I want} [\textit{feature: something, requirement}] & \\
\textbf{So that} [\textit{benefit: achieve some business goal}] & \\
\hline
\multicolumn{2}{l}{} \\[-1.5ex]
\textbf{Background:} & \multirow[c]{3}{*}{\textbf{Common Precondition}}\\
\textbf{Given} [\textit{some condition}] & \\
\textbf{And} [\textit{one more thing}] & \\
\hline
\multicolumn{2}{l}{} \\[-1.5ex]
\textbf{Scenario: Title} & \textbf{Scenario (Acceptance Criteria)} \\
\textbf{Given} [\textit{context}] & \multirow[c]{2}{*}{\textbf{Precondition}}\\
\textbf{And} [\textit{some more context}] &  \\
\hdashline
\textbf{When} [\textit{event}] & \multirow[c]{2}{*}{\textbf{Precondition}} \\
\textbf{And} [\textit{some more conditions}] &  \\
\hdashline
\textbf{Then} [\textit{outcome}]& \multirow[c]{2}{*}{\textbf{Behavior Action}} \\
\textbf{And} [\textit{another outcome}] & \\
\hdashline
\multicolumn{2}{l}{} \\[-1.5ex]
\textbf{Scenario: ...} & \textbf{Another Scenario} \\
\hline
\multicolumn{2}{l}{} \\[-1.5ex]
\textbf{Scenario Outline:} & \multirow[c]{7}{*}{\textbf{Same Scenario with Different Parameters}}\\ 
\textbf{Given} I have [\textit{<something>}] & \\
\textbf{And} I also have [\textit{<number> <thing>}] &  \\
\textbf{Examples:} & \\
\textbf{| something | number | thing |} & \\
\textbf{| … | … | … |} & \\
\textbf{| … | … | … |} & \\
\bottomrule
\end{tabular}
\end{table}

We argue that Gherkin, a domain-specific language (DSL) used in Behavior-Driven Development (BDD), holds potential promise (\textbf{S1}). BDD is introduced by Dan North in 2003\footnote{\url{https://dannorth.net/introducing-bdd/}}, emphasizes capturing requirements through collaboration between technical and business stakeholders, using scenario-based descriptions that serve both as specifications and acceptance criteria \cite{bergsmann2024first}.
Gherkin’s structured syntax helps both technical and non-technical stakeholders collaborate effectively \cite{horina2023advantages}, clarifies feasibility, and elicits implicit needs \cite{parsa2023acceptance}. Moreover, Gherkin scenarios can be turned into executable tests, reducing misalignment between implementation and goals (\textbf{G1}), and are widely used in agile development \cite{farooq2023behavior}.
Gherkin offers the following benefits as a requirement expression language:

1) Structured description of behavior-driven scenarios.

2) A shared understanding of expected software behavior among stakeholders.

3) Different stakeholders collaboratively conceptualize software interactions to uncover implicit requirements.

Table~\ref{tab:Gherkin_example} shows Gherkin syntax (left) and the corresponding terminology used in this paper (right). Elements such as "And", "Background", and "Scenario Outline" are optional, with "And" used to extend conditions or actions as needed.
A challenge in BDD is stakeholders’ limited writing experience, as shown in Fig.\ref{fig:sec_2_fig1_b}, making it hard to write high-quality acceptance criteria \cite{santos4933832perception}. LLMs are increasingly used across the software lifecycle for their strong text/code generation capabilities \cite{10.1145/3695988}, helping reduce human effort.  Recent research such as AgileGen \cite{zhang2024empowering} has begun to explore the use of Gherkin as a bridging language between user narratives and code implementations. However, these methods rely on prompting LLMs with predefined templates, such as shown in Table~\ref{tab:Gherkin_example}, without constraining the causal relationship between preconditions and behavior action (\textbf{C1}). This leads to unclear requirement logic, as discussed in Section~\ref{sec:LLMq}.

\subsection{Clarifying Requirement Logic through Causal-Effect Graphs}
Even experienced requirement analysts often produce vague or incomplete specifications within development teams (\textbf{Q2}) \cite{northrop2006ultra}. Detecting and resolving unclear logic in the early stages of requirement design remains a key challenge in software engineering \cite{white2024chatgpt}. Causal-effect graphs (CEGs) \cite{elmendorf1970automated} provide a graphical representation of the logical relationships between input conditions (Causes) and output behaviors (Effects). Commonly used in black-box test design \cite{10155063}, CEGs use Boolean logic (AND, OR, NOT, DIR) to link precondition nodes to behavior actions. By making causal relationships explicit, CEGs help identify incompleteness and ambiguities in requirement specifications (\textbf{S2}).

\subsubsection{Definition of Causal-Effect Graphs}
 \label{subsubsection:define of CEG}
\begin{figure}[h]
	\centering
	\includegraphics[width=0.75\textwidth]{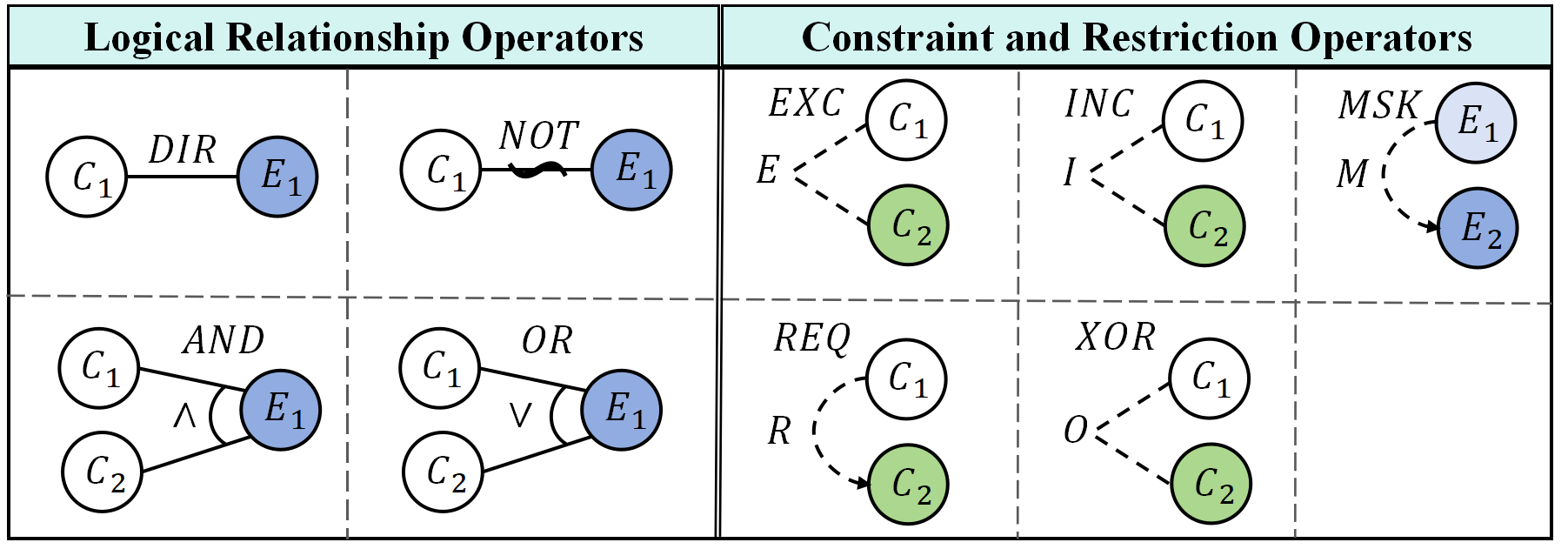}
\caption{\label{fig:CEGdef} Illustration of Logical relationship Operators and Constraint Symbols in Causal-Effect Graphs.
}
\end{figure}

A causal-effect graph (CEG) consists of three elements \cite{10051799} (see Fig.~\ref{fig:CEGdef}): nodes (Causes ($C$), Effects ($E$)), logical relationships between different node types, and constraints (restrictions) among nodes of the same type. The links between causes and effects are logical relationships, dependencies among causes are constraints, and dependencies among effects are restrictions. The operators that express these links—DIR, AND, OR, NOT—together with the constraint and restriction operators, are listed in Table \ref{tab:operaters}.

\begin{table}[ht]
  \centering
  \small
  \caption{Formal semantics and interpretations of operators}
  \label{tab:operaters}
  \begin{tabular}{@{}lll@{}}
    \toprule
    Operator & Formal definition & Interpretation \\ \midrule
    $DIR(C_1)=E_1$      & $(C_1\!\rightarrow\!E_1)\land(\lnot C_1\!\rightarrow\!\lnot E_1)$     & $E_1$ occurs iff $C_1$ occurs \\
    $AND(C_1,C_2)=E_1$  & $(C_1\land C_2\!\rightarrow\!E_1)\land(\lnot C_1\lor\lnot C_2\!\rightarrow\!\lnot E_1)$ & $E_1$ occurs iff both conditions hold \\
    $OR(C_1,C_2)=E_1$   & $((C_1\lor C_2)\!\rightarrow\!E_1)\land(\lnot C_1\land\lnot C_2\!\rightarrow\!\lnot E_1)$ & $E_1$ occurs iff at least one condition holds \\
    $NOT(C_1)=E_1$      & $(C_1\!\rightarrow\!\lnot E_1)\land(\lnot C_1\!\rightarrow\!E_1)$      & $E_1$ occurs exactly when $C_1$ does not \\
    $EXC(C_1,C_2)$      & $\lnot(C_1\land C_2)$                                                   & $C_1$ and $C_2$ cannot both be true (they may both be false) \\
    $INC(C_1,C_2)$      & $C_1 \lor C_2$                                                          & At least one of $C_1$ or $C_2$ must be true \\
    $REQ(C_1,C_2)$      & $C_1 \rightarrow C_2$                                                   & $C_2$ must hold whenever $C_1$ does \\
    $XOR(C_1,C_2)$      & $(C_1\lor C_2)\land\lnot(C_1\land C_2)$                                 & Exactly one of $C_1$ or $C_2$ is true \\
    $MSK(E_1,E_2)$      & $E_1 \rightarrow \lnot E_2$                                             & $E_2$ is forbidden when $E_1$ occurs; otherwise $E_2$ is unrestricted \\ 
    \bottomrule
  \end{tabular}
  \vspace{-0.6\baselineskip}
\end{table}

\subsubsection{CEGs Help Identify Ambiguities in Requirement Logic}
Causal-effect graphs (CEGs) can effectively identify vague or inconsistent issues in requirement specifications through clear, logical expressions. This explicit modeling helps teams quickly identify and correct ambiguities in the requirements.
Taking the requirement example shown in Fig.~\ref{fig:fig0}, identifying the atomic conditions ($C$) and effects ($E$) in the requirement.
Atomic Conditions($C$): 
$C_{AE}$: the user selects the Ancient Egypt time period; 
$C_{Ren}$: the user selects the Renaissance time period; 
$C_{IE}$: the user selects Industrial Era; 
$C_{intro}$: the user has completed the introductory tutorial; 
$C_{pre}$: the user has completed previous time periods; 
Atomic Effects($E$): 
$E_{AEChar}$: show historical characters from Ancient Egypt; 
$E_{PyraGame}$: play a mini-game related to pyramid building; 
$E_{Leon}$: learn about Leonardo da Vinci; 
$E_{PaintD}$: paint a digital masterpiece; 
$E_{InventC}$: play invention challenge; 

The mapping of scenario preconditions and behavior actions to the logical relationships in a Causal-Effect Graph (CEG) is as follows: $Scenario\ 1:
    AND(C_{AE},\ C_{intro}) = E_{AEChar}; 
    AND(C_{AE},\ C_{intro}) = E_{PyraGame}; 
    REQ(C_{AE},\ C_{intro}).$
$Scenario\ 2: 
    DIR(C_{Ren}) = E_{Leon}; 
    DIR(C_{Ren}) = E_{PaintD}.$
$Scenario\ 3:
    AND(C_{IE},\ C_{pre}) = E_{Edison}; 
    AND(C_{IE},\ C_{pre}) = E_{InventC}; 
    REQ(C_{IE},\ C_{pre}).$

\textbf{1) Identifying Inconsistent Preconditions Using CEGs.}
CEGs offer a clear and structured expression format, making it easier to spot inconsistencies by comparing variations in their logical representations. For example:
In Scenario 2, the user directly accesses the Renaissance ($C_{Ren}$) and triggers actions without any preconditions. In contrast, Scenarios 1 and 3 require preconditions completion of the tutorial ($C_{intro}$) or a prior period ($C_{pre}$) before conducting actions. 
If the stakeholder requires entry conditions for all scenarios, before entering any time period $C_i$, at least one of the following must hold: $\forall i,\ \Bigl(REQ(C_i, C_{\mathrm{intro}}) \lor REQ(C_i, C_{\mathrm{pre}})\Bigr).$ Then Scenario 2 lacks this critical constraint: $REQ(C_{Ren}, C_{intro})$ or  $REQ(C_{Ren}, C_{pre})$. The inconsistency exposed here needs to be clarified with stakeholders: "Should users be allowed direct access to the Renaissance period without completing the tutorial or a previous period?" If not, the requirement scenario should be updated to explicitly include the semantics of $REQ(C_{Ren}, C_{intro})$ or $REQ(C_{Ren}, C_{pre})$.

\textbf{2) Identifying Missing Edge Cases Using CEGs.}
CEGs help uncover missing edge cases by explicitly enumerating all input conditions ($C$) and their possible combinations, along with the corresponding output effects ($E$). When certain negated conditions appear in the CEGs but have no associated effects, this indicates missing edge scenarios in the requirement.
Example: The Gherkin requirement defines system behavior when the user selects Ancient Egypt ($C_{AE}$), the Renaissance ($C_{Ren}$), or the Industrial Era ($C_{IE}$), but does not specify what happens if no period is selected.
If we define the condition where the user selects no time period as $C_{NoSelection}$, then: $C_{NoSelection} \coloneqq \neg C_{AE} \wedge \neg C_{Ren} \wedge \neg C_{IE}$. Since $C_{NoSelection}$ has no corresponding effects in the CEGs, it indicates a missing edge case in the requirement.

\textbf{3) Identifying Missing Design Details and Evaluation Criteria Using CEGs.}
When constructing a CEG, the team maps textual requirements into specific conditions, effect nodes, and logical relationships. If specific nodes lack sufficient definition or logical constraints during modeling, it reveals issues such as missing design details or unclear evaluation criteria.
Example: In Scenario 1, the user is instructed to "play a mini-game related to pyramid building" ($E_{PyraGame}$). However, the Gherkin requirement does not specify what follows, such as whether the user earns a reward ($E_{Score}$) or can retry after failing ($E_{FailRetry}$). An intended CEG expression might be:
 $AND(C_{AE}, C_{intro}) = E_{pyraGame} \wedge {(DIR(C_{GameWin} = E_{Score}) \lor (DIR(C_{GameFail}) = E_{FailRetry}))}$
However, the requirement lacks definitions for $C_{GameWin}$ and $C_{GameFail}$, making it impossible to model these outcomes. This exposes missing design logic and undefined evaluation criteria in the requirement.

\subsubsection{Challenges in Constructing Causal-Effect Graphs (CEGs)}
While constructing CEGs offer advantages in identifying ambiguities in requirements, the manual process poses several challenges. First, identifying atomic conditions ($C$) and effects ($E$) from natural language requires business-domain knowledge and practiced requirement decomposition skills from team members \cite{jang2022automatic}. Second, representing causal relationships using Boolean logic demands logical reasoning and modeling expertise. These challenges are amplified when dealing with complex requirements, making it harder to accurately model causality during CEGs construction.
Some studies have explored automating CEGs generation using traditional machine learning \cite{krupalija2023usage} or lexical analysis techniques \cite{jang2022automatic}, but these methods are often domain-specific and lack generalizability \cite{10155063}. Additionally, the limited availability of CEGs-related datasets constrains their performance on unseen scenarios \cite{krupalija2024etf}.
In contrast, LLMs demonstrate strong capabilities in natural language understanding and reasoning, potentially reducing manual effort and improving generalization through their broad pretraining. However, ensuring that LLM-generated CEGs accurately capture the intended causal relationships is challenging (\textbf{C2}) due to the hallucination issue inherent in LLMs. Guaranteeing both the formal correctness and semantic consistency of these generated CEGs with the original requirements is therefore crucial (\textbf{G2}).

\subsection{Functional Diversity Using Feature Tree}
In software development, once initial user narratives are gathered, they are typically further refined and analyzed to define the scope of the diversity feature clearly (\textbf{Q3}) \cite{wiegers2013software}. Feature Tree is commonly used in this process to represent and manage evolving functional requirements. Typically created in the early stages of a project, a feature tree organizes system functions hierarchically, using a top-down structure that clarifies the relationship between core features and their subcomponents. This structure enables stakeholders to understand the role of each feature and facilitates clear, effective communication \cite{FeatureTree}. Due to these benefits, we argue that feature trees could support functional diversification analysis, helping teams refine and expand functionality during requirements engineering (\textbf{S3}).

\subsubsection{Definition of Feature Tree}
 \label{subsec:featuretree definition}
\begin{figure}[t]
	\centering
	\subfigure[Feature Tree.]{%
		\centering
		\includegraphics[width=0.49\textwidth]{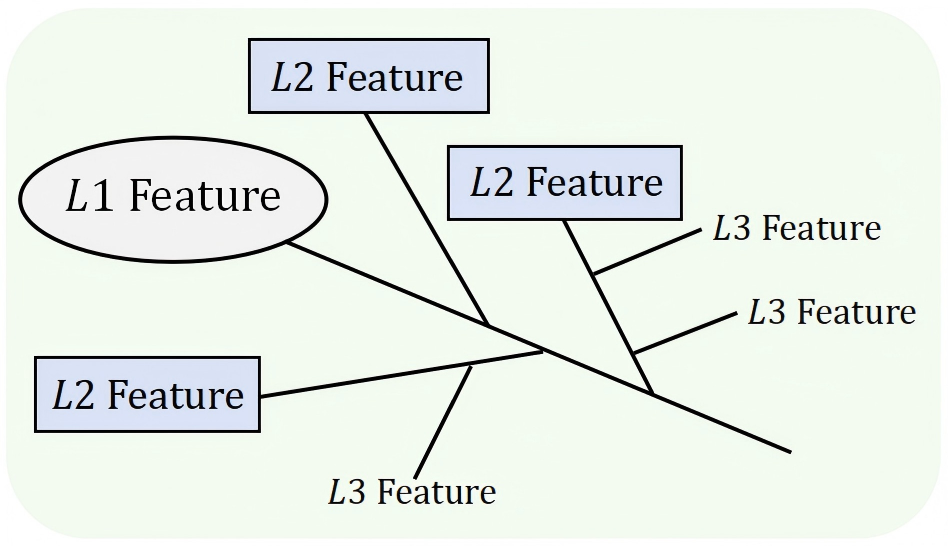}
		\label{fig:FT}
	}%
	\subfigure[Example for "Time Travel Adventure App"]{%
		\centering
		\includegraphics[width=0.49\textwidth]{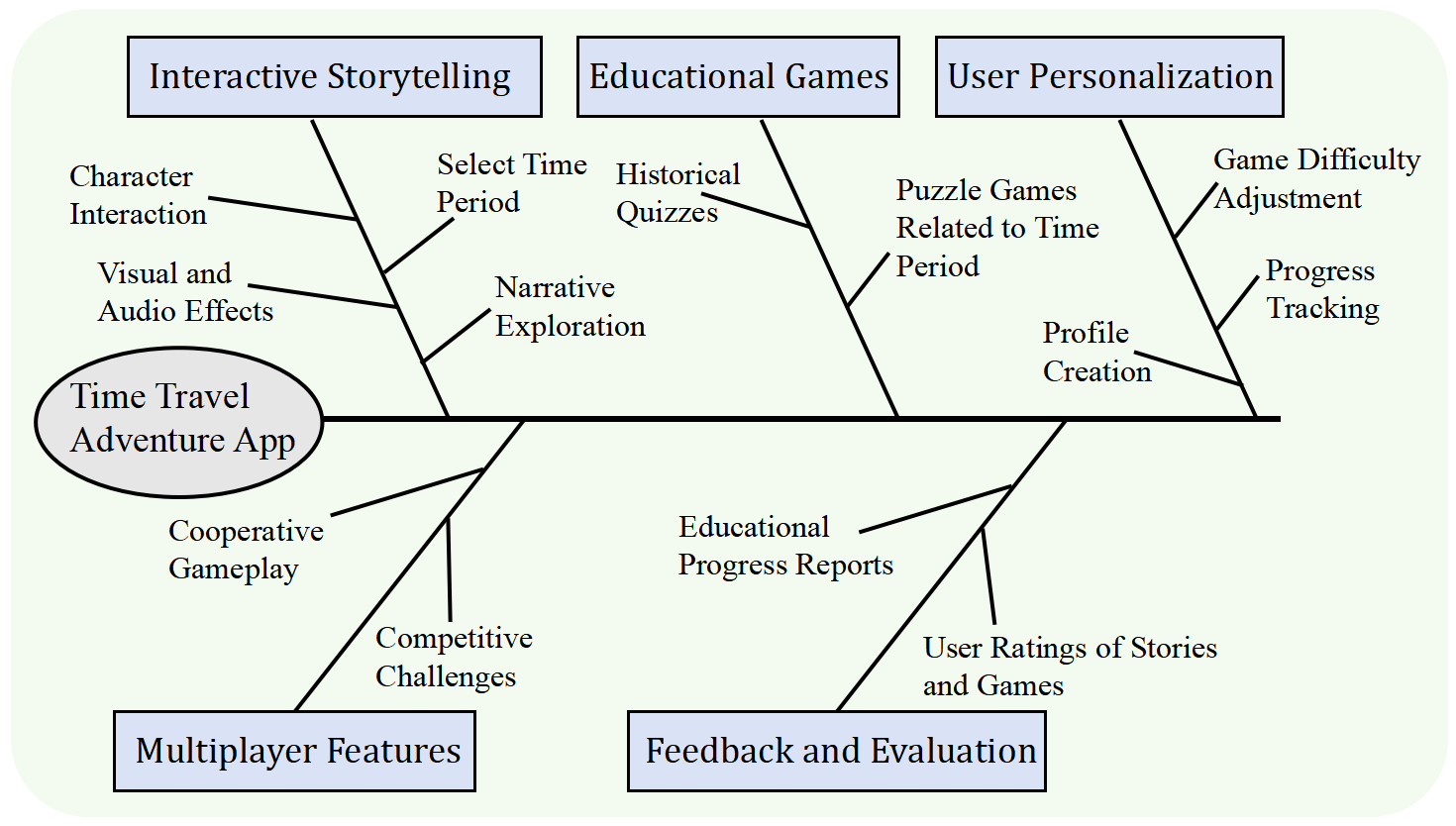}
		\label{fig:FTE}
	}
        \label{fig:featuertree}
	\caption{Feature Tree Grouping and Example for "Time Travel Adventure App"}
\end{figure}

A feature tree resembles a fishbone (Ishikawa) diagram, organizing features into logical groupings based on functional relationships. As shown in Fig.~\ref{fig:FT}, it typically consists of up to three hierarchical levels: Level 1 ($L1$), Level 2 ($L2$), and Level 3 ($L3$).
$L2$ features are sub-features of $L1$ features.
$L3$ features are sub-features of $L2$ features.
Not all feature trees include all three levels. The depth depends on the system's complexity and the desired level of detail.

\subsubsection{Feature Tree Helps Functional Diversification}
When constructing a feature tree from user requirements, the team begins by identifying the product name and then explores potential functionalities at various levels surrounding it. The hierarchical structure not only visualizes the internal relationships between features but also helps uncover missing or redundant functionalities.
As illustrated in Fig.~\ref{fig:FTE}, the partial feature tree for the "Time Travel Adventure" application shows the "tree trunk" as a central line, with an ellipse at the end representing the product name. Level 2 ($L2$) features, depicted as blue boxes, include core functionalities such as "Interactive Storytelling". These branch into Level 3 ($L3$) sub-features like "Character Interaction", which further specify and expand the core features. Through ongoing collaboration with stakeholders, the features of the project can be continually refined and expanded, enabling functional diversification.

\subsubsection{Challenges in Constructing Feature Trees}
While the hierarchical structure of feature trees supports the collaborative refinement of functional requirements among diverse stakeholders, their construction often depends on the domain expertise of the production manager. As system complexity increases, the number of nodes and branches can grow rapidly, making it difficult to maintain consistency and readability through manual methods alone.
Existing tools like the open-source FeatureIDE\footnote{\url{https://www.featureide.de/}} still rely on traditional, manual workflows, which are time-consuming and resource-intensive. In contrast, LLMs offer strong capabilities in text comprehension and planning, supported by broad pretraining \cite{achiam2023gpt}. They have the potential to enhance feature scoping, reduce manual effort, and promote functional diversity by drawing on learned knowledge and patterns (\textbf{G3}). However, to achieve this, LLMs must be able to infer logical relationships between potential features from natural language requirements, which are often unstructured and ambiguous. Bridging this gap between natural language flexibility and the formal structure of software components remains a key challenge (\textbf{C3}) in applying LLMs to automate feature tree construction.

\section{RequireCEG: A CEGs-driven Agent for Requirement Elicitation and Review}
\subsection{Overview}

\begin{figure}[t]
	\centering
	\includegraphics[width=1\textwidth]{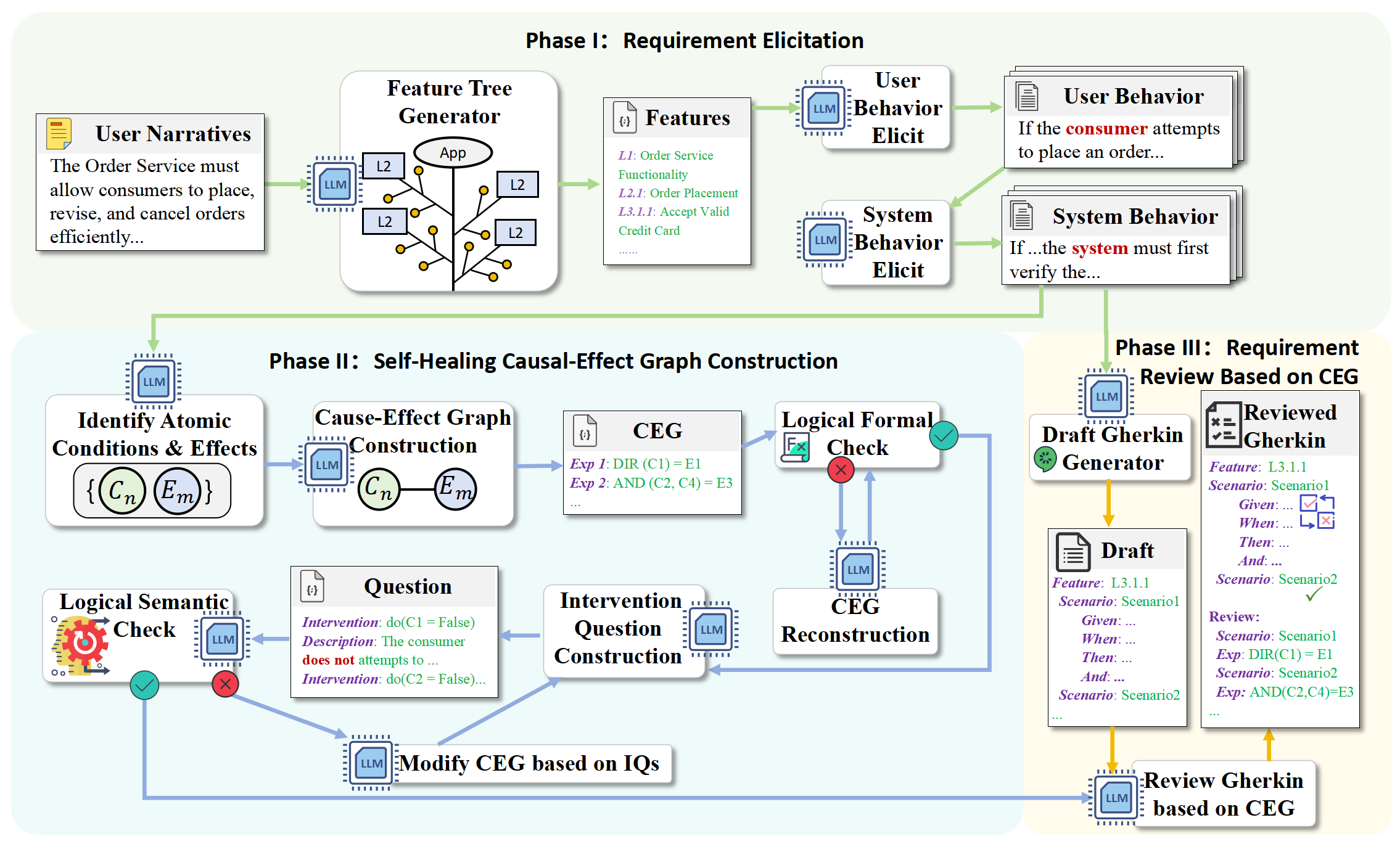}
\caption{\label{fig:Overview} RequireCEG Overview Diagram. This is a Causal-Effect Graphs (CEGs)-driven multi-agent framework that takes a user narrative as input and produces Gherkin requirements reviewed by the CEGs as output.}
\end{figure}

This section presents the framework of RequireCEG (Fig.~\ref{fig:Overview}), which enhances Large Language Models (LLMs) with capabilities similar to those of a human requirements analyst for the autonomous elicitation and review of generated requirements from user narratives. Gherkin is adopted as the requirement representation language to support practical end-user software engineering workflows.

\textbf{Phase I: Requirement Elicitation}. RequireCEG first extracts a hierarchical structure of software components from natural language narratives using a feature tree. It helps bridge the gap between the flexibility of natural-language narratives and the formal logic of software components while enhancing functional diversity. For each component, it then describes behavior from both user operation and system response perspectives to support later review.

\textbf{Phase II: Self-Healing Causal-Effect Graph Construction}. RequireCEG constructs and iteratively refines the Causal-Effect Graphs (CEGs) based on system behavior requirements, ensuring that its logical form and semantics accurately reflect the implicit causal relationships in the system behavior requirement, thereby improving reliability. 

\textbf{Phase III: Requirement Review Based on CEG}. The CEGs are then used to review the previously generated Gherkin scenarios. Any scenario that violates the CEGs logic is revised or extended accordingly. This formal causal review clarifies the relationship between preconditions and behavior actions, ensuring consistency between natural-language system behavior requirements and formal Gherkin requirements.
The full process is illustrated as algorithm~\ref{alg:multi-dim-ceg-check}:

\begin{algorithm}[h]
\footnotesize
\caption{RequireCEG (Pseudocode)}
\label{alg:multi-dim-ceg-check}
\begin{algorithmic}[1]
\Require \textit{Narratives} \quad \Comment{User Narrative}
\Ensure  \textit{ReviewedGherkin} \quad \Comment{Gherkin of Reviewed Scenarios}
\State \textbf{//Phase I: Requirement Elicitation}
\State $featureTree \gets \mathrm{FeatureTreeGenerator}(\textit{Narratives})$ \Comment{Feature Tree Generator}
\For{$feature_n \in featureTree$}
    \State $UserBehavior \gets \mathrm{AnalyzeUserBehavior}(feature_n)$   \Comment{User Behavior Elicit}
    \State $systemBehavior \gets \mathrm{AnalyzeSystemBehavior}(UserBehavior)$ \Comment{System Behavior Elicit}

    \State \textbf{// Phase II: Self-Healing Causal-Effect Graph Construction}
    \State $\{Conditions, Effects\} \gets \mathrm{IdentifyCAndE}(systemBehavior)$ \Comment{Identify Atomic Conditions \& Effects}
    \State $CEG_{feature_n} \gets \mathrm{BuildCEG}(\{Conditions, Effects\}, systemBehavior)$ \Comment{Cause-Effect Graph Construction}

    \State // Logical Formal Check Process
    \Repeat
        \State $errList \gets \emptyset$
        \State $errList \gets \mathrm{LogicalFormCheck}(CEG_{feature_n})$
        \If{$errList \neq \emptyset$} \Comment{CEG Reconstruction}
            \State $CEG_{feature_n} \gets \mathrm{ReconstructCEG}(\{Conditions, Effects\}, systemBehavior, errList)$
        \EndIf
    \Until{$errList = \emptyset$}

    \State // Logical Semantic Check Process
    \Repeat
        \State $IssueSet \gets \emptyset$
        \For{$c \in Conditions$}
            \State $IQs \gets \mathrm{ConstructIQ}(c=\text{FALSE}, \{Conditions, Effects\}, CEG_{feature_n})$ \Comment{Intervention Questions Construction}
        \EndFor
        \For{$Iq \in IQs$}
            \If{$\mathrm{ReasoningIQ}(systemBehavior, Iq) = \mathrm{TRUE}$} \Comment{Logical Semantic Check}
                \State $IssueSet \gets IssueSet \cup \{c, Iq, CEG_{feature_n}\}$
            \EndIf
        \EndFor
        \If{$IssueSet \neq \emptyset$} \Comment{Modify CEG based on IQs}
            \State $CEG_{feature_n} \gets \mathrm{ModifyCEG}(\{Conditions, Effects\}, systemBehavior, IssueSet)$
        \EndIf
    \Until{$IssueSet = \emptyset$}

    \State \textbf{// Phase III: Requirement Review Based on CEG}
    \State $draftGherkin \gets \mathrm{GenerateGherkin}(systemBehavior)$ \Comment{Draft Gherkin Generator}
    \State $ReviewedGherkin_{feature_n} \gets \mathrm{Review}(draftGherkin, CEG_{feature_n})$ \Comment{Review Gherkin based on CEG}
\EndFor

\State \textbf{return} $ReviewedGherkin$
\end{algorithmic}
\end{algorithm}

We adopt an LLM-based agent chaining framework, where each subtask, such as "FeatureTreeGenerator", "AnalyzeUserBehavior", "AnalyzeSystemBehavior", "GenerateGherkin", "IdentifyCAndE", "BuildCEG", "ReconstructCEG", "ReasoningIQ", "ModifyCEG", and "Review" are encapsulated as LLM-driven agents. Carefully designed prompts guide each agent to ensure accurate and task-specific execution.
The overall process is decomposed into modular, independently executable subtasks and composed into an integrated pipeline, forming a neuro-symbolic collaboration architecture. In this framework, the LLMs handle high-level semantic reasoning, while the Causal-Effect Graphs (CEGs) support symbolic planning and formal logic review, and the two complement each other. 
\subsection{Requirement Elicitation}
This section presents a requirement elicitation method that extracts structured requirements from user narratives, addressing the challenge of limited functional diversity in existing requirements (\textbf{Q3}).
As illustrated in Phase I of Fig.~\ref{fig:Overview}, the method bridges the gap between the flexibility of natural language and the formal logic of software components (\textbf{C3}) through hierarchical analysis. First, the user narrative is decomposed into a structured hierarchy of software components. Next, a Large Language Model (LLM) is guided to define user-operating conditions for each component from the user's perspective. Based on these conditions, the LLM is further prompted to generate corresponding system behavior actions (\textbf{M3}).
The resulting system behavior requirements serve as input for both the construction of the Causal-Effect Graphs (CEGs) and the pre-generation of draft Gherkin.

\textbf{Feature Tree Generator.}
Hierarchical extraction of user narratives into a feature tree is a key step in bridging the gap between the flexibility of natural language and the formal logic of software components. This module employs a top-down layered reasoning approach to decompose and organize potential functions implicit in the user narrative. The definition of the feature tree (Section~\ref{subsec:featuretree definition}) guides the decomposition \cite{beatty2012visual}, which serves as contextual knowledge for the LLM to ensure the specification of analysis. The refinement across three levels helps clarify functional boundaries and define the scope of software features. Notably, our feature tree generation process also takes into account various categories of user satisfaction \cite{kano1984attractive}, which classifies software functions into must-be, one-dimensional, attractive, and indifferent features. This satisfaction perspective enhances functional diversity and supports the uncovering of user needs for software, laying an input for subsequent user and system behavior requirement design.

\textbf{User Behavior Elicitation.}
Each leaf node of the feature tree represents a concrete software component and is used to elicit user behavior. This module takes as input: (1) a feature leaf node ($feature_n$) and (2) the user narrative.
By mapping each leaf node back to its corresponding segment in the narrative, we clarify the functional role of each component, helping to avoid disconnects with the user narrative during requirement elicitation.
During the implementation, this module leverages user behavior examples to guide the LLM in generating component-specific triggering conditions from the user’s perspective, for example, \textit{"if the user uploads file 'A' or 'B'..."}. These outputs clarify user intent and prepare for atomic condition extraction in CEGs construction.

\textbf{System Behavior Elicitation}
\label{sec:sys behavior}
After designing the user behavior requirements, this module proceeds to explore how the system should respond under triggering conditions in user behavior requirements. Using system behavior examples, the module adopts a condition–response mapping strategy to guide the LLM in tightly coupling user operation with system responses. For instance, when a user uploads a specific file 'A', the system may respond by displaying a notification or executing a computation task. This system behavior elicitation process not only clearly defines the relationship between triggering conditions and system responses but also facilitates the identification of atomic conditions and effects.

\subsection{Self-Healing Causal-Effect Graph Construction} 
This section presents the autonomous construction method for Causal-Effect Graphs (CEGs), designed to address unclear logic in requirements (\textbf{Q2}). As shown in Phase II of Fig.~\ref{fig:Overview}, the proposed method aims to mitigate the challenge of ensuring that CEGs generated by LLMs faithfully represent the implicit causal relationships within requirements, given the susceptibility of LLMs to generate hallucinated content (\textbf{C2}).
We start by identifying atomic conditions ($C$) and atomic effects ($E$) from the output of the System Behavior Elicitation module. These serve as the foundation for prompting the LLM to infer cause-effect relationships and construct an initial CEG.
We then perform iterative logical form checks, followed by semantic checks using the LLM's reasoning capabilities to refine the CEG. The result is a self-checked CEG (\textbf{M2}) that symbolically represents the requirements and makes their causal logic explicit, enabling the detection of ambiguities and inconsistencies early in the process.

\subsubsection{Initial Construction of the Causal-Effect Graph}
 \label{sec:CEG_construction}
This subsection explains how system behavior requirements are further mapped into the Causal-Effect Graphs (CEGs).
The construction of the CEGs involves decomposing and organizing the triggering conditions and system responses within the system behavior requirements. It consists of two key steps: 1) Identify Atomic Conditions \& Effects. 2) Cause-Effect Graph Construction.

\textbf{Identify Atomic Conditions \& Effects.}
This module extracts atomic triggering conditions ($C$) and system responses ($E$) from the system behavior requirements, focusing on identifying elements that are semantically distinct and independently meaningful. This decomposition reduces ambiguity and redundancy in the subsequent construction of the Causal-Effect Graphs (CEGs).
The extraction is made easier by the prior two-stage elicitation process: User Behavior Elicitation defines the triggering conditions for each software component. And System Behavior Elicitation specifies the system responses to each triggering condition.
With these foundations, the process of identifying atomic $C$ and $E$ elements becomes more tractable.
During the implementation, we guide the LLM using examples from an "Atomic Identify Example" prompt to identify atomic-level conditions and effects from complex system behavior requirements.

An Atomic Identify Example like: \textit{Given the system behavior requirement: "A file is considered updated if the character in the first column is "A" or "B" and the second column is a number. If the first character is wrong, the message x should be printed. If the second column is not a number, the message y should be printed.”}
The following atomic conditions and effects can be extracted:
$C_1$: The character in column 1 is "A", 
$C_2$: The character in column 1 is "B", 
$C_3$: The character in column 2 is a number, 
$E_1$: The file update is complete, 
$E_2$: Prints the message x, 
$E_3$: Print the message y.
In this way, system behavior requirements are mapped into a set of atomic elements, providing clear node definitions for the subsequent construction of the Causal-Effect Graph.

\textbf{Cause-Effect Graph Construction.}
Based on atomic conditions and actions, the Causal-Effect Graph will be constructed by choosing appropriate logical operators and constraint or restriction relationships to explicitly represent the causal relationships in system behavior requirements.
The construction focuses on two key parts: Logical relationship operators: Operators such as $AND$, $OR$, $NOT$, and $DIR$ are used to model logical patterns like "multiple preconditions → single action" or "single condition → single action." Constraint and restriction operators: To capture dependencies or conflicts among conditions or effects, constraints such as $ EXC$, $ REQ$, $ INC$, and $ XOR$ are introduced, as well as restrictions $MSK$. (see the definition of Section~\ref{subsubsection:define of CEG}). In implementing, we provide the CEGs construction examples guide for LLMs to build CEGs. A Cause-Effect Graph Example is as follows:
\begin{equation}
\begin{aligned}
    &Exp 1: AND(OR(C_1, C_2), C_3) = E_1 \quad Exp 2: NOT(C_3) = E_3 \\
    &Exp 3: NOT(OR(C_1, C_2)) = E_2 \quad Exp 4: EXC(C_1, C_2)
\end{aligned}
\end{equation}

Through the CEGs construction process, the relationships among atomic elements in system behavior requirements can be more clearly and explicitly represented. This also provides formal evidence for subsequent review of requirements.
However, due to the inherent hallucination problem in LLMs, ensuring that the generated CEGs expressions are both logically well-formed and semantically faithful to the requirements remains a challenge. Therefore, after the initial construction of the CEGs, it is necessary to perform an autonomous check of its logical format and semantics.

\subsubsection{Autonomous Logical Formal Check}
This subsection introduces the method for autonomously checking the logical formal of the initially constructed Causal-Effect Graphs (CEGs). The process consists of two steps: 1) Logic Formal Check. 2) CEGs Reconstruction. The goal is to ensure that all CEGs expressions conform to the formal definitions provided in Section~\ref{subsubsection:define of CEG}. By performing formal checking and reconstructing the graph when necessary, this process helps prevent incorrect mappings caused by syntax errors.

\textbf{Logic Formal Check.}
This phase adopts principles from Neuro-Symbolic collaboration to automatically check the syntactic correctness of CEGs logical expressions through pattern matching. Specifically, the RequireCEG agent first detects predefined logical operators (e.g., $DIR, AND, OR, NOT$) and extracts their atomic conditions and effects to verify correct usage and argument count. For constraint and restriction operators (REQ, EXC, INC, XOR, MSK), the check involves scanning for invalid symbols (e.g., misuse of =) to ensure syntactic compliance. Any expression that fails to meet the expected structure is recorded in an $Error List$ for further correction. If no errors are found, the process proceeds directly to the logical semantic check.

\textbf{CEGs Reconstruction.}
If the $Error List$ is non-empty, this phase automatically repairs the malformed CEG expressions. It begins by removing invalid expressions from the graph and then prompts the LLM to reanalyze the System Behavior Requirement and the extracted atomic conditions and effects to identify any unmet logical relationships.
Based on this analysis, the RequireCEG agent reconstructs the missing or invalid segments by selecting appropriate logical operators and atomic elements, following the CEG expression definitions in Section~\ref{subsubsection:define of CEG}. The revised CEG is then re-checked through another round of Logic Formal Check, and the process iterates until all expressions pass the formal check.
\subsubsection{Autonomous Logical Semantic Check}

This section presents a method for autonomously checking the semantic consistency between a Causal-Effect Graph (CEG) and the system behavior requirements. The approach leverages a chained LLM agent structure to simulate a causal intervention process. It involves three main steps: 1) Intervention Question (IQ) Construction. 2) Logical Semantic Check. 3) Modify CEG Based on IQ. The method utilizes the reasoning capabilities of LLMs to iteratively check and modify CEG expressions, thereby enhancing their semantic reliability.

\textbf{Intervention Question Construction}. 
This module applies the idea of causal intervention by designing a structured approach to generate intervention questions using LLMs. Initially, all condition nodes $C_n$ in the CEG are assumed to be true. Then, in each intervention step, one condition is set to false (i.e., $do(C_n = False) $). RequireCEG identifies all expressions in the CEGs affected by this intervention—denoted as $AE_n$—based on their logical dependency. Each affected expression is then mapped into a natural language intervention question $ Iq_n$, based on the semantics of its atomic elements. The result is a collection of intervention questions: $IQ = \{Iq_1, Iq_2, Iq_3, \dots, Iq_n\}$ This set serves as the foundation for reasoning checks in the next phase.

\textbf{Logical Semantic Check.}
This phase introduces a intervention-based semantic check mechanism. For each question in the intervention set $IQ$, the system prompts the LLM to answer:
\textit{"Can the system behavior requirement logically derive the behavior described by this intervention question?"}
Before giving a binary response (Yes or No), the LLM must first provide detailed reasoning, explaining the logical basis and justification for its conclusion. This process ensures transparency and trustworthiness in the check process. If inconsistencies are detected, the agent compiles an $Issue Set$ containing all semantically misaligned expressions and their corresponding intervention questions. If all intervention questions pass, the CEG is considered semantically consistent, and the checked CEG is the evidence for review Gherkin.

\textbf{Modify CEG based on IQs}. 
Upon analyzing the reasoning results from the intervention questions, RequireCEG isolates and extracts the semantically misaligned expressions and their IQs from the $Issue Set$.
If the set is non-empty, the agent revisits the semantics of atomic elements from earlier steps and applies logic corrections. These modifications aim to preserve the original intent of the system behavior requirement while satisfying the formal definitions defined in Section~\ref{subsubsection:define of CEG}.
The corrected CEG is then fed back into the Logical Semantic Check module for another validation cycle. This iterative check continues until all intervention questions pass, resulting in a semantically consistent CEG constructed from the system behavior requirements.

\subsection{Requirement Review Based on CEG}
This section introduces a CEG-based requirement review method designed to address inconsistencies between generated draft Gherkin scenarios and the system behavior logic (\textbf{Q1}). As shown in the Phase III in Fig.~\ref{fig:Overview}. To clarify the relationships between preconditions and behavior actions in the Gherkin scenario (\textbf{C1}), the RequireCEG agent performs an automated review of these drafts using the CEG (\textbf{M1}). This process ensures that the reviewed Gherkin scenarios are consistent with the original system behavior requirements, thereby strengthening Gherkin’s role as a shared language in End-User Software Engineering.

\textbf{Draft Gherkin Generator.}
In this phase, we use Gherkin-format examples (as shown in Table~\ref{tab:Gherkin_example} in section~\ref{sec:2.2}) to guide the LLM in expressing system behavior requirements with greater structural clarity. The Gherkin begins by establishing the overall intent and goal in a "Narrative." Then, a shared "Background" section defines the common initial system state, reducing repetition across scenarios. Each scenario is described using the "Given–When–Then" format to link preconditions to system behavior actions.
The generated Gherkin serves as a preliminary draft that would be reviewed and consistent with the causal logic defined in the CEG.

\textbf{Review Gherkin based on CEG.}
This module performs an automated review of Gherkin scenarios based on CEG, reviewing for consistency between each scenario and the formalized logical expressions in the graph. Specifically, the agent interprets each CEG expression and reviews the corresponding Gherkin scenarios, identifying and correcting logical mismatches within the scenarios. For causal relationships not yet covered by any scenario, the agent proactively generates new scenarios to ensure completeness.
Each scenario adheres strictly to the Given–When–Then structure, ensuring the presence of well-defined preconditions and behavior actions. This method leverages the explicit causal relationship encoded in the CEG to detect and resolve inconsistencies caused by ambiguous language, requirement misinterpretation, or reasoning errors—resulting in more semantically aligned Gherkin requirement specifications.

\section{Experimental Setup}
This section outlines the research questions (RQs), describes the datasets used in our experiments, and provides detailed implementation settings.

\subsection{Research Questions}
RequireCEG is the first requirement-elicitation and self-review neuro-symbolic collaboration framework that explicitly encodes the causal relationships between pre-conditions and behavioural actions.  As summarised in Table \ref{tab:summary}, it weaves together three complementary strategies—\textbf{S1} adopting Gherkin as a shared end-user language, \textbf{S2} a self-healing Causal-Effect Graph (CEG) that clarifies requirement logic, and \textbf{S3} feature tree guided elicitation that boosts functional diversity.
We explore the following research questions:

\textbf{RQ1}: How are RequireCEG’s Gherkin scenarios in terms of syntactic compliance and readability?

\textbf{RQ2}: How effective is RequireCEG in generating Gherkin scenario requirements?

\textbf{RQ3}: What are the main factors behind RequireCEG’s effectiveness in improving Gherkin scenario requirements?

Specifically, \textbf{RQ1} examines whether the generated Gherkin scenarios exhibit high syntactic compliance and readability, validating their potential as a shared language between users and AI agents (→ \textbf{G1}). \textbf{RQ2} leverages the INVEST principle, functional diversity entropy, and consistency metrics to evaluate the quality of the Gherkin requirements, their functional diversity, and the consistency improvements enabled by the CEG-driven self-review (→ \textbf{G3}, \textbf{G1}). \textbf{RQ3} performs ablation studies to assess the contribution of each strategy and further verifies the role of the self-healing CEG in the reliability of expressing logical relationships (→ \textbf{G2}). 

\subsection{Datasets}
The fundamental task addressed in this paper is the generation of Gherkin requirements from user narratives. The input is a natural language user narrative, and the output is a corresponding set of Gherkin requirements.
Currently, there is a lack of publicly available, high-quality datasets tailored to this task \cite{karpurapu2024comprehensive}. While some existing datasets focus on the structural or specification quality of Gherkin in open-source projects \cite{chandorkar2022exploratory}, they lack the paired narratives needed for our evaluation. As a result, they are not suitable for assessing narrative-to-Gherkin generation. To enable a comprehensive evaluation of RequireCEG, we curate two datasets: RGPair and Mini-RG.

\subsubsection{RGPair Dataset}
 The RGPair dataset consists of natural language requirements and their corresponding Gherkin requirements (i.e., Feature files). In this dataset, a single narrative may correspond to one or more Feature files.
RGPair is collected from 40 high-quality open-source repositories on GitHub. It includes a total of 413 Feature files paired with 40 natural language narratives. The names of the GitHub repositories used in this dataset are listed in Table~\ref{tab:RGPair}.

\begin{table}[htbp]
\footnotesize 
\centering
\caption{Table of repositories for the RGPair dataset.}
\label{tab:RGPair}
\resizebox{\textwidth}{!}{
\begin{tabular}{p{0.8cm}p{5.8cm}p{0.8cm}p{5.8cm}}
\hline
\textbf{ID} & \textbf{Repos Name} & \textbf{ID} & \textbf{Repos Name} \\ \hline
1  & eugenp/tutorials                                                  & 21 & gojuno/jrg                                                         \\
2  & neo4j/neo4j                                                       & 22 & Magicianred/React\_Examples                                        \\
3  & geoserver/geoserver                                               & 23 & vivekaka/testleaf\_code                                            \\
4  & apache/servicecomb-pack                                           & 24 & deGov/deGov                                                        \\
5  & microservices-patterns/ftgo-application                           & 25 & samvloeberghs/protractor-gherkin-cucumberjs-angular                \\
6  & iotaledger/iri                                                    & 26 & spirosikmd/cucumber-puppeteer-example                              \\
7  & SmartBear/soapui                                                  & 27 & sabre1041/ose-bdd-demo                                             \\
8  & w3c/epubcheck                                                     & 28 & JephWilli/tdd-bdd-final-project                                    \\
9  & aws/aws-sdk-java-v2                                               & 29 & CodemateLtd/Android-Cucumber-BDD-Sample                            \\
10 & bugsnag/bugsnag-android                                           & 30 & marinasundstrom/todo-app                                           \\
11 & blox/blox                                                         & 31 & telekom/bdd-web-app                                                \\
12 & ddd-by-examples/factory                                           & 32 & ChineDmitri/Behavior-Driven-Development\_ScrutinMajoritaireProject \\
13 & FluentLenium/FluentLenium                                         & 33 & aliraiki/bdd-project                                               \\
14 & AppiumTestDistribution/AppiumTestDistribution                     & 34 & ibm-developer-skills-network/xgcyk-tdd-bdd-final-project-template  \\
15 & mzheravin/exchange-core                                           & 35 & case451/hw3\_rottenpotatoes                                        \\
16 & iriusrisk/bdd-security                                            & 36 & naveenanimation20/LatestCucumber6WithPOM                           \\
17 & jbangdev/jbang                                                    & 37 & nyu-devops/lab-flask-bdd                                           \\
18 & rejeep/ruby-tools.el                                              & 38 & serenity-bdd/serenity-cucumber.calculator                          \\
19 & camiloribeiro/cucumber-gradle-parallel                            & 39 & serenity-bdd/serenity-cucumber.web                                 \\
20 & ShittySoft/symfony-live-berlin-2018-doctrine-tutorial             & 40 & selenium-cucumber/selenium-cucumber-java                           \\ \hline
\end{tabular}
}
\end{table}

\textbf{Dataset Collection.}
We began by filtering a high-quality dataset originally curated for studying the specification quality of Gherkin in open-source projects \cite{chandorkar2022exploratory}. Specifically, we excluded repositories that contained over 100 Feature files, as these are often used for teaching and lack internal relevance, unable to form a system narrative. 
Additionally, we removed two repositories whose Feature files contained extensive implementation-level code, as RequireCEG focuses on the business level, not the implementation.
After filtering, we retained 17 repositories and further validated their Feature files, removing duplicates and semantically meaningless scenarios.
We then expanded our dataset by collecting open-source projects hosted on GitHub. Using the search term "language:Gherkin" and sorting results by star count, we retrieved approximately 11.2k repositories (as of March 25, 2025). We manually reviewed those with more than 20 stars to identify repositories that used Gherkin as functional criteria. To narrow the search to behavior-driven development (BDD) projects, we also used the keyword "BDD driven project".
After cross-verifying the results and filtering out false positives, we successfully collected 13 additional repositories, contributing 110 more Feature files to our dataset.

\textbf{Dataset Construction.}
Our task requires paired Feature files and natural language narratives, which are generally unavailable in GitHub open-source projects. To overcome this limitation, we employed ChatGPT~\footnote{\url{https://chatgpt.com/}} to generate a concise project-level summary narrative based on the content of Feature files.
The prompt used to instruct ChatGPT to generate project narratives from Feature files is shown below:

\begin{tcolorbox}[
    colback=white,           
    colframe=black!75!white,   
    rounded corners,           
    boxrule=1pt,              
    top=2pt, bottom=2pt,       
    left=3pt, right=3pt        
]
You're a requirements analyst, and you're very good at summarizing the Gherkin scenario into user narrative.

Gherkin Features: [\{ Gherkin features Replacement Flag \}]

Summarize the feature content as a user narrative based on the "Gherkin Features".

\end{tcolorbox}

The placeholder [\{ Gherkin features Replacement Flag \}] is used as a substitution token, which is replaced at runtime with the full content of all Feature files from a given repository.
During dataset construction, we retained the data structure format defined in \cite{chandorkar2022exploratory} while extending it by adding a new keyword at the end of each repository entry:
"narrative" which stores the summarized project narratives generated from the Feature files.

\subsubsection{Mini-RG Dataset}

To evaluate RequireCEG on authentic narratives, we additionally collected the Mini-RG dataset. This dataset is derived from a recent study that explores the generation of test scenarios and cases using GPT-3.5 \cite{adu2024artificial}, where the authors fine-tuned the model using a small set of curated examples. Within their fine-tuning data~\footnote{\url{https://github.com/Gilbert136/Test-scenario-and-case-generation-with-AI/blob/master/training_data/fine-tuning-data-1719165598971.json}}, we identified several data pairs that align well with our task. From this resource, we collected a total of 12 data instances, each consisting of a natural language user requirement and its corresponding Gherkin Feature file.

\subsection{Implementation Details}

 Our proposed method, RequireCEG, is built upon GPT-4o-mini as the base model, chosen for its broad applicability and cost-effectiveness within the OpenAI family. To ensure stability in the automated framework, we set the temperature parameter, which controls output randomness, is set to 0.5. RequireCEG involves an iterative process during CEG validation and correction. To prevent excessive loops, we cap the maximum number of iterations at 5. Additionally, since some baseline methods are not originally designed to generate Gherkin, we enhance their task prompts by including Gherkin examples. This adjustment aims to improve their ability to produce Gherkin outputs and ensures a fairer comparison in our experiments.

\section{Experimental Design and Results}
\subsection{RQ1: How are RequireCEG’s Gherkin scenarios in terms of syntactic compliance and readability?}
\label{RQ1}
\subsubsection{\textbf{Motivation}}

RequireCEG is designed to autonomously elicit and review requirements, ultimately producing well-structured Gherkin scenarios. The quality of these scenarios directly affects the correctness and readability of the requirement specification. Therefore, evaluating the syntactic correctness and semantic clarity of the Gherkin scenarios generated by RequireCEG is essential. This ensures that various stakeholders can accurately interpret Gherkin and reduces the risk of misunderstanding during understanding.

\subsubsection{\textbf{Methodology}}

\label{sec:characters_syn}
To explore RQ1, we will introduce the approach step by step according to a. Evaluation Metrics, b. Baselines, and c. RQ Experimental Setup.
\paragraph{\textbf{a. Evaluation Metrics.}} 
This section evaluates the syntactic and semantic quality of the Gherkin scenarios generated by RequireCEG using three categories of metrics: Gherkin keyword statistics, syntactic accuracy, and readability indicators.

\textbf{Gherkin Keyword Statistics.}
This set of metrics measures the structural characteristics of the generated Gherkin across different methods. It includes the average number of "Features" per project (F@Num). Average feature length in lines (Avg\_Loc). The number of scenarios per "Feature" (F@Sce). Keyword frequency counts for Given (Key@Given), When (Key@When), Then (Key@Then), and Examples (Key@Examples)
\textbf{Syntactic Accuracy (Acc@Syn)~\cite{karpurapu2024comprehensive}.}
This metric measures the proportion of Feature files that are fully syntactically correct. We use the tool gherkin-lint~\footnote{\url{https://github.com/gherkin-lint/gherkin-lint}}, which validates each generated Feature against 35 predefined Gherkin syntax rules.
Let $|F|$ be the total number of generated Feature files for a project. Acc@Syn is defined as:

\[
\operatorname{Acc@Syn}
= \frac{1}{|F|}\,
  \sum_{i=1}^{|F|}
  \delta\bigl(s_i\bigr),
\qquad
\delta(s_i)=
  \begin{cases}
    1, & \text{if Feature } i \text{ has no syntax errors}, \\[6pt]
    0, & \text{otherwise}.
  \end{cases}
\]

\textbf{Readability~\cite{eleyan2020enhancing}.}
To compare the readability of the generated Gherkin requirements for users, we evaluate their readability using two established metrics: the Gunning Fog Index, which focuses on sentence length and the proportion of complex words, to assess how easily a text can be understood. The Linear Write Index~\footnote{\url{https://en.wikipedia.org/wiki/Linsear_Write}}, which is designed for technical documents, places more emphasis on syllable weighting to evaluate text clarity.
Both metrics return scores in terms of U.S. education grade levels, where lower values indicate that fewer years of formal education are needed to read the text fluently, implying higher readability.

\paragraph{\textbf{b. Baselines.}}
\label{sec:baselines}

We compare RequireCEG against six state-of-the-art requirement generation methods on both the RGPair and Mini-RG datasets. The baselines include AgileGen, MetaGPT, RequireLite, Chain-of-Thought (CoT), GPT-o3-mini, and Gemini-2.5-Pro.

\begin{itemize}
 \item \textbf{AgileGen}~\footnote{\url{https://huggingface.co/spaces/HarrisClover/AgileGen}}~\cite{zhang2024empowering} is a state-of-the-art Human-AI collaborative framework for requirement elicitation. It uses Gherkin to convert vague user requirements into clear, testable scenario descriptions. We evaluate its \textit{Scenarios Design} module in a fully automated setup, excluding memory pool reuse and human-in-the-loop decisions, for a fair comparison.
\item \textbf{MetaGPT}~\footnote{\url{https://mgx.dev/}}~\cite{hong2023metagpt} represents a leading multi-agent framework for software development, simulating roles such as product managers, architects, and project managers. We evaluate the Product Requirement Document (PRD) produced by its product manager agent, denoted as \textbf{MetaGPT (PRD)}. Additionally, to reduce noise from unrelated content, we isolate and evaluate the "User Story" sections only, denoted as \textbf{MetaGPT (Story)}.
\item \textbf{RequireLite} is a simplified agent variant of RequireCEG, also based on the GPT-4o-mini model. It directly generates Gherkin from the user narrative and analyzes whether there is an unclear logical relationship between user behavior and system response within scenarios, revising them accordingly. We compare the revised output of RequireLite with the full RequireCEG pipeline.
\item \textbf{Chain-of-Thought (CoT)}~\cite{wei2022chain} is a widely adopted prompting strategy that guides large language models (LLMs) to reason step-by-step. We apply CoT prompting to GPT-4o-mini by appending "Let’s think step by step." to the instruction to enhance reasoning before generating Gherkin scenarios.

\item \textbf{Reasoning LLMs}~\footnote{\url{https://platform.openai.com/docs/guides/reasoning?api-mode=chat}}: These are advanced large language models designed for autonomous high-level reasoning. They feature built-in chains of thought that enable them to analyze and respond to user queries deeply. Our experiments include the following models:
    \begin{itemize}
     \item \textbf{GPT-o3-mini}~\footnote{\url{https://openai.com/index/openai-o3-mini/}}: Released by OpenAI in 2025. It is designed to "think more carefully" when facing complex tasks and represents one of the most advanced reasoning capabilities among OpenAI’s models.
     \item \textbf{Gemini-2.5-Pro (Preview 05-06)}~\footnote{\url{https://aistudio.google.com/prompts/new_chat}}: Introduced by Google, this model excels in high-level reasoning tasks and achieves state-of-the-art performance on several benchmark tests. As of May 30, 2025, it ranks first on the LMArena leaderboard~\footnote{\url{https://huggingface.co/spaces/lmarena-ai/chatbot-arena-leaderboard}}, making it one of the most advanced reasoning LLMs currently available.
     \end{itemize}
 \end{itemize}

\paragraph{\textbf{c. RQ Experimental Setup.}}
In the RGPair dataset, we use the value with the key "narrative" as input, while in the Mini-RG dataset, we use the value of "narrative" as input. To fairness across all methods, we apply the same \textit{Gherkin Example} prompt template used in our approach to those that do not inherently generate Gherkin scenarios, such as CoT and reasoning LLMs, guiding them towards Gherkin-style output. Importantly, we ensure that all methods are executed fully automatically, from input to output, without any human intervention during the process. 

\subsubsection{Result Analysis.}

\begin{table}[thbp]
  \caption{Performance comparison on RGPair and Mini-RG datasets (higher $\uparrow$ / lower $\downarrow$ is better).}
  \label{tab:syn_read}
  \centering
  \large
  \begin{adjustbox}{max width=\linewidth}
    \begin{tabular}{@{}lcccc|cccc@{}}
      \toprule
      \multirow{2}{*}{\textbf{Method}} &
        \multicolumn{4}{c|}{\textbf{RGPair Dataset}} &
        \multicolumn{4}{c}{\textbf{Mini-RG Dataset}} \\
      \cmidrule(lr){2-5}\cmidrule(lr){6-9}
        & \textbf{F@Num}\,$\uparrow$ & \textbf{Acc@Syn}\,$\uparrow$ & \textbf{Gunning Fog}\,$\downarrow$ & \textbf{Linsear Write}\,$\downarrow$
        & \textbf{F@Num}\,$\uparrow$ & \textbf{Acc@Syn}\,$\uparrow$ & \textbf{Gunning Fog}\,$\downarrow$ & \textbf{Linsear Write}\,$\downarrow$ \\
      \midrule
      AgileGen           & 1.000 & 95.00\% & $7.09\!\pm\!1.610$ & $5.37\!\pm\!1.197$ & 1.000 & \textbf{100.00\%} & $6.71\!\pm\!2.137$ & $4.43\!\pm\!1.184$ \\
      MetaGPT(PRD)       & \underline{4.250} & / & $7.75\!\pm\!0.642$ & $6.77\!\pm\!1.382$ & \underline{4.000} & / & $7.81\!\pm\!0.770$ & $5.73\!\pm\!0.672$ \\
      MetaGPT(Story)     & \underline{4.250} & / & $14.37\!\pm\!2.468$ & $14.62\!\pm\!2.208$ & \underline{4.000} & / & $13.63\!\pm\!1.734$ & $14.66\!\pm\!1.836$ \\
      RequireLite        & 1.000 & 97.50\% & $7.31\!\pm\!1.624$ & \underline{$4.86\!\pm\!0.798$} & 1.000 & \textbf{100.00\%} & $5.83\!\pm\!0.907$ & \underline{$4.41\!\pm\!0.820$} \\
      CoT                & 1.300 & 88.46\% & \underline{$6.88\!\pm\!1.656$} & $5.46\!\pm\!1.168$ & 1.167 & \underline{92.86\%} & $\mathbf{5.26\!\pm\!0.907}$ & $4.52\!\pm\!1.246$ \\
      GPT-o3-mini        & 1.325 & 90.57\% & $7.94\!\pm\!2.039$ & $6.11\!\pm\!1.193$ & 1.000 & 91.67\% & \underline{$5.63\!\pm\!0.709$} & $4.69\!\pm\!0.925$ \\
      Gemini-2.5-Pro     & 3.550 & \underline{97.89\%} & $7.04\!\pm\!1.154$ & $5.14\!\pm\!0.956$ & 1.167 & \textbf{100.00\%} & $5.95\!\pm\!0.858$ & $4.68\!\pm\!0.881$ \\
      \textbf{RequireCEG} & \textbf{16.925} & \textbf{99.85\%} & $\mathbf{6.47\!\pm\!1.041}$ & $\mathbf{4.53\!\pm\!0.495}$
                         & \textbf{13.083} & \textbf{100.00\%} & $5.68\!\pm\!0.451$ & $\mathbf{4.40\!\pm\!0.481}$ \\
      \bottomrule
    \end{tabular}
  \end{adjustbox}
\end{table}

\begin{figure*}[h]
	\centering
	\includegraphics[width=0.96\textwidth]{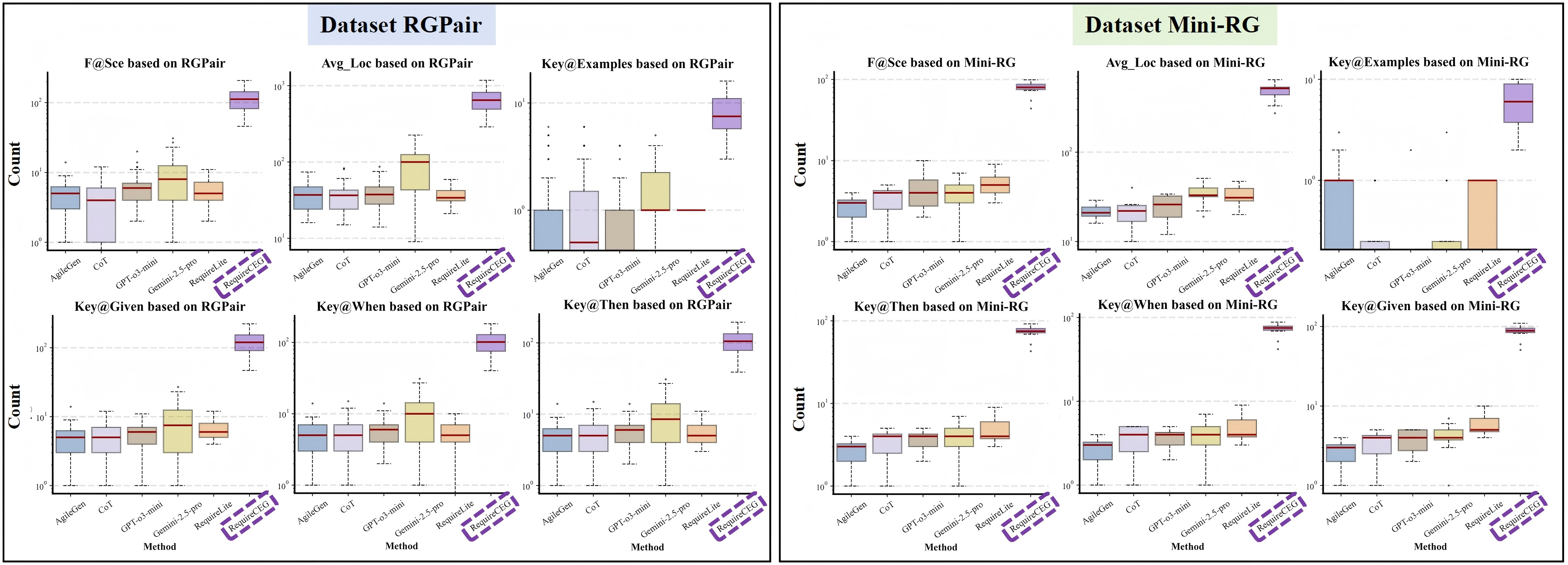}
\caption{\label{fig:Gherkin_Characters} Gherkin keyword Statistics Across Methods. The vertical axis uses a logarithmic scale, and the horizontal axis represents different methods.}
\end{figure*}

\textbf{1. Analysis of Gherkin Keyword Statistics.}
Across both datasets, we compare six methods for generating Gherkin scenarios from input narratives using structural keyword metrics. As shown in Table~\ref{tab:syn_read}, \textbf{F@Num} measures the average number of Features per project—each representing a distinct functionality. RequireCEG generates the most Features on average, which is attributed to its feature tree module that systematically plans functional components. Fig.~\ref{fig:Gherkin_Characters} shows other keyword statistics (on a logarithmic scale), where RequireCEG (purple) consistently outperforms others and shows less variation, indicating more stable outputs.
Overall, RequireCEG produces more comprehensive and consistent Gherkin scenarios, generating at least 11 times more scenarios than other methods.

\textbf{2. Syntactic Accuracy Analysis.}
As shown in Table~\ref{tab:syn_read}, although RequireCEG generates significantly more Gherkin content than other methods, it still maintains a high level of syntactic accuracy at 99.85\%. Among all syntax errors, the most frequent type is "no-multiline-steps" (50.8\%), followed by "unexpected-error" and "no-files-without-scenarios", caused mainly by missing keywords during generation.

\textbf{3. Semantic Readability Analysis.}
To explore whether this increase in content affects semantic clarity. As reported in Table~\ref{tab:syn_read}, RequireCEG achieves comparable scores to other Gherkin-generation methods on both the \textbf{Gunning Fog Index} and \textbf{Linsear Write}. It even achieves the best readability scores on the RGPair dataset. In contrast, MetaGPT performs worse on readability, primarily due to its use of longer, more complex sentences in user story descriptions. This further supports the idea that using Gherkin as a shared language helps reduce the cognitive load required to understand requirements.

\begin{tcolorbox}[
    colback=gray!10,           
    colframe=black!75!white,   
    rounded corners,           
    boxrule=1pt,               
    top=2pt, bottom=2pt,       
    left=3pt, right=3pt        
]
\textbf{Findings in RQ1}: The Gherkin requirements generated by RequireCEG contain at least 11 times more content than those produced by existing SOTA methods. Despite this larger scale, RequireCEG maintains high syntactic accuracy and achieves a readability equivalent to that of around a 5th-grade education, indicating that increased content volume does not negatively impact the clarity or correctness of the output.
\end{tcolorbox}

\subsection{RQ2: How effective is RequireCEG in generating Gherkin scenario requirements?}

\label{RQ2}
\subsubsection{\textbf{Motivation}}
RequireCEG is designed to autonomously elicit and review requirements, utilizing Causal-Effect Graphs to identify and resolve unclear logic, thereby enhancing both functional diversity and consistency in the resulting requirements. To evaluate the effectiveness of RequireCEG in generating high-quality Gherkin scenarios, we apply the INVEST principle and compare the results with those of existing state-of-the-art methods using metrics for requirement quality, functional diversity, and consistency.

\subsubsection{\textbf{Methodology}}
\label{sec:invest_fundiv}
To explore RQ2, we will introduce the approach step by step according to a. Evaluation Metrics, b. Baselines, and c. RQ Experimental Setup.

\paragraph{\textbf{a. Evaluation Metrics.}} 
We evaluate the effectiveness of different methods in generating Gherkin requirements using three criteria: the INVEST principle, functional diversity, and consistency metrics.

\textbf{INVEST Principle}~\cite{eijkel2024exploring}:
The INVEST principle is a widely used guideline in Agile development for assessing the quality of user stories. It was introduced by Bill Wake in 2003~\footnote{\url{https://xp123.com/invest-in-good-stories-and-smart-tasks/}}, with each letter representing a key characteristic: \textbf{I}ndependent, \textbf{N}egotiable, \textbf{V}aluable, \textbf{E}stimatable, \textbf{S}mall, and \textbf{T}estable.
To evaluate how well the generated Gherkin scenarios adhere to these six qualities, we use GPT-4.1-mini as an automated evaluator. Following the three-level scoring strategy proposed in~\cite{xing2025prompt}, each dimension is rated as~\footnote{\url{https://openai.com/index/improving-model-safety-behavior-with-rule-based-rewards/}}: ideal (3 grade), less\_good (2 grade), unacceptable (1 grade). The evaluation criteria for each INVEST dimension are defined as follows:
\begin{itemize}
    \item \textbf{Independent}: \textit{Unacceptable}: The scenario heavily depends on other scenarios and cannot be described or tested independently of them. \textit{Less\_good}: The scenario has partial independence, but some key prerequisites or functions depend on other scenarios. \textit{Ideal}: The scenario has complete triggering conditions and clear boundaries, allowing it to be understood, executed, and validated independently.
    \item \textbf{Negotiable}: \textit{Unacceptable}: The scenario places too much emphasis on technical and data details, leaving no room for flexible adjustments. \textit{Less\_good}: The scenario’s core objective is clear, but only a limited scope is left for negotiation or iterative improvements. \textit{Ideal}: The scenario clearly defines its core objective while allowing considerable freedom in technical choices and detail adjustments. 
    \item \textbf{Valuable}: \textit{Unacceptable}: The scenario’s business value is unclear, offering no solid justification or necessary purpose. \textit{Less\_good}: The scenario has some potential value, but it is not sufficiently concrete or requires further clarification and confirmation. \textit{Ideal}: The scenario addresses core user needs or business goals, demonstrating significant, verifiable value.
    \item \textbf{Estimable}: \textit{Unacceptable}: The scenario is too vague or overly complex, lacking sufficient details for effective estimation. \textit{Less\_good}: The scenario has a preliminary implementation path, but certain uncertainties remain, requiring more detail to refine the estimate. \textit{Ideal}: The scenario is clearly and specifically described, with a well-defined implementation path that allows accurate estimation based on the available information.
    \item \textbf{Small}: \textit{Unacceptable}: The scenario is overly complex, incorporating multiple large or independent functionalities, and cannot be completed in a short cycle. \textit{Less\_good}: The scenario’s scope is somewhat controlled and potentially completable in a short time but still has room for further subdivision. \textit{Ideal}: The scenario precisely describes a single function or specific goal, making it easy to implement and deliver quickly.
    \item \textbf{Testable}: \textit{Unacceptable}: Lacks clear acceptance criteria or testing scenarios, making it difficult to determine whether requirements are met. \textit{Less\_good}: Provides basic test conditions, but acceptance criteria remain partially unclear or incomplete. \textit{Ideal}: Offers explicit, detailed acceptance criteria and testing standards, enabling quick and straightforward validation of results.
\end{itemize}

\textbf{Functional Diversity (Fundiv):}
We measure the diversity of functionality in generated requirements using the concept of information entropy~\cite{shannon1948mathematical}. A higher entropy score indicates a more diverse distribution of functional types, while a lower score suggests that the functions are concentrated within a narrow scope.
\begin{equation}
    H(X)=-\sum_{i=1}^n p({x_i}) \log_2p({x_i})
\end{equation}
Here, $x_i$ denotes the $i$-th category of functionality. We use the FURPS model~\footnote{\url{https://en.wikipedia.org/wiki/FURPS}} (Functionality, Usability, Reliability, Performance, Supportability) to define functional categories, following the approach in ~\cite{puspita2024analysis}. A few-shot fine-tuned large language model is used to classify the generated functionalities into these categories.
$p({x_i})$ represents the proportion of functionality instances that fall under the category $x_i$.

\begin{table}[ht]
\caption{The grading scale used for evaluating the requirement along with their interpretations~\cite{krishna2024using}.}
\label{tab:Likert-5}
\centering\footnotesize
\begin{tabular}{|c|p{12cm}|}
\hline
\textbf{Rating} & \textbf{Interpretation} \\ \hline
1 & \textbf{Strongly Disagree}: Falls far below the expected standards for the particular parameter being evaluated. \\ 
2 & \textbf{Disagree}: Requires significant improvement to meet the expected standards. \\ 
3 & \textbf{Neutral}: Meets the expected standards for the particular parameter being evaluated, but the document misses some details. \\ 
4 & \textbf{Agree}: Generally meets or slightly exceeds the expected standards with minor areas for improvement. \\ 
5 & \textbf{Strongly Agree}: Excellent and fully meets or exceeds the expected standards for the parameter being evaluated. \\ \hline
\end{tabular}
\end{table}

\textbf{Consistency:}
Our consistency assessment includes two sub-metrics, Internal Consistency and Business Consistency, which respectively measure logical consistent within the requirements and alignment with business goals. We adopt an LLM-as-a-Judge strategy~\cite{gu2024survey} for automated evaluation, using a 5-point Likert scale (1 - Strongly Disagree, 5 - Strongly Agree). The explanation of each Likert score level is provided in Table~\ref{tab:Likert-5}. Definitions of the two metrics are as follows~\cite{krishna2024using}:
\begin{itemize}
    \item \textbf{Internal Consistency (Consis@In)}: A requirement is internally consistent if and only if no subsets of individual requirements are logically contradictory.
    \item \textbf{Business Consistency (Consis@Bus)}: These functional descriptions in the requirements document closely align with and support the established business goals, and there are no requirements that deviate from or conflict with these goals.
\end{itemize}

\paragraph{\textbf{b. Baselines.}}
 As described in Section~\ref{sec:baselines}, we compare RequireCEG with six state-of-the-art requirement generation methods on both the RGPair and Mini-RG datasets.

\paragraph{\textbf{c. RQ Experimental Setup.}}
 For requirement quality evaluation using the LLM-as-a-Judge strategy, we follow best practices from \cite{gu2024survey} to ensure reliability:
\textbf{1) Robust Evaluation Protocol}: Each project is independently scored in 3 rounds using GPT-4.1-mini, and the final score is determined by majority voting. In the event of a tie, a more advanced evaluator (GPT-4.1) is used to break the tie through independent judgment.
\textbf{2) Improved LLM Task Understanding}: We clearly define evaluation criteria and shuffle input order during evaluation to mitigate positional bias.
\textbf{3) Structured Output Format}: The LLM is instructed to return results in JSON format, along with a justification for each judgment to ensure transparency.
We follow the same process when classifying functions for the functional diversity metric.

\subsubsection{\textbf{Result Analysis.}}

\begin{table}[ht]
  \caption{Performance comparison on RGPair and Mini-RG datasets (higher is better).}
  \label{tab:test1}
  \centering
  \small                
  \begin{adjustbox}{max width=\linewidth} 
    \begin{tabular}{@{}lcccc|cccc@{}}
      \toprule
      \multirow{2}{*}{\textbf{Method}} &
        \multicolumn{4}{c|}{\textbf{RGPair Dataset}} &
        \multicolumn{4}{c}{\textbf{Mini-RG Dataset}} \\
      \cmidrule(lr){2-5}\cmidrule(lr){6-9}
        & \textbf{INVEST}\,$\uparrow$ & \textbf{Fundiv}\,$\uparrow$ & \textbf{Consis@bus}\,$\uparrow$ & \textbf{Consis@In}\,$\uparrow$
        & \textbf{INVEST}\,$\uparrow$ & \textbf{Fundiv}\,$\uparrow$ & \textbf{Consis@bus}\,$\uparrow$ & \textbf{Consis@In}\,$\uparrow$ \\
      \midrule
      AgileGen           & $2.50\!\pm\!0.298$ & 0.634 & 3.425 & 3.300 & $2.50\!\pm\!0.391$ & 0.414 & 2.833 & 3.250 \\
      MetaGPT(PRD)       & $2.39\!\pm\!0.466$ & \underline{0.933} & 3.500 & 3.475 & $2.35\!\pm\!0.343$ & \underline{0.836} & 3.333 & 3.583 \\
      MetaGPT(Story)     & $2.36\!\pm\!0.353$ & \underline{0.933} & 3.125 & \underline{4.025} & $2.36\!\pm\!0.282$ & \underline{0.836} & 2.917 & \underline{3.667} \\
      RequireLite        & $2.55\!\pm\!0.305$ & 0.619 & 3.125 & 3.525 & $2.47\!\pm\!0.344$ & 0.414 & \underline{3.417} & 3.417 \\
      CoT                & $2.52\!\pm\!0.232$ & 0.650 & 3.350 & 3.425 & $2.49\!\pm\!0.410$ & 0.372 & 3.083 & 3.250 \\
      GPT-o3-mini        & $2.54\!\pm\!0.235$ & 0.641 & \underline{3.575} & 3.525 & \underline{$2.50\!\pm\!0.354$} & 0.414 & 3.167 & \underline{3.667} \\
      Gemini-2.5-Pro     & \underline{$2.58\!\pm\!0.232$} & 0.742 & 3.500 & 3.575 & $2.46\!\pm\!0.288$ & 0.372 & 3.333 & \underline{3.667} \\
      \textbf{RequireCEG} &
        $\mathbf{2.80\!\pm\!0.165}$ & \textbf{1.417} & \textbf{4.450} & \textbf{4.475} &
        $\mathbf{2.78\!\pm\!0.251}$ & \textbf{1.371} & \textbf{4.750} & \textbf{4.667} \\
      \bottomrule
    \end{tabular}
  \end{adjustbox}
\end{table}

\begin{figure}[t]
	\centering
	\includegraphics[width=0.9\textwidth]{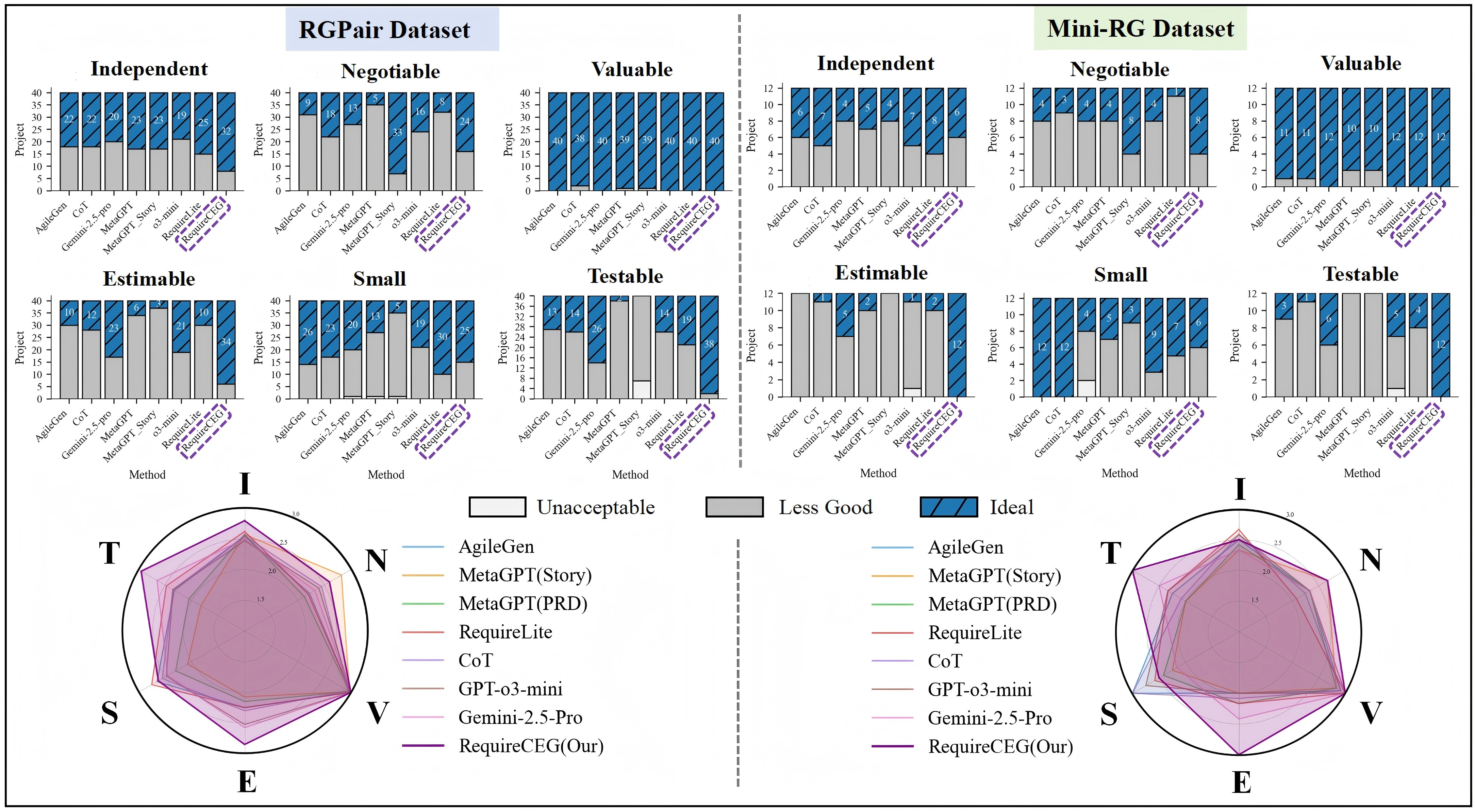}
\caption{\label{fig:invest} Stacked bar chart and radar chart showing the detailed ratings of different methods across the INVEST principle.}
\end{figure}

\begin{figure}[t]
	\centering
	\subfigure[Diversity comparisons based on RGPair dataset.]{%
		\centering
		\includegraphics[width=0.45\textwidth]{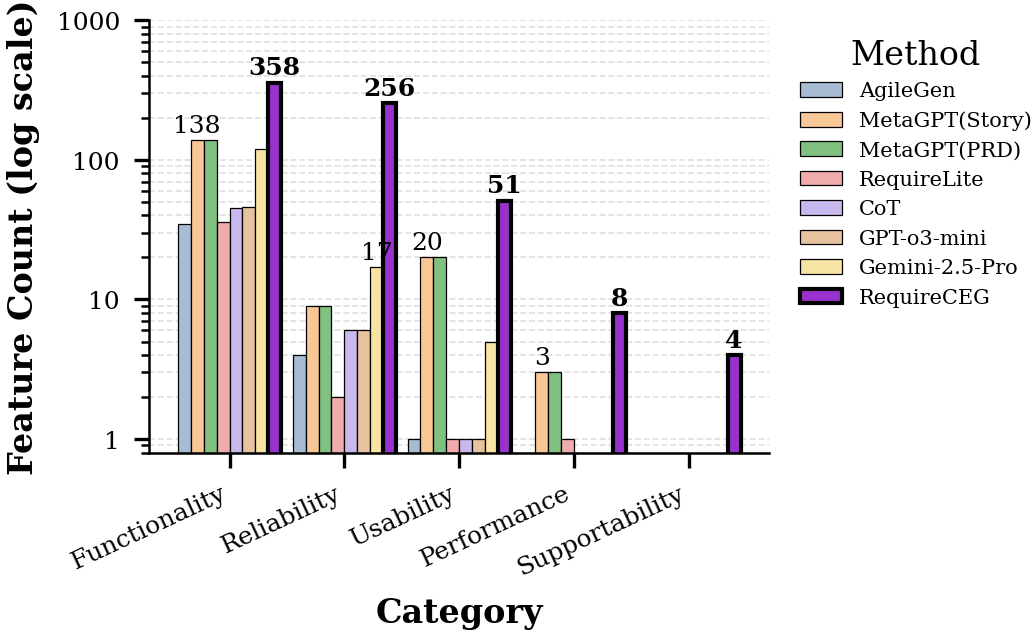}
		\label{fig:diver_total_d1}
	}%
	\subfigure[Diversity comparisons based on Mini-RG dataset.]{%
		\centering
		\includegraphics[width=0.45\textwidth]{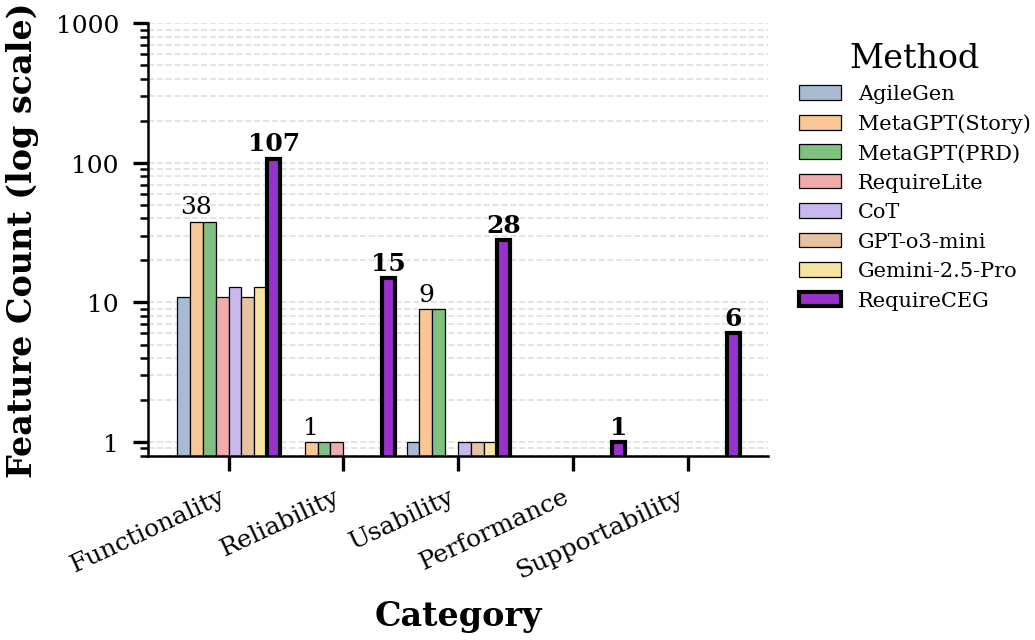}
		\label{fig:diver_total_d2}
	}
	\caption{Comparison of the number of features across different methods by category. The horizontal axis represents functional categories based on the FURPS model, and the vertical axis shows the feature count on a logarithmic scale.}
    \label{fig:diver_total}
\end{figure}
\textbf{1. INVEST Metrics Analysis.}
We evaluate the quality of Gherkin requirements generated by RequireCEG and six baseline methods on the RGPair and Mini-RG datasets using the INVEST principle. Each metric is rated on a scale from 1 to 3, and Table~\ref{tab:test1} shows the mean and standard deviation across the six INVEST metrics. RequireCEG achieves the highest average INVEST score, outperforming the second-best method by 14.8\% on RGPair and 8.5\% on Mini-RG. It also shows lower variance, indicating consistent performance across all six criteria. As visualized in Fig.~\ref{fig:invest}, RequireCEG (dashed line) performs excels in "Estimable", "Testable," and "Valuable" due to its structured and detailed functional planning. It also scores higher in "Negotiable", indicating that the generated Gherkin scenarios leave more flexibility in implementation while still maintaining meaningful constraints. 
In contrast, MetaGPT (Story) shows high "Negotiable" but lacks "Testable" and "Estimable." due to its natural language descriptions.
RequireCEG also slightly outperforms others in "Small" and "Independent". While its higher number of generated scenarios may introduce complexity, the results remain well-balanced across metrics.
Overall, as shown in the radar chart (Fig.~\ref{fig:invest}), RequireCEG (in purple) covers the largest area across all six INVEST metrics, with 24.90\% and 17.97\% higher coverage than the second-best method on each dataset. This demonstrates that RequireCEG consistently delivers the highest-quality Gherkin requirements following the INVEST principle.

\textbf{2. Fundiv Metric Analysis.}
The Fundiv metric reflects the diversity of functionality distribution, with values ranging from [0, $\log_25 \approx 2.322$]. A higher score indicates more diverse functional coverage, while a lower score means that features are concentrated in only a few categories.
As shown in Table~\ref{tab:test1}, RequireCEG outperforms the second-best method, MetaGPT~\footnote{MetaGPT (PRD) and MetaGPT (Story) share the same diversity score since both are evaluated using the same User Story section.}, by approximately 51.88\% on the RGPair dataset and 64.00\% on the Mini-RG dataset.
Fig.~\ref{fig:diver_total} provides a more detailed view. Most methods focus heavily on the "Functionality" category, with limited coverage of other FURPS categories. In contrast, RequireCEG (shown in purple) produces a more balanced distribution across all five categories. This is attributed to its use of a feature tree, which allows for more comprehensive functional planning.
In summary, RequireCEG demonstrates a clear advantage in overall functional diversity compared with baseline methods.

\begin{figure}[t]
	\centering
	\includegraphics[width=\textwidth]{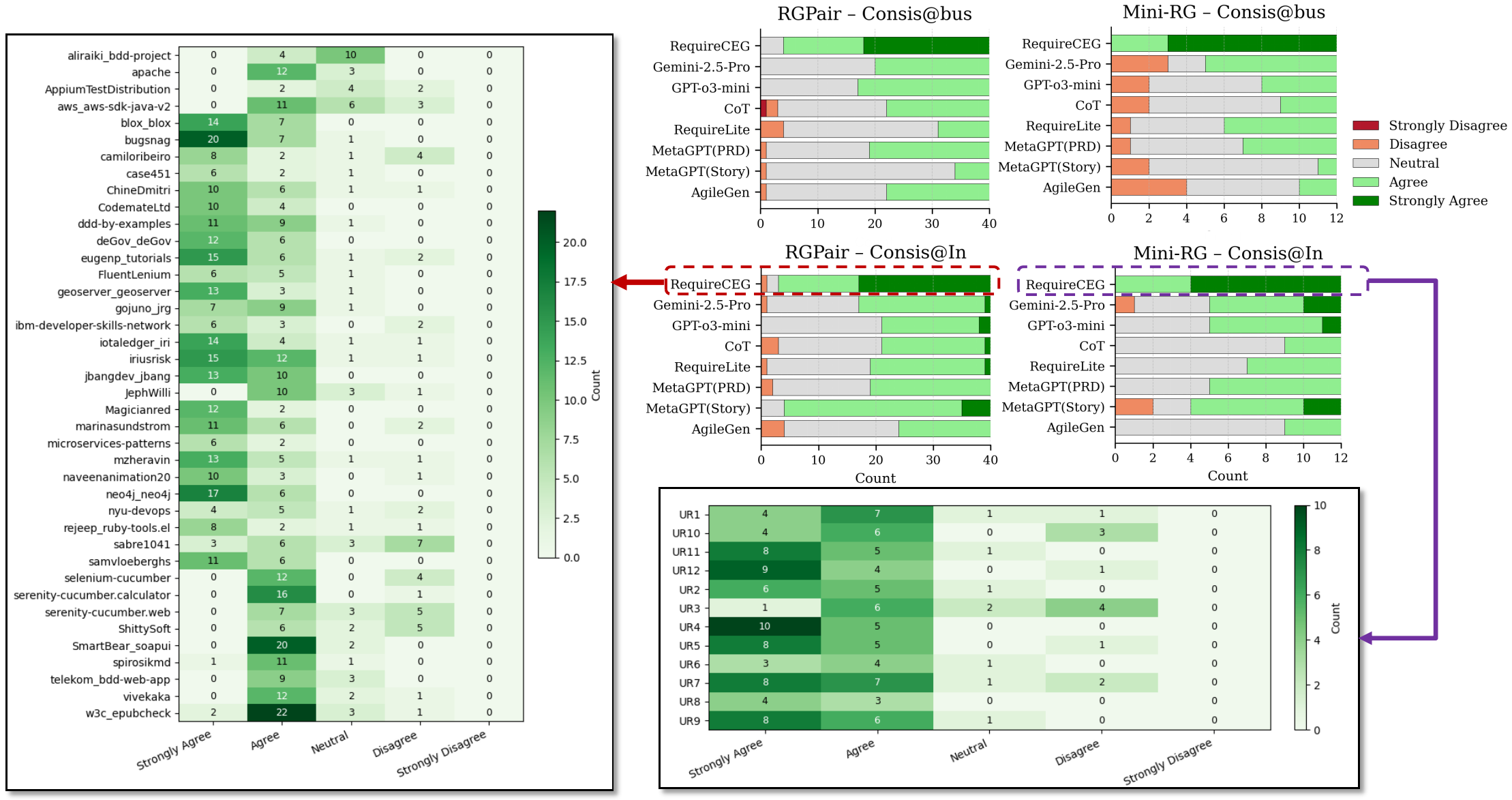}
\caption{\label{fig:consis} Comparison of rating distributions across different state-of-the-art (SOTA) methods on consistency metrics.}
\end{figure}

\textbf{3. Consistency Metric Analysis.}
As shown in Table~\ref{tab:test1}, RequireCEG outperforms all other methods in both business and internal consistency. On the Consis@Bus metric, it scores 24.48\% and 39.01\% higher than the second-best method on RGPair and Mini-RG, respectively. For Consis@In, it leads by 11.18\% and 27.27\%. Fig.~\ref{fig:consis} shows that most RequireCEG results are rated as "Strongly Agree" or "Agree", highlighting the effectiveness of its CEG-based review in resolving requirement ambiguity. In contrast, other methods are often rated "Neutral" or "Disagree", especially on the Mini-RG dataset, where input narratives are vaguer.
A heatmap in Fig.~\ref{fig:consis} further confirms that, although our method is rated as "Disagree" like the "sabre1041" project, half of our outputs still achieve high internal consistency ratings. In conclusion, RequireCEG ensures higher business alignment and logical consistency in the generated requirements compared with other methods.

\begin{tcolorbox}[
    colback=gray!10,           
    colframe=black!75!white,   
    rounded corners,           
    boxrule=1pt,               
    top=2pt, bottom=2pt,       
    left=3pt, right=3pt        
]
\textbf{Findings in RQ2}: 
RequireCEG outperforms other methods by at least 24.90\% in overall requirement quality, particularly in the "Testable" and "Estimable" categories, indicating more detailed and comprehensive requirements. Despite generating more scenarios, it maintains good independence and feasibility. It also produces more diverse functionalities while ensuring business and internal consistency. In summary, RequireCEG is a more effective solution for generating high-quality Gherkin requirements.
\end{tcolorbox}

\subsection{RQ3: What is the main reason for RequireCEG to improve the generation effectiveness of the Gherkin?}
\subsubsection{Motivation.}
Previous results have shown that RequireCEG is effective at improving both the diversity and consistency of generated requirements. This section examines how various components of our method contribute to enhancing these performance aspects. We conduct an ablation study by removing the following: 1) The requirement elicitation module (Our w/o Elicitation) to assess its impact on diversity. 2) The causal-effect graph (Our w/o CEG) to assess its impact on consistency. 3) The logical form and semantic check steps of CEG (Our w/o refine-CEG) to evaluate their effect on reliability.

\subsubsection{Methodology.}
To investigate RQ3, we compare our method with its ablated versions on both datasets. First, we evaluate the impact of each component using quality, diversity, and consistency metrics. Then, we conduct a case study for deeper analysis.

\paragraph{\textbf{a. Evaluation Metrics.}}
We use the same set of metrics as in Section~\ref{sec:invest_fundiv}, including the INVEST quality metric, the Fundiv diversity metric, and the consistency metrics Consis@Bus and Consis@In. Additionally, the Gherkin keyword statistics described in Section~\ref{sec:characters_syn} are used to highlight structural differences in the generated Gherkin requirements.

\paragraph{\textbf{b. Baselines.}}
We evaluate three ablated versions of RequireCEG on both datasets:
\begin{itemize}
    \item \textbf{Our w/o Elicitation}: Removes the requirement elicitation stage. The CEG is constructed directly from the input narrative, and Draft Gherkin is generated and reviewed without feature tree analysis. As shown in Algorithm~\ref{alg:multi-dim-ceg-check}, this version skips the "Feature Tree Hierarchical Analysis" module and uses $Narratives$ to replace $systemBehavior$.
    \item \textbf{Our w/o CEG}: Removes the CEG-based review. Draft Gherkin is generated directly from the system behavior requirements and used as the final output. According to Algorithm~\ref{alg:multi-dim-ceg-check}, this version skips everything under "Initial Construct Causal-Effect Graph" and outputs the unreviewed $draftGherkin$.
    \item \textbf{Our w/o Refine-CEG}: Removes the formal and semantic check of the CEG. The initially constructed CEG is used directly for reviewing Gherkin scenarios without undergoing the "Logical Formal Check" and "Logical Semantic Check" modules. The resulting $CEG_{feature}$ is passed directly to the review stage.
\end{itemize}

\paragraph{\textbf{c. RQ Experimental Setup.}}
The setup follows the same process as in RQ1 (see Section~\ref{sec:baselines}). All ablation methods are executed fully autonomously without any manual tuning to ensure fairness. Evaluation criteria and scoring procedures follow those of RQ2 (see Section~\ref{sec:invest_fundiv}). In this section, F@Num and F@Sce are reported as project-level averages.

\begin{table}[t]
    \centering
    \small
    \setlength{\tabcolsep}{4pt}
    \renewcommand{\arraystretch}{1.15}
    \caption{Detailed ablation results (\,$\uparrow$ indicates that higher values are better).}
    \label{tab:ablation}
   {
    \begin{tabular}{lcccccc}
        \toprule
        & \multicolumn{6}{c}{\textbf{RGPair Dataset}} \\ 
        \cmidrule(lr){2-7}
        \textbf{Methods} & \textbf{F@Num$\uparrow$} & \textbf{F@Sec$\uparrow$} & \textbf{INVEST$\uparrow$} & \textbf{Fundiv$\uparrow$} & \textbf{Consis@bus$\uparrow$} & \textbf{Consis@In$\uparrow$} \\
        \midrule
        Our w/o Elicitation & 1.000 & 7.175 & $2.39\!\pm\!0.2504$ & 0.550 & 3.225 & 4.100 \\
        Our w/o CEG & \underline{16.475} & 88.125 & $2.48\!\pm\!0.4191$ & \underline{0.886} & \underline{3.850} & 3.150 \\
        Our w/o RefineCEG & \underline{16.475} & \underline{107.550} & \underline{$2.52\!\pm\!0.3344$} & 0.853 & 3.675 & \underline{3.500} \\
        \textbf{RequireCEG(Our)} & \textbf{16.925} & \textbf{114.150} & $\mathbf{2.80\!\pm\!0.1646}$ & \textbf{1.417} & \textbf{4.450} & \textbf{4.475} \\
        \midrule
        & \multicolumn{6}{c}{\textbf{Mini-RG Dataset}} \\ 
        \cmidrule(lr){2-7}
        \textbf{Methods} & \textbf{F@Num$\uparrow$} & \textbf{F@Sec$\uparrow$} & \textbf{INVEST$\uparrow$} & \textbf{Fundiv$\uparrow$} & \textbf{Consis@bus$\uparrow$} & \textbf{Consis@In$\uparrow$} \\
        \midrule
        Our w/o Elicitation & 1.000 & 5.750 & $2.57\!\pm\!0.3226$ & 0.414 & 3.250 & 4.333 \\
        Our w/o CEG & \underline{10.583} & 48.833 & $2.54\!\pm\!0.2566$ & 0.658 & \underline{3.667} & 3.083 \\
        Our w/o RefineCEG & \underline{10.583} & \underline{62.750} & \underline{$2.65\!\pm\!0.2198$} & \underline{0.763} & 3.500 & \underline{3.583} \\
        \textbf{RequireCEG(Our)} & \textbf{13.083} & \textbf{77.917} & $\mathbf{2.78\!\pm\!0.2509}$ & \textbf{1.371} & \textbf{4.750} & \textbf{4.667} \\
        \bottomrule
    \end{tabular}}
\end{table}

\begin{figure*}[t]
	\centering
	\includegraphics[width=\textwidth]{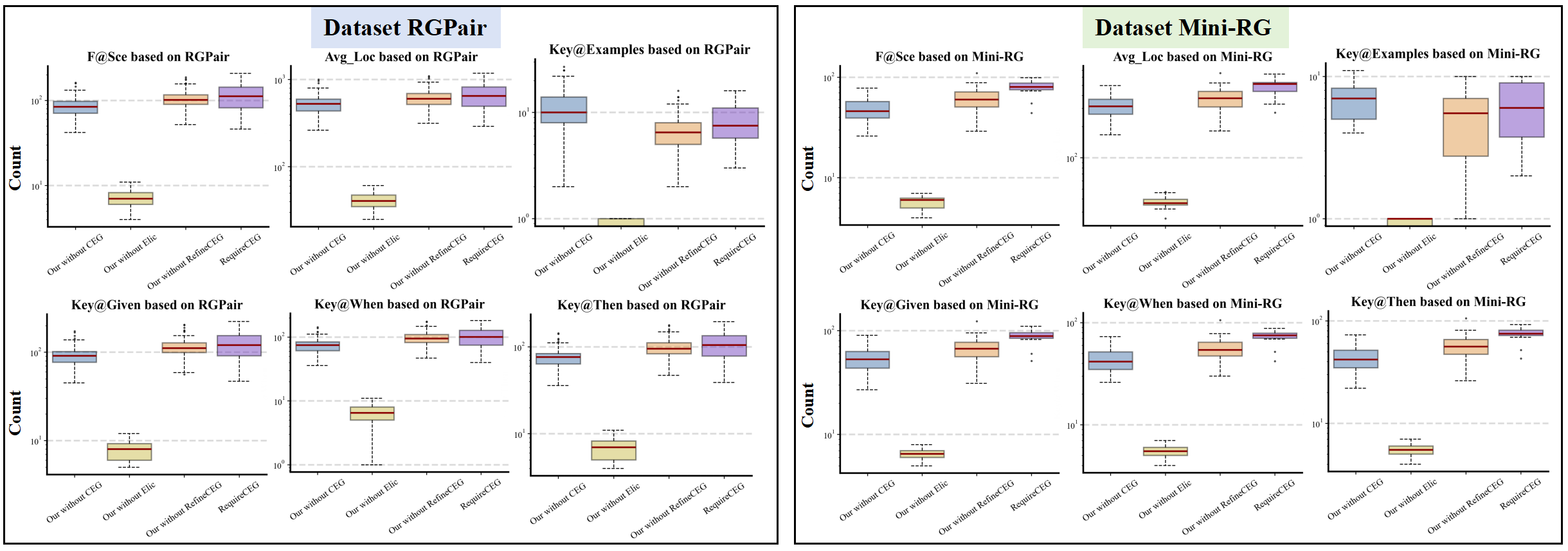}
\caption{\label{fig:Gherkin_Characters_ab} Gherkin keyword Statistics Generated by Ablation Methods. The vertical axis uses a logarithmic scale, and the horizontal axis represents different methods.}
\end{figure*}

\begin{figure}[t]
	\centering
	\includegraphics[width=\textwidth]{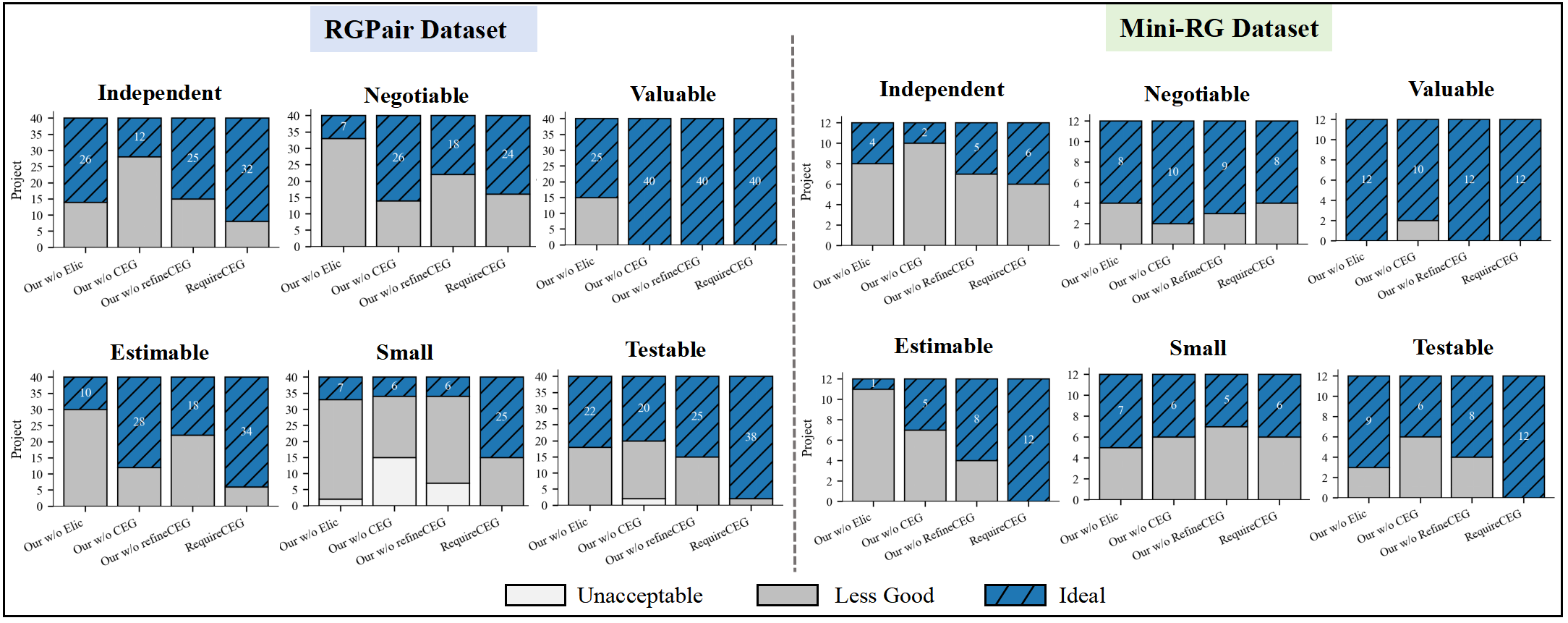}
\caption{\label{fig:INVEST_ab} Stacked Bar Chart of INVEST Ratings for Ablation Methods.}
\end{figure}

\begin{figure*}[t]
	\centering
	\subfigure[Diversity comparisons based on RGPair dataset.]{%
		\centering
		\includegraphics[width=0.49\textwidth]{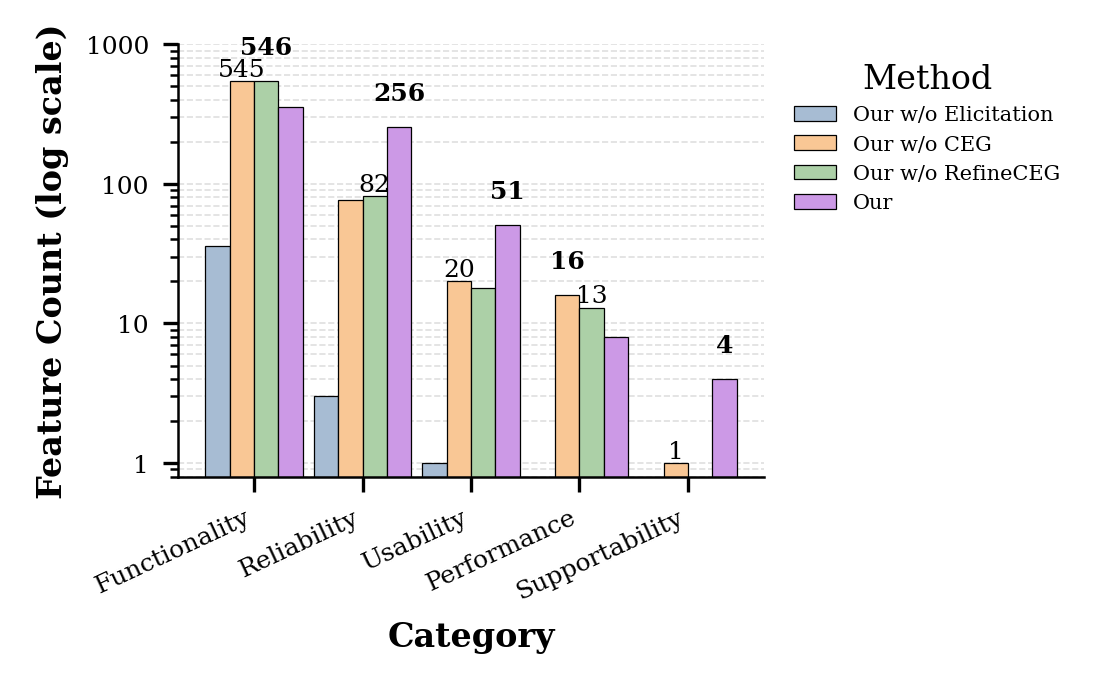}
		\label{fig:diver_total_d1_ab}
	}%
	\subfigure[Diversity comparisons based on Mini-RG dataset.]{%
		\centering
		\includegraphics[width=0.49\textwidth]{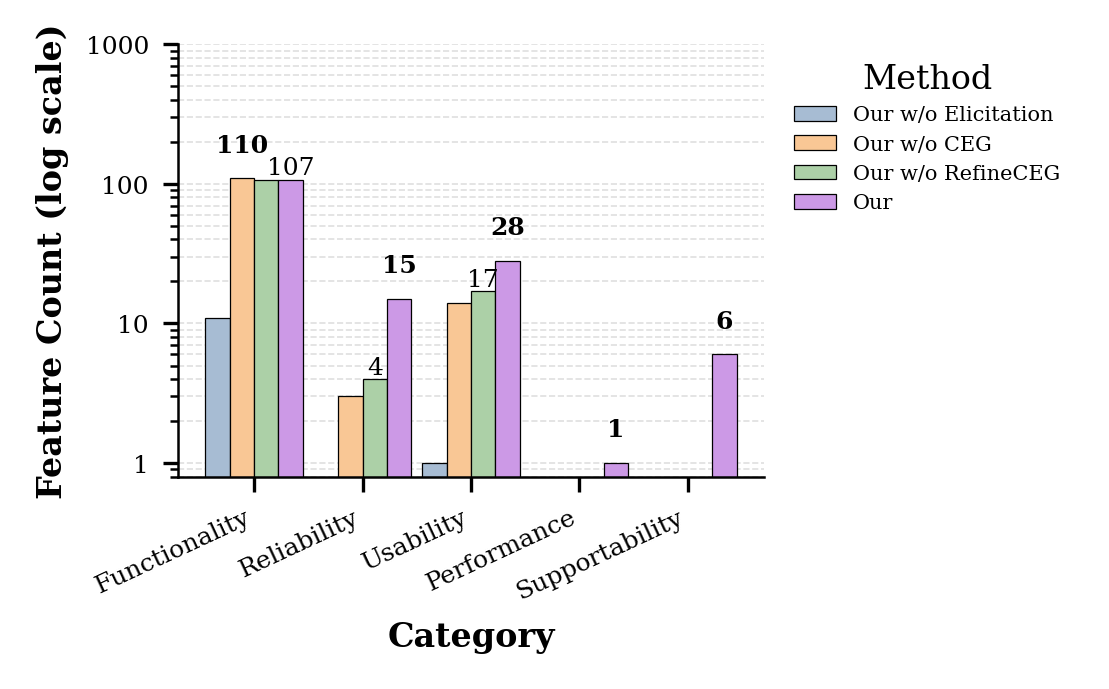}
		\label{fig:diver_total_d2_ab}
	}
	\caption{Comparison of Feature Counts by FURPS Categories for Ablation Methods.}
    \label{fig:diver_total_ab}
\end{figure*}

\begin{figure}[t]
	\centering
	\includegraphics[width=\textwidth]{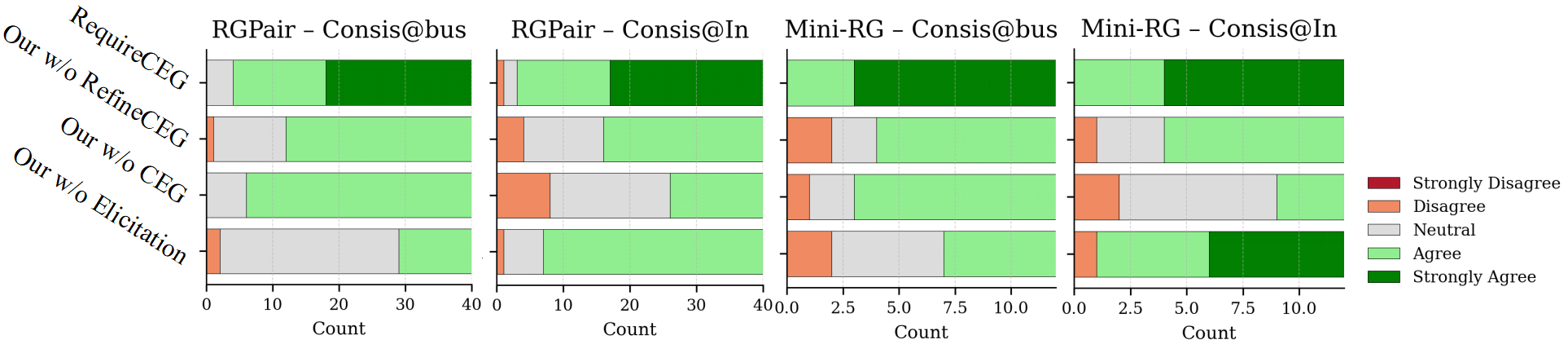}
\caption{\label{fig:consis_ab} Comparison of Rating Distributions on Consistency Metrics for Ablation Methods.}
\end{figure}

\begin{figure}[t]
	\centering
	\includegraphics[width=0.93\textwidth]{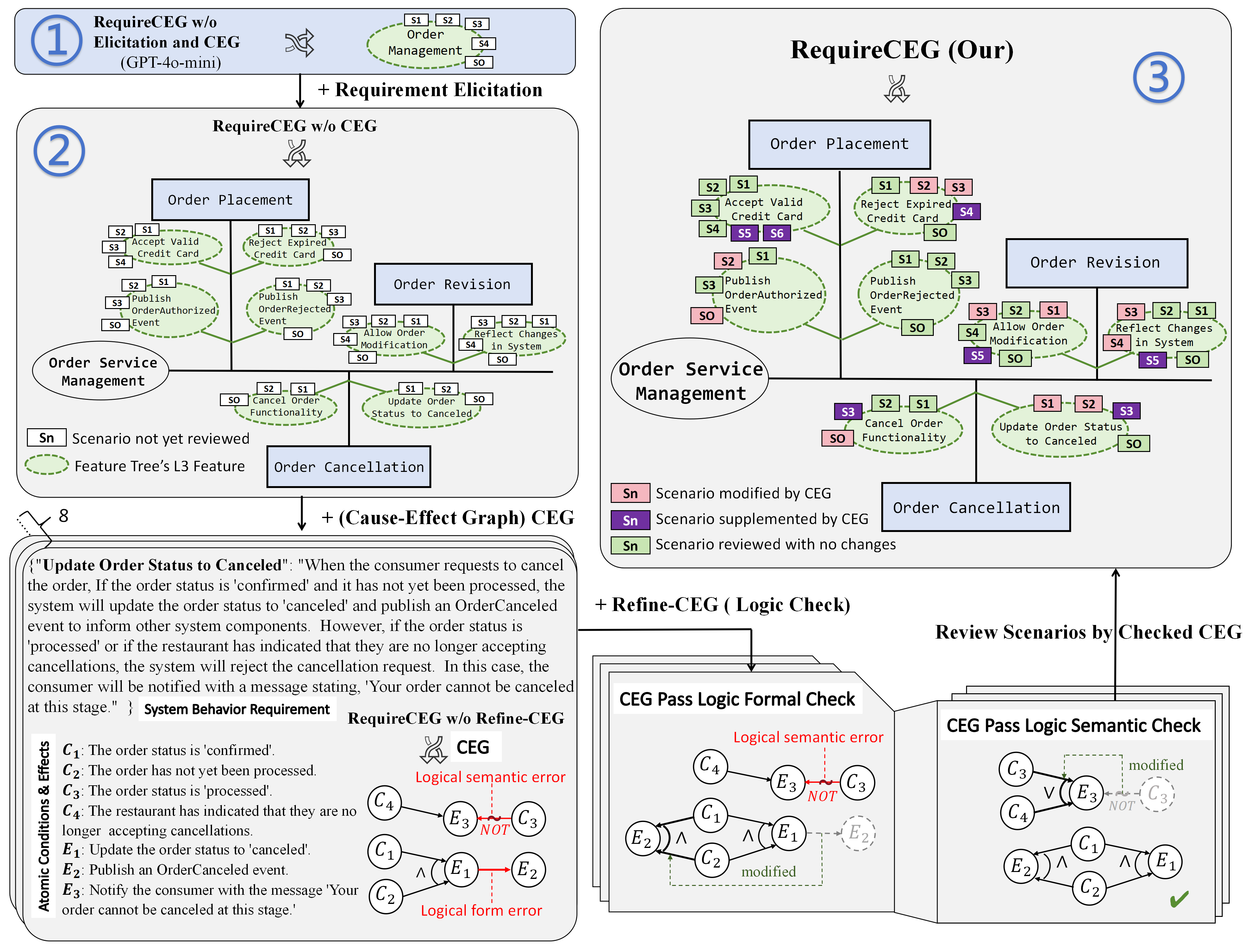}
\caption{\label{fig:case} Visualization of the outputs at different stages of the RequireCEG method using the "microservices-patterns/ftgo-application" project from the RGPair dataset as an example.}
\end{figure}

\subsubsection{Result Analysis.}
The ablation results are summarized in Table~\ref{tab:ablation}, with detailed Gherkin keyword statistics shown in Fig.~\ref{fig:Gherkin_Characters_ab}. Diversity and consistency results are visualized in Fig.~\ref{fig:diver_total_ab} and~\ref{fig:consis_ab}, respectively.

\textbf{(1) Our w/o Elicitation}: Removing the requirement elicitation module leads to a significant reduction in the number of generated scenarios. The Features tend to describe generalized functionality, resulting in a diversity score reduction of over 61\%. Both the quality and consistency of requirements also decline. Notably, business consistency (Consis@Bus) drops by at least 27.5\%, as the generated scenarios miss business details. Without structured elicitation, it becomes harder to construct meaningful CEGs, preventing effective correction of such gaps.

\textbf{(2) Our w/o CEG}: Removing the CEG-based review stage reduces scenario count by at least 22.8\% due to the inability to detect and supplement missing scenarios. As shown in Fig.~\ref{fig:INVEST_ab}, this version scores well on "Negotiable", since it lacks constraints from CEG checking. However, it performs worse on "Independent", "Small", and "Testable", as some scenarios contain dependencies or lack clear acceptance criteria. Diversity and consistency also decrease, with internal consistency dropping by at least 29.6\%, highlighting the importance of CEG in resolving logical ambiguity.

\textbf{(3) Our w/o Refine-CEG}: This version does not significantly reduce the number of scenarios but impacts scenario quality. Fig.~\ref{fig:INVEST_ab} shows that scores for "Small", "Estimable", and "Testable" are nearly halved compared with RequireCEG, especially on "Small" in the RGPair dataset. Business consistency even drops below the "Our w/o CEG" version due to incorrect or misleading logic in unchecked CEGs. This confirms the critical role of formal and semantic checks in building a reliable CEG for effective requirement review.

\begin{figure*}[t]
  \centering
  \subfigure[Missing "And" Preconditions]{%
    \includegraphics[width=0.49\textwidth]{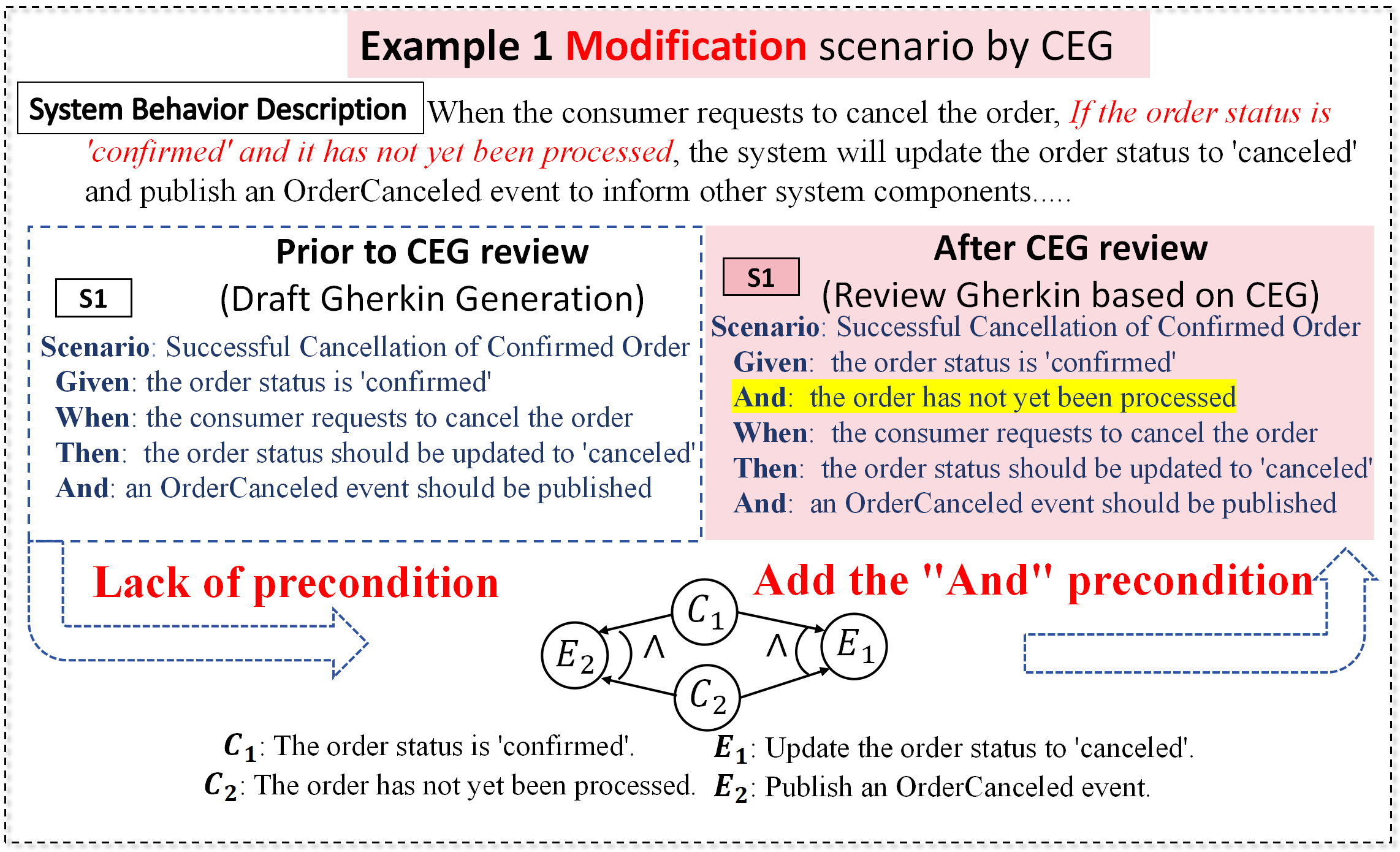}
    \label{fig:sub_ex1}}
  \subfigure[Missing "And" User-Experience-Related Actions]{%
    \includegraphics[width=0.49\textwidth]{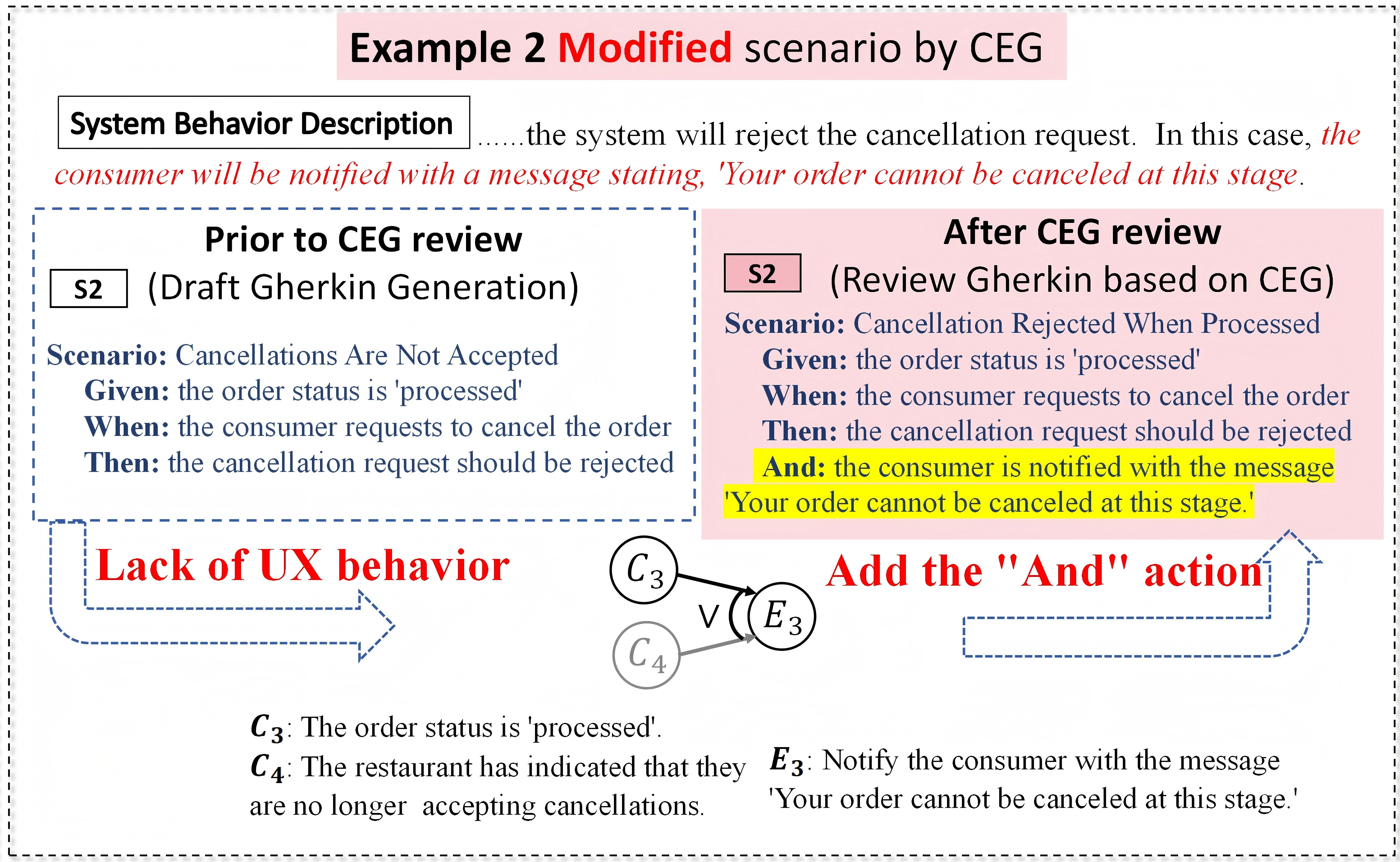}
    \label{fig:sub_ex2}}
  \subfigure[Missing "OR" Branch Scenarios]{%
        \includegraphics[width=0.5\textwidth]{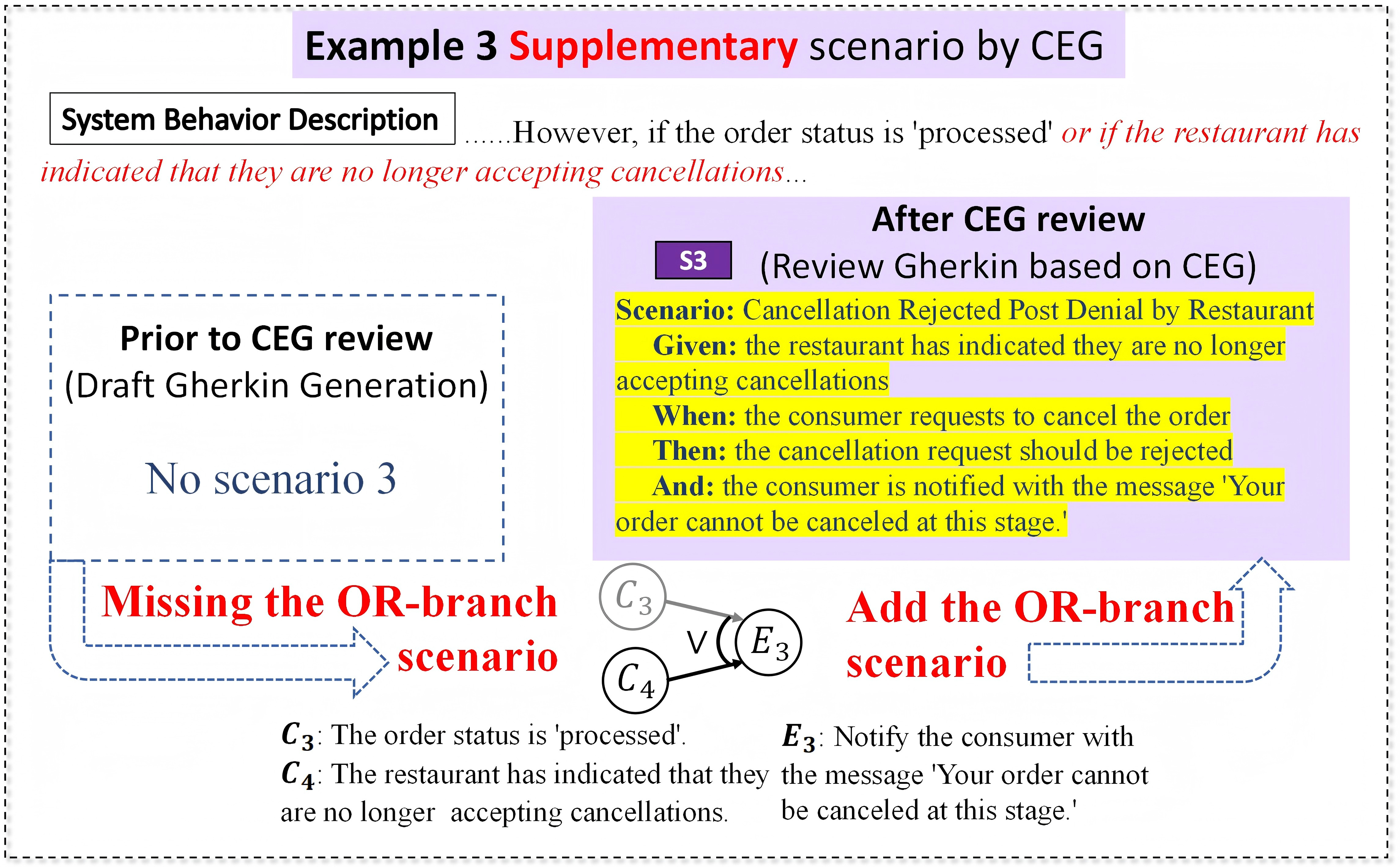}
        \label{fig:sub_ex3}}
  \caption{Illustrative examples of using the Causal-Effect Graph (CEG) to review and correct three common types of missing elements in Gherkin scenarios.}
  \label{fig:combined_examples}
\end{figure*}

We use the "ftgo-application" project from the RGPair dataset as a case study. This project focuses on order creation, modification, and cancellation. Fig.~\ref{fig:case} illustrates the outputs at different stages of the RequireCEG. In the top-left section (\ding{202}), RequireCEG w/o Elicitation and CEG represent the base model (GPT-4o-mini), which directly generates Gherkin. It produces a single Feature called "Order Management" with four scenarios (S1–S4) and one scenario outline (SO). 
With the addition of the Requirement Elicitation module (RequireCEG w/o CEG), the model generates 8 Features and a total of 32 scenarios (3–5 per Feature), but without any causal reviewing.
In section \ding{203}, we add the Causal-Effect Graph (CEG) module (RequireCEG w/o Refine-CEG). This module builds CEGs from system behavior requirements by identifying atomic conditions and effects. However, due to LLM hallucinations, the initial CEGs often contain syntax or semantic errors—less than 30\% are correct. These errors can lead to incorrect scenario review and affect the final Gherkin quality.
Finally, by enabling our full Refine-CEG module (RequireCEG) the system performs logical formal and semantic checks. It fixes incorrect expressions like $DIR(E_1)=E_2$ or $NOT(C3)=E_3$. As shown in section \ding{204} of Fig.~\ref{fig:case}, after applying the complete RequireCEG process, 65.6\% of the original 32 scenarios were retained, 34.4\% were modified based on CEG, and 7 new scenarios were added. The final output includes 8 Features and a total of 39 well-reviewed Gherkin scenarios.
Next, we take the Feature "Update Order Status to Canceled" as a concrete example to illustrate how the CEG is used to review and modify the scenario.

\textbf{Example 1: Lack of preconditions:}
As shown in Fig.~\ref{fig:sub_ex1}, the generated draft Gherkin for scenario S1 lacks the essential precondition "it has not yet been processed" before updating the order to a canceled status. This omission leads to a critical business logic error. After a CEG-based review, the scenario is corrected by adding the missing precondition based on the CEG expressions ($AND(C1, C2) = E1$, $AND(C1, C2) = E2$), ensuring logical consistency with the system behavior requirement.

\textbf{Example 2: Lack of UX Behavior:}
In Fig.~\ref{fig:sub_ex2}, scenario S2 initially omits the user-facing system feedback: "The consumer will be notified with a message stating, 'Your order cannot be canceled at this stage'." This negatively impacts user experience. After review, the system behavior is revised to include the missing message according to the CEG expressions ($DIR(C3) = E3$, $OR(C3, C4) = E3$).

\textbf{Example 3: Missing OR-Branch Scenario:}
As shown in Fig.~\ref{fig:sub_ex3}, scenario S3 was missing entirely in the draft Gherkin generation, leaving out the OR-branch condition from the System Behavior Requirement: "if the order status is 'processed' or if the restaurant has indicated they are no longer accepting cancellations...". The CEG-based review identifies this gap and supplements the missing OR-branch scenario using the logical expression ($DIR(C4) = E3$, $OR(C3, C4) = E3$).

\begin{tcolorbox}[
    colback=gray!10,          
    colframe=black!75!white,  
    rounded corners,          
    boxrule=1pt,              
    top=2pt, bottom=2pt,      
    left=3pt, right=3pt       
]
\textbf{Findings in RQ3}: 
To explore why RequireCEG is effective, we performed ablation studies. Removing the elicitation module resulted in a reduction of functional diversity by over 61.2\%. Without the CEG-based review, internal consistency dropped by 29.6\%. Removing the CEG check further lowered business and internal consistency by 17.4\% and 21.7\%, respectively. These results demonstrate that elicitation enhances diversity, review improves consistency, and CEG checks ensure reliable scenario validation.
\end{tcolorbox}

\section{Field Study}

Previous experimental results have shown that RequireCEG improves Gherkin quality on the RGPair and Mini-RG datasets, generating diverse Features while maintaining consistency. This section examines the practical utility of RequireCEG by assessing whether the generated Gherkin scenarios align with real-world needs and whether generated functionalities are beneficial to users.

\subsection{Preparations}
 To assess the utility of RequireCEG, we collected a real-world dataset (Pub-website Dataset) consisting of five publicly available websites. We perform human evaluations to assess scenario coverage.

\subsubsection{Dataset}
To facilitate human evaluation, we selected five publicly available AI-related websites that are domain-specific, user-preferred, and relatively low in complexity. From each website’s homepage, we collected the introduction narrative text and manually identified their main functional scopes for reference. This dataset is referred to as the Pub-website dataset. It contains a total of five narratives and 96 primary functions. The five websites cover various business domains, as shown in Table \ref{tab:real-web}.

\begin{table}[h]
  \small
  \centering       
  \caption{Details of the Pub‑website Dataset}
  \label{tab:real-web}
  \begin{tabular}{clllc}
    \toprule
    \textbf{ID} & \textbf{Name}   & \textbf{Link}  & \textbf{Domain}  &\textbf{Number of Functions}         \\ \midrule
    1           & Lyrebird health & \url{https://www.lyrebirdhealth.com/au} & Medical Consultation  &22    \\
    2           & STORM           & \url{https://storm.genie.stanford.edu/}       & Article Creation &18         \\
    3           & Notebook LM     & \url{https://notebooklm.google/}              & Summary of Notes   &21       \\
    4           & CourtAid        & \url{https://courtaid.ai/}                    & Legal Aid  &16               \\
    5           & Image Prompt    & \url{https://imageprompt.org/image-to-prompt} & Image to Prompt Generator &19 \\ \bottomrule
  \end{tabular}
\end{table}

\subsubsection{Evaluation Metrics}
We employ a binary evaluation approach to manually assess functional coverage. If a generated Feature has a similar core function on the real website, it is marked as "Yes"; otherwise, it is marked as "No." A "Yes" does not require exact matching—any related functionality is acceptable. Participants selecting "Yes" must provide the corresponding reference function from the Pub-website Dataset and function web link as justification; otherwise, they must explain why it was marked "No."

\begin{itemize}
    \item \textbf{Cover@Our}: This metric measures how many of the generated Gherkin Features align with actual functionalities implemented on the real websites, reflecting the realism of the generated requirements. Three participants independently evaluated each generated function. If at least two mark it as "Yes," it is considered covered.
    \begin{equation}
    \text { Cover@Our }=\frac{\text { Number of covered generated functions }}{\text { Total number of generated functions }}
    \end{equation}
    \item \textbf{Cover@Real\_Func}: This metric assesses the extent to which the reference functions identified in the Pub-website Dataset are covered by the generated Gherkin Features, indicating the completeness of the generated requirements for development. Coverage is determined based on the explanations provided by participants.
    \begin{equation}
    \text { Cover@Real\_Func }=\frac{\text { Number of covered pub-website functions }}{\text {Total number of pub-website}}
    \end{equation}
\end{itemize}

\subsubsection{Field Study Setup}
\textbf{Participants.} To evaluate the functional coverage of Gherkin requirements generated by our method, we invited three graduate students as participants: two males and one female. One is from Computer Science and Technology, and two are from Electronic Information majors. All have more than three years of programming experience. None of them were involved in any stage of our system development, ensuring objective and unbiased evaluations.

\textbf{Procedure.} To ensure participants fully understood the evaluation criteria, we held a training session. Using a website not included in the evaluation dataset as an example, we demonstrated how RequireCEG generates requirements and how Gherkin-described features map to actual website functionalities. During training, we practiced with a binary evaluation table, selecting "Yes" only if the generated function matched a labeled real-world function and providing the function name and website link for reference. The session included a Q\&A to ensure clarity and understanding. After training, each participant was assigned five projects, each with a set of Gherkin requirements, a binary evaluation form, and a corresponding real website. Participants independently determined whether each generated function matched any real functionality on the site. For each "Yes" judgment, they noted the corresponding labeled function and link; for "No," they provided a reason. Afterward, we conducted interviews to gather feedback on the usefulness of the generated Gherkin scenarios. To ensure fairness, all evaluated Gherkin requirements were generated automatically from website narratives using RequireCEG, with no human intervention during the generation process.

\subsection{Field Study Analysis.}

\begin{table}[]
\small
\caption{Human-evaluated coverage table on the pub-website dataset.}
\label{tab:Real-test}
{
\begin{tabular}{clcc}
\toprule
\textbf{ID} & \multicolumn{1}{c}{\textbf{Name}} & \textbf{Cover@Our} & \textbf{Cover@Real\_Func} \\ \midrule
1 & Lyrebird health & 84.71\% & 68.18\% \\
2 & STORM           & 83.33\% & 83.33\% \\
3 & Notebook LM     & 88.89\% & 66.67\% \\
4 & CourtAid        & 90.48\% & 68.75\% \\
5 & Image Prompt    & 87.50\% & 63.16\% \\ \bottomrule
\end{tabular}
}
\end{table}

\begin{figure}[h]
	\centering
	\includegraphics[width=\textwidth]{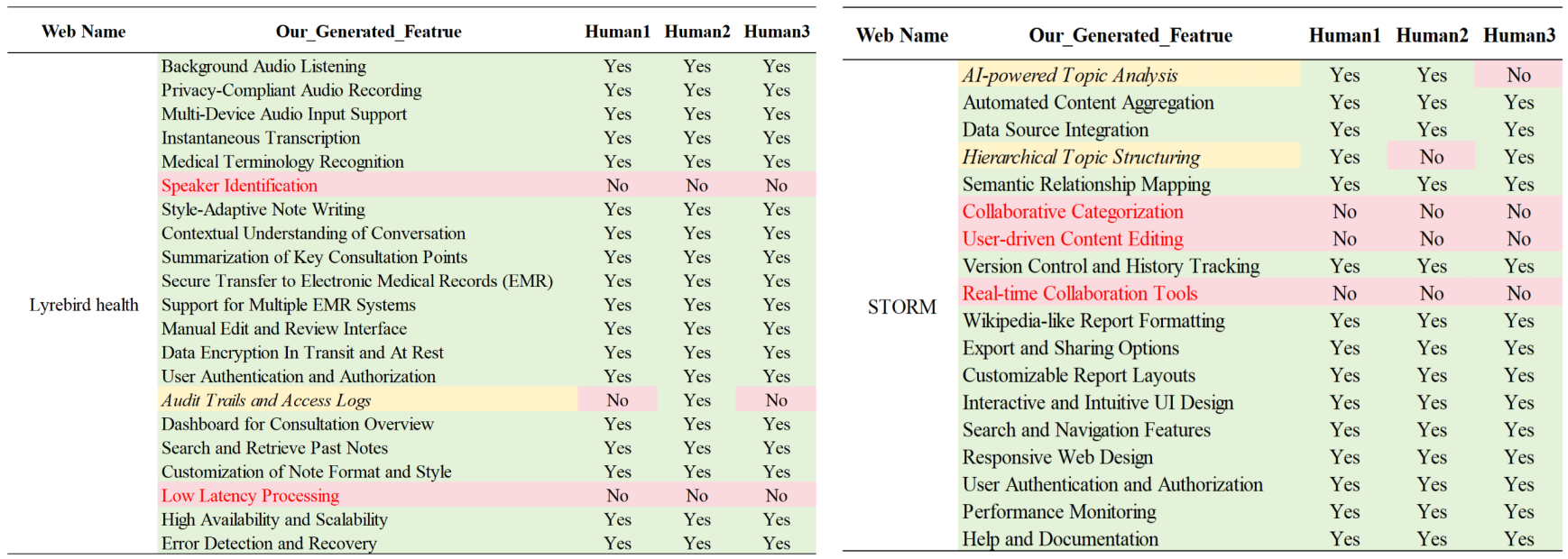}
\caption{\label{fig:Human_test} A detailed visualization of the human evaluation is shown for the projects "Lyrebird Health" and "STORM." Green indicates functions that all three participants evaluated as "Yes," yellow represents functions with conflicting evaluations, and red indicates functions that all three participants evaluated as "No".}
\end{figure}

Through human evaluation, the utility and completeness of the functional requirements generated by our method are demonstrated in Table~\ref{tab:Real-test}, with a Cohen's $\kappa$ > 0.82 among the three participants, indicating strong inter-rater agreement. The results are analyzed as follows:

\textbf{Cover@Our} measures the proportion of our generated features that are covered by similar real functionalities on the websites. For the projects "Lyrebird Health" and "STORM", three features are not covered, whereas all other projects had only two uncovered features. Overall, the feature coverage ranged from a minimum of 83.33\% to a maximum of 90.48\% (achieved by "CourtAid"), demonstrating strong utility. As shown in Fig.~\ref{fig:Human_test}, we detail the evaluation results for the two projects with the lower coverage:
\begin{itemize}
    \item \textbf{Lyrebird Health} is an AI scribe for healthcare professionals that transcribes doctor-patient conversations in real-time. The uncovered features generated by RequireCEG include "Speaker Identification," "Low Latency Processing," and "Audit Trails and Access Logs." For "Low Latency Processing," participants note that low latency is essential for transcription but hard to assess directly. "Speaker Identification" is not found on the website, and participants explained the system transcribes without identifying different speakers. The feature "Audit Trails and Access Logs" is disputed: two participants said it was not visible in normal user accounts, while one believed it would exist for admin roles and marked it as covered.
    \item \textbf{STORM} is a research prototype web system that generates topic reports (similar to Wikipedia) using AI, supporting interactive knowledge management. Among the 18 features generated by RequireCEG, three were unanimously marked as uncovered: "Collaborative Categorization," "User-driven Content Editing," and "Real-time Collaboration Tools." These features focused on multi-user collaboration and post-generation editing. Participants commented that while these functions were not present on the site, they were well-conceived and highly desirable.
\end{itemize}

\textbf{Cover@Real\_Func} measures the proportion of real website functionalities that are covered by features similar to those generated by our method. The average coverage is 70\%, with the highest at 83.3\% for the "STORM" project, indicating the completeness of our method. The remaining 30\% of uncovered pub-website functionalities mainly fall into three categories: 
(1) Account \& Support Functions (38\%): Examples include profile editing, preference settings, and subscription management in "Lyrebird Health," as well as login and subscription features in "CourtAid".
(2) Additional Features (34\%): These go beyond the core functions described on the websites. For instance, in "Notebook LM," while the primary focus is on summarizing and querying uploaded documents and generating audio overviews, some features, such as "Mind Map Generation" and "Learning Guide Generation from documents," are involved. Similarly, the "ImagePrompt" project description primarily focuses on understanding images and generating prompts, with AI optimizing the prompts. However, it also involves features such as "text-to-video prompt generation" and "image-to-video prompt conversion."
(3) User Experience Enhancements (28\%): These include auxiliary features such as help guides in "Lyrebird Health," the ability to view the brainstorming process in "STORM," or inspirational image creation examples in "ImagePrompt."

\begin{tcolorbox}[
    colback=gray!10,           
    colframe=black!75!white,   
    rounded corners,           
    boxrule=1pt,               
    top=2pt, bottom=2pt,       
    left=3pt, right=3pt        
]
\textbf{Findings in Field Study}: 
To evaluate the practical utility of RequireCEG in meeting real-world requirements, we invited three participants to assess its functional coverage. Results show that 87\% of the features generated by RequireCEG are reflected in real websites, indicating that most generated requirements are realistic and align with real website implementations. Additionally, 70\% of the real website functions are covered by our method, demonstrating that our approach approximately matches the real website functions. Overall, RequireCEG shows strong utility in handling real-world requirement scenarios.

\end{tcolorbox}

\section{Discussion}
RequireCEG is a neuro-symbolic collaboration agent for requirement elicitation and review, capable of transforming vague narratives into detailed Gherkin-based requirements. Experimental results show that RequireCEG produces higher-quality and more diverse Gherkin requirements. Its ability to design functions is approximately comparable to that of the real website. This section first discusses the sources of error and limitations.
\begin{itemize}
    \item \textbf{Requirement Quality}: RequireCEG generates about 17 features per project, each with 6–7 scenarios. This large scale reduces its relative advantage on the INVEST principles of "Small" and "Independent". Although RequireCEG explicitly defines preconditions and behavior actions for each scenario, some similar preconditions with constraints INC or REQ may lead to scenario coupling risks. In contrast, baseline methods generate fewer features and scenarios, naturally reducing complexity and dependency but at the cost of significantly lower testability and estimability. Despite the scale, RequireCEG maintains strong readability and syntactic correctness, indicating that its output remains easy to understand.
    \item \textbf{Functional Diversity}: RequireCEG outperforms others in terms of diversity, but the distribution is not very even. As shown in Fig.~\ref{fig:diver_total}, RequireCEG generates fewer functionalities related to "Performance" and "Supportability." This may be because the current approach focuses more on user satisfaction, emphasizing functional design, user experience, and system robustness while paying relatively less attention to operational aspects such as system efficiency and maintainability.
    \item \textbf{Requirement Consistency}: RequireCEG utilizes causal-effect graphs (CEGs) to review requirements, demonstrating clear advantages in both business and internal consistency. However, in the project "sabre1041\_ose-bdd-demo", both consistency scores were low. For instance, in 19 features, 7 were rated "Disagree" and 3 "Neutral" for Consis@In. The root cause is ambiguity in business rules about shipping fees at boundary conditions (e.g., "Total cart price $\leq$ 25\$ gets \$3.99 freight", but "Total cart price $=$25\$ gets \$4.99 freight"). According to the logs, the CEG modification reached a maximum of 5 iterations for 11 features in this project, and 10 still failed the check, resulting in poor consistency scores.
    \item \textbf{Functional Coverage}: Regarding Cover@Our, about 87\% of the generated features align with real website functions. The remaining 13\% may still be useful as suggestions for potential improvements—e.g., "Real-time Collaboration Tools" in the STORM project. Of course, real implementations consider trade-offs like cost and feasibility. Regarding Cover@Real\_Func, our method covers approximately 70\% of website functions, indicating that our approach closely aligns with the actual website functions. Lower coverage was primarily found in "Account \& Support," "UX Enhancements," and so on. Tasks like "user subscriptions" are typically omitted unless explicitly mentioned in the input narrative.
\end{itemize}

Next, this section discusses the value for generative software development.

\begin{itemize}
    \item \textbf{The value of using Gherkin as a generative software requirements.} This method condenses requirements, code, and tests into a single, readable, and executable ".feature" file. Its standardized syntax is easy for different stakeholders to understand, and it also helps LLMs treat each Scenario as a sub-task in the overall business flow, naturally forming a "chain-of-thought" approach to step-by-step code generation. Because these scenarios are BDD acceptance criteria, pairing them with mature frameworks like Cucumber or Behave yields immediate green-light/red-light feedback. If a test fails, feed the error stack back to the generative model for iteration; the objective signal suppresses hallucinations and continually ensures that the implementation stays aligned with the requirements.
    \item \textbf{The value of neuro-symbolic collaboration for generative software development.} RequireCEG leverages neural-symbolic collaboration by combining semantic reasoning from LLMs with symbolic review through CEGs to systematically transform user narratives into testable system behavior scenarios in Gherkin. At the top level, a Feature Tree decomposes user narratives into hierarchical components. Then, CEGs map each atomic condition-behavior chain into Boolean logic, capturing causal relationships. These symbolic graphs not only enable the automatic detection of conflicts and gaps, but also form a traceable "behavior map." In generative software development, any test failure can be traced step by step back to the originating conditions and business-level requirements via the graph. Overall, RequireCEG provides a traceable, verifiable, and evolvable safety net for generative software development through executable causal-effect graphs, surpassing the limitations of natural language descriptions alone.
\end{itemize}

\section{Related work}
This paper presents an autonomous AI agent, RequireCEG, for requirement elicitation and review to support end-user software engineering. RequireCEG is designed with the capabilities of a requirements analyst, capable of transforming vague user narratives into detailed, structured requirements described in Gherkin. The related work is as follows:

Large Language Models (LLMs) have shown remarkable capabilities in automatically generating code based on provided natural language requirements. However, user requirements are often ambiguous, resulting in generated code that may deviate from user expectations. In recent years, clarifying ambiguous requirements has attracted growing attention~\cite{vogelsang2025impact}. In July 2024, Mu et al.~\cite{10.1145/3660810} proposed ClarifyGPT, which performs consistency checks on code units to determine whether requirements are ambiguous and guides LLMs to generate targeted clarification questions. In April 2025, Wu et al.~\cite{wu2025clarifycoder} proposed ClarifyCoder, which identifies ambiguous requirements using data synthesis techniques before generating code. In May 2025, Jia et al.~\cite{jia2025automated} introduced SPECFIX, a method that decomposes vague requirements into a program problem and maps back to the original requirement for automated correction after solving the program. However, these methods primarily target function-level code units with clear input and output. The natural language requirements they address typically describe specific and observable program functionality.

LLM-based development enables users to build apps from natural language, but vague and imprecise input remains a key challenge.
Hong et al. proposed MetaGPT in 2023, which collaborates LLMs acting in different software development roles. In February 2025, they released a natural language programming platform~\footnote{\url{https://mgx.dev/}}. Around the same time, Wu et al.~\cite{wu2023autogen} introduced AutoGen, a framework that enables multiple intelligent agents to collaborate on completing application development tasks. In 2024, Qian et al. ~\cite{qian2024chatdev} designed ChatDev, a multi-agent collaborative software development method guided by dialogue chains. Subsequent iterations utilized experience learned from historical trajectories in software development ~\cite{qian2023experiential} and directed acyclic graphs to support large-scale agent collaboration ~\cite{qian2024scaling}. Also, in 2024, Nguyen et al.~\cite{nguyen2024agilecoder} proposed AgileCoder, which organizes agent workflows into development sprints to enhance productivity. These approaches adopt an end-to-end modal, but their intermediate outputs, such as agent conversations, are not well-suited for end-user understanding or participation. While MetaGPT provides requirement analysis documents through its analyst role, these documents often contain simplistic user stories, making them less effective for applications with complex functionalities.

Recent studies have also focused on supporting end users in interacting with intelligent agents. In 2024, Arteaga Garcia et al.~\cite{10.1145/3597503.3608130} designed conversational intelligent agents to assist end users in interacting with machine learning tasks. In January 2025, Xing et al.~\cite{xing2025prompt} introduced CNL-P, a controlled natural language prompting method that incorporates precise grammatical structures and strict semantic constraints to eliminate ambiguities in natural language, thereby helping LLMs better understand and execute user prompts. In April 2025, Ma et al.~\cite{10.1145/3731756} proposed ROPE, a requirement-oriented prompt optimization approach that helps inexperienced end users express their needs to LLMs more accurately. These studies primarily aim to facilitate interaction between end users and LLM Agents for general-purpose tasks.

End-user software development requires a shared language, allowing users to articulate their requirements clearly. The domain-specific language Gherkin, used in behavior-driven development (BDD), offers this with its clear structure and understanding among stakeholders. However, research on the automatic generation of Gherkin from user requirements remains limited.
In 2024, Karpurapu et al.~\cite{karpurapu2024comprehensive} explored using LLMs for Gherkin generation based on User Stories in terms of grammatical accuracy.
In 2025, Zhang et al.~\cite{zhang2024empowering} proposed human-AI collaboration to empower agile generative software development. They made the first attempt to generate Gherkin from end-user requirements for software development.
Yet, current approaches often rely on few-shot prompting to guide LLMs in mimicking Gherkin templates. These methods lack deep requirement analysis and fail to leverage reasoning capabilities, making it difficult to uncover implicit requirements or ensure coherent logic. In contrast, RequireCEG defines function scope via a feature tree and empowers LLMs to construct and check Causal-Effect Graphs (CEGs) autonomously for requirement review. This injects analyst-level reasoning into the process, ensuring that the generated Gherkin scenarios reflect user intent.

\section{Conclusion}
This paper presents RequireCEG, a neuro-symbolic collaboration framework driven by LLMs that autonomously elicits and reviews software requirements for end-user generative software development. It translates informal user narratives into Gherkin scenarios with explicit acceptance criteria and applies Causal-Effect Graphs (CEGs) to review the causal links between preconditions and actions, ensuring internal and business consistency.
To address the common issue of limited functional diversity, RequireCEG employs a Feature Tree to plan the functional scope and elicit system behavior requirements from both the user operation and system response perspectives. We further propose a self-healing CEG construction process for effectively uncovers and represents causal relations within system behavior requirements. This process combines formal and semantic checking, enabling the automatic healing of the CEG through a generate-check-modify loop.
We construct an RGPair benchmark dataset (40 GitHub projects, 413 feature files). Experimental results demonstrate that RequireCEG outperforms state-of-the-art methods in both requirement quality and functional diversity, approaching the level of product-grade analysis and design.
Beyond software requirements, RequireCEG could apply to general natural language requirement clarification tasks. By leveraging CEG construction, our method offers a transparent and explainable reasoning path for LLMs, showcasing their potential to discover and interpret causal logic. In future work, we aim to explore further how LLMs can more precisely express and refine Gherkin requirements to support end-user software development better.

\bibliographystyle{ACM-Reference-Format}
\bibliography{sample-base}


\begin{thebibliography}{53}


\ifx \showCODEN    \undefined \def \showCODEN     #1{\unskip}     \fi
\ifx \showDOI      \undefined \def \showDOI       #1{#1}\fi
\ifx \showISBNx    \undefined \def \showISBNx     #1{\unskip}     \fi
\ifx \showISBNxiii \undefined \def \showISBNxiii  #1{\unskip}     \fi
\ifx \showISSN     \undefined \def \showISSN      #1{\unskip}     \fi
\ifx \showLCCN     \undefined \def \showLCCN      #1{\unskip}     \fi
\ifx \shownote     \undefined \def \shownote      #1{#1}          \fi
\ifx \showarticletitle \undefined \def \showarticletitle #1{#1}   \fi
\ifx \showURL      \undefined \def \showURL       {\relax}        \fi
\providecommand\bibfield[2]{#2}
\providecommand\bibinfo[2]{#2}
\providecommand\natexlab[1]{#1}
\providecommand\showeprint[2][]{arXiv:#2}

\bibitem[Achiam et~al\mbox{.}(2023)]%
        {achiam2023gpt}
\bibfield{author}{\bibinfo{person}{Josh Achiam}, \bibinfo{person}{Steven Adler}, \bibinfo{person}{Sandhini Agarwal}, \bibinfo{person}{Lama Ahmad}, \bibinfo{person}{Ilge Akkaya}, \bibinfo{person}{Florencia~Leoni Aleman}, \bibinfo{person}{Diogo Almeida}, \bibinfo{person}{Janko Altenschmidt}, \bibinfo{person}{Sam Altman}, \bibinfo{person}{Shyamal Anadkat}, {et~al\mbox{.}}} \bibinfo{year}{2023}\natexlab{}.
\newblock \showarticletitle{Gpt-4 technical report}.
\newblock \bibinfo{journal}{\emph{arXiv preprint arXiv:2303.08774}} (\bibinfo{year}{2023}).
\newblock


\bibitem[Adu(2024)]%
        {adu2024artificial}
\bibfield{author}{\bibinfo{person}{Gilbert Adu}.} \bibinfo{year}{2024}\natexlab{}.
\newblock \emph{\bibinfo{title}{Artificial Intelligence in Software Testing: Test scenario and case generation with an AI model (gpt-3.5-turbo) using Prompt engineering, Fine-tuning and Retrieval augmented generation techniques}}.
\newblock \bibinfo{thesistype}{Master's\ thesis}. \bibinfo{school}{It{\"a}-Suomen yliopisto}.
\newblock


\bibitem[Arteaga~Garcia et~al\mbox{.}(2024)]%
        {10.1145/3597503.3608130}
\bibfield{author}{\bibinfo{person}{Emily~Judith Arteaga~Garcia}, \bibinfo{person}{Jo\~{a}o~Felipe Nicolaci~Pimentel}, \bibinfo{person}{Zixuan Feng}, \bibinfo{person}{Marco Gerosa}, \bibinfo{person}{Igor Steinmacher}, {and} \bibinfo{person}{Anita Sarma}.} \bibinfo{year}{2024}\natexlab{}.
\newblock \showarticletitle{How to Support ML End-User Programmers through a Conversational Agent}. In \bibinfo{booktitle}{\emph{Proceedings of the IEEE/ACM 46th International Conference on Software Engineering}} (Lisbon, Portugal) \emph{(\bibinfo{series}{ICSE '24})}. \bibinfo{publisher}{Association for Computing Machinery}, \bibinfo{address}{New York, NY, USA}, Article \bibinfo{articleno}{53}, \bibinfo{numpages}{12}~pages.
\newblock
\showISBNx{9798400702174}
\urldef\tempurl%
\url{https://doi.org/10.1145/3597503.3608130}
\showDOI{\tempurl}


\bibitem[Barricelli et~al\mbox{.}(2019)]%
        {barricelli2019end}
\bibfield{author}{\bibinfo{person}{Barbara~Rita Barricelli}, \bibinfo{person}{Fabio Cassano}, \bibinfo{person}{Daniela Fogli}, {and} \bibinfo{person}{Antonio Piccinno}.} \bibinfo{year}{2019}\natexlab{}.
\newblock \showarticletitle{End-user development, end-user programming and end-user software engineering: A systematic mapping study}.
\newblock \bibinfo{journal}{\emph{Journal of Systems and Software}}  \bibinfo{volume}{149} (\bibinfo{year}{2019}), \bibinfo{pages}{101--137}.
\newblock


\bibitem[Beatty and Chen(2012)]%
        {beatty2012visual}
\bibfield{author}{\bibinfo{person}{J. Beatty} {and} \bibinfo{person}{A. Chen}.} \bibinfo{year}{2012}\natexlab{}.
\newblock \bibinfo{booktitle}{\emph{Visual Models for Software Requirements}}.
\newblock \bibinfo{publisher}{Microsoft Press}.
\newblock
\showISBNx{9780735667723}
\showLCCN{2012939549}
\urldef\tempurl%
\url{https://books.google.com.au/books?id=9hdaiCMmMiQC}
\showURL{%
\tempurl}


\bibitem[Bergsmann et~al\mbox{.}(2024)]%
        {bergsmann2024first}
\bibfield{author}{\bibinfo{person}{Severin Bergsmann}, \bibinfo{person}{Alexander Schmidt}, \bibinfo{person}{Stefan Fischer}, {and} \bibinfo{person}{Rudolf Ramler}.} \bibinfo{year}{2024}\natexlab{}.
\newblock \showarticletitle{First Experiments on Automated Execution of Gherkin Test Specifications with Collaborating LLM Agents}. In \bibinfo{booktitle}{\emph{Proceedings of the 15th ACM International Workshop on Automating Test Case Design, Selection and Evaluation}}. \bibinfo{pages}{12--15}.
\newblock


\bibitem[Burnett et~al\mbox{.}(2004)]%
        {burnett2004end}
\bibfield{author}{\bibinfo{person}{Margaret Burnett}, \bibinfo{person}{Curtis Cook}, {and} \bibinfo{person}{Gregg Rothermel}.} \bibinfo{year}{2004}\natexlab{}.
\newblock \showarticletitle{End-user software engineering}.
\newblock \bibinfo{journal}{\emph{Commun. ACM}} \bibinfo{volume}{47}, \bibinfo{number}{9} (\bibinfo{year}{2004}), \bibinfo{pages}{53--58}.
\newblock


\bibitem[Burnett and Myers(2014)]%
        {burnett2014future}
\bibfield{author}{\bibinfo{person}{Margaret~M Burnett} {and} \bibinfo{person}{Brad~A Myers}.} \bibinfo{year}{2014}\natexlab{}.
\newblock \showarticletitle{Future of end-user software engineering: beyond the silos}.
\newblock In \bibinfo{booktitle}{\emph{Future of Software Engineering Proceedings}}. \bibinfo{pages}{201--211}.
\newblock


\bibitem[Chandorkar et~al\mbox{.}(2022)]%
        {chandorkar2022exploratory}
\bibfield{author}{\bibinfo{person}{Adwait Chandorkar}, \bibinfo{person}{Nitish Patkar}, \bibinfo{person}{Andrea Di~Sorbo}, {and} \bibinfo{person}{Oscar Nierstrasz}.} \bibinfo{year}{2022}\natexlab{}.
\newblock \showarticletitle{An exploratory study on the usage of gherkin features in open-source projects}. In \bibinfo{booktitle}{\emph{2022 IEEE International Conference on Software Analysis, Evolution and Reengineering (SANER)}}. IEEE, \bibinfo{pages}{1159--1166}.
\newblock


\bibitem[Eijkel(2024)]%
        {eijkel2024exploring}
\bibfield{author}{\bibinfo{person}{Stefan van~den Eijkel}.} \bibinfo{year}{2024}\natexlab{}.
\newblock \emph{\bibinfo{title}{Exploring the INVEST Model in Agile Software Development: An In-Depth Analysis}}.
\newblock \bibinfo{thesistype}{Master's\ thesis}.
\newblock


\bibitem[Eleyan et~al\mbox{.}(2020)]%
        {eleyan2020enhancing}
\bibfield{author}{\bibinfo{person}{Derar Eleyan}, \bibinfo{person}{Abed Othman}, {and} \bibinfo{person}{Amna Eleyan}.} \bibinfo{year}{2020}\natexlab{}.
\newblock \showarticletitle{Enhancing software comments readability using flesch reading ease score}.
\newblock \bibinfo{journal}{\emph{Information}} \bibinfo{volume}{11}, \bibinfo{number}{9} (\bibinfo{year}{2020}), \bibinfo{pages}{430}.
\newblock


\bibitem[Elmendorf(1970)]%
        {elmendorf1970automated}
\bibfield{author}{\bibinfo{person}{William~R Elmendorf}.} \bibinfo{year}{1970}\natexlab{}.
\newblock \showarticletitle{Automated design of program test libraries}.
\newblock \bibinfo{journal}{\emph{IBM Technial Report TR 00.2089}} (\bibinfo{year}{1970}).
\newblock


\bibitem[Farooq et~al\mbox{.}(2023)]%
        {farooq2023behavior}
\bibfield{author}{\bibinfo{person}{Muhammad~Shoaib Farooq}, \bibinfo{person}{Uzma Omer}, \bibinfo{person}{Amna Ramzan}, \bibinfo{person}{Mansoor~Ahmad Rasheed}, {and} \bibinfo{person}{Zabihullah Atal}.} \bibinfo{year}{2023}\natexlab{}.
\newblock \showarticletitle{Behavior driven development: A systematic literature review}.
\newblock \bibinfo{journal}{\emph{IEEE Access}} (\bibinfo{year}{2023}).
\newblock


\bibitem[Gu et~al\mbox{.}(2024)]%
        {gu2024survey}
\bibfield{author}{\bibinfo{person}{Jiawei Gu}, \bibinfo{person}{Xuhui Jiang}, \bibinfo{person}{Zhichao Shi}, \bibinfo{person}{Hexiang Tan}, \bibinfo{person}{Xuehao Zhai}, \bibinfo{person}{Chengjin Xu}, \bibinfo{person}{Wei Li}, \bibinfo{person}{Yinghan Shen}, \bibinfo{person}{Shengjie Ma}, \bibinfo{person}{Honghao Liu}, {et~al\mbox{.}}} \bibinfo{year}{2024}\natexlab{}.
\newblock \showarticletitle{A survey on llm-as-a-judge}.
\newblock \bibinfo{journal}{\emph{arXiv preprint arXiv:2411.15594}} (\bibinfo{year}{2024}).
\newblock


\bibitem[Hong et~al\mbox{.}(2023)]%
        {hong2023metagpt}
\bibfield{author}{\bibinfo{person}{Sirui Hong}, \bibinfo{person}{Xiawu Zheng}, \bibinfo{person}{Jonathan Chen}, \bibinfo{person}{Yuheng Cheng}, \bibinfo{person}{Jinlin Wang}, \bibinfo{person}{Ceyao Zhang}, \bibinfo{person}{Zili Wang}, \bibinfo{person}{Steven Ka~Shing Yau}, \bibinfo{person}{Zijuan Lin}, \bibinfo{person}{Liyang Zhou}, {et~al\mbox{.}}} \bibinfo{year}{2023}\natexlab{}.
\newblock \showarticletitle{Metagpt: Meta programming for multi-agent collaborative framework}.
\newblock \bibinfo{journal}{\emph{arXiv preprint arXiv:2308.00352}} (\bibinfo{year}{2023}).
\newblock


\bibitem[Horina and Oleksandr(2023)]%
        {horina2023advantages}
\bibfield{author}{\bibinfo{person}{Kseniia Horina} {and} \bibinfo{person}{Karatanov Oleksandr}.} \bibinfo{year}{2023}\natexlab{}.
\newblock \showarticletitle{Advantages of Automated Testing of Medical Applications and Information Systems Using Gherkin and Behavior-Driven Development}. In \bibinfo{booktitle}{\emph{Conference on Integrated Computer Technologies in Mechanical Engineering--Synergetic Engineering}}. Springer, \bibinfo{pages}{379--391}.
\newblock


\bibitem[Hou et~al\mbox{.}(2024)]%
        {10.1145/3695988}
\bibfield{author}{\bibinfo{person}{Xinyi Hou}, \bibinfo{person}{Yanjie Zhao}, \bibinfo{person}{Yue Liu}, \bibinfo{person}{Zhou Yang}, \bibinfo{person}{Kailong Wang}, \bibinfo{person}{Li Li}, \bibinfo{person}{Xiapu Luo}, \bibinfo{person}{David Lo}, \bibinfo{person}{John Grundy}, {and} \bibinfo{person}{Haoyu Wang}.} \bibinfo{year}{2024}\natexlab{}.
\newblock \showarticletitle{Large Language Models for Software Engineering: A Systematic Literature Review}.
\newblock \bibinfo{journal}{\emph{ACM Trans. Softw. Eng. Methodol.}} \bibinfo{volume}{33}, \bibinfo{number}{8}, Article \bibinfo{articleno}{220} (\bibinfo{date}{Dec.} \bibinfo{year}{2024}), \bibinfo{numpages}{79}~pages.
\newblock
\showISSN{1049-331X}
\urldef\tempurl%
\url{https://doi.org/10.1145/3695988}
\showDOI{\tempurl}


\bibitem[Jang and Kim(2022)]%
        {jang2022automatic}
\bibfield{author}{\bibinfo{person}{Woo~Sung Jang} {and} \bibinfo{person}{R~Young~Chul Kim}.} \bibinfo{year}{2022}\natexlab{}.
\newblock \showarticletitle{Automatic cause--effect graph tool with informal Korean requirement specifications}.
\newblock \bibinfo{journal}{\emph{Applied Sciences}} \bibinfo{volume}{12}, \bibinfo{number}{18} (\bibinfo{year}{2022}), \bibinfo{pages}{9310}.
\newblock


\bibitem[Jia et~al\mbox{.}(2025)]%
        {jia2025automated}
\bibfield{author}{\bibinfo{person}{Haoxiang Jia}, \bibinfo{person}{Robbie Morris}, \bibinfo{person}{He Ye}, \bibinfo{person}{Federica Sarro}, {and} \bibinfo{person}{Sergey Mechtaev}.} \bibinfo{year}{2025}\natexlab{}.
\newblock \showarticletitle{Automated Repair of Ambiguous Natural Language Requirements}.
\newblock \bibinfo{journal}{\emph{arXiv preprint arXiv:2505.07270}} (\bibinfo{year}{2025}).
\newblock


\bibitem[Joy~Beatty({[n.\,d.]})]%
        {FeatureTree}
\bibfield{author}{\bibinfo{person}{Karl~Wiegers Joy~Beatty}.} \bibinfo{year}{[n.\,d.]}\natexlab{}.
\newblock \bibinfo{booktitle}{\emph{{Using Feature Trees to Depict Scope}}}.
\newblock
\urldef\tempurl%
\url{https://www.modernanalyst.com/Resources/Articles/tabid/115/ID/6061/Using-Feature-Trees-to-Depict-Scope.aspx}
\showURL{%
\tempurl}
\newblock
\shownote{(2022)}.


\bibitem[Kano et~al\mbox{.}(1984)]%
        {kano1984attractive}
\bibfield{author}{\bibinfo{person}{Noriaki Kano}, \bibinfo{person}{Nobuhiku Seraku}, \bibinfo{person}{Fumio Takahashi}, {and} \bibinfo{person}{Shinichi Tsuji}.} \bibinfo{year}{1984}\natexlab{}.
\newblock \showarticletitle{Attractive quality and must-be quality}.
\newblock  (\bibinfo{year}{1984}).
\newblock


\bibitem[Karpurapu et~al\mbox{.}(2024)]%
        {karpurapu2024comprehensive}
\bibfield{author}{\bibinfo{person}{Shanthi Karpurapu}, \bibinfo{person}{Sravanthy Myneni}, \bibinfo{person}{Unnati Nettur}, \bibinfo{person}{Likhit~Sagar Gajja}, \bibinfo{person}{Dave Burke}, \bibinfo{person}{Tom Stiehm}, {and} \bibinfo{person}{Jeffery Payne}.} \bibinfo{year}{2024}\natexlab{}.
\newblock \showarticletitle{Comprehensive Evaluation and Insights into the Use of Large Language Models in the Automation of Behavior-Driven Development Acceptance Test Formulation}.
\newblock \bibinfo{journal}{\emph{IEEE Access}} (\bibinfo{year}{2024}).
\newblock


\bibitem[Ko et~al\mbox{.}(2011)]%
        {ko2011state}
\bibfield{author}{\bibinfo{person}{Amy~J Ko}, \bibinfo{person}{Robin Abraham}, \bibinfo{person}{Laura Beckwith}, \bibinfo{person}{Alan Blackwell}, \bibinfo{person}{Margaret Burnett}, \bibinfo{person}{Martin Erwig}, \bibinfo{person}{Chris Scaffidi}, \bibinfo{person}{Joseph Lawrance}, \bibinfo{person}{Henry Lieberman}, \bibinfo{person}{Brad Myers}, {et~al\mbox{.}}} \bibinfo{year}{2011}\natexlab{}.
\newblock \showarticletitle{The state of the art in end-user software engineering}.
\newblock \bibinfo{journal}{\emph{ACM Computing Surveys (CSUR)}} \bibinfo{volume}{43}, \bibinfo{number}{3} (\bibinfo{year}{2011}), \bibinfo{pages}{1--44}.
\newblock


\bibitem[Krishna et~al\mbox{.}(2024)]%
        {krishna2024using}
\bibfield{author}{\bibinfo{person}{Madhava Krishna}, \bibinfo{person}{Bhagesh Gaur}, \bibinfo{person}{Arsh Verma}, {and} \bibinfo{person}{Pankaj Jalote}.} \bibinfo{year}{2024}\natexlab{}.
\newblock \showarticletitle{Using LLMs in software requirements specifications: an empirical evaluation}. In \bibinfo{booktitle}{\emph{2024 IEEE 32nd International Requirements Engineering Conference (RE)}}. IEEE, \bibinfo{pages}{475--483}.
\newblock


\bibitem[Krupalija et~al\mbox{.}(2022)]%
        {10051799}
\bibfield{author}{\bibinfo{person}{Ehlimana Krupalija}, \bibinfo{person}{Emir Cogo}, \bibinfo{person}{Šeila Bećirović}, \bibinfo{person}{Irfan Prazina}, {and} \bibinfo{person}{Ingmar Bešić}.} \bibinfo{year}{2022}\natexlab{}.
\newblock \showarticletitle{Cause-effect Graphing Technique: A Survey of Available Approaches and Algorithms}. In \bibinfo{booktitle}{\emph{2022 IEEE/ACIS 23rd International Conference on Software Engineering, Artificial Intelligence, Networking and Parallel/Distributed Computing (SNPD)}}. \bibinfo{pages}{162--167}.
\newblock
\urldef\tempurl%
\url{https://doi.org/10.1109/SNPD54884.2022.10051799}
\showDOI{\tempurl}


\bibitem[Krupalija et~al\mbox{.}(2023a)]%
        {10155063}
\bibfield{author}{\bibinfo{person}{Ehlimana Krupalija}, \bibinfo{person}{Emir Cogo}, \bibinfo{person}{Šeila Bećirović}, \bibinfo{person}{Irfan Prazina}, \bibinfo{person}{Damir Pozderac}, {and} \bibinfo{person}{Ingmar Bešić}.} \bibinfo{year}{2023}\natexlab{a}.
\newblock \showarticletitle{CEGSet: Collection of standardized cause-effect graph specifications}. In \bibinfo{booktitle}{\emph{2023 12th Mediterranean Conference on Embedded Computing (MECO)}}. \bibinfo{pages}{1--4}.
\newblock
\urldef\tempurl%
\url{https://doi.org/10.1109/MECO58584.2023.10155063}
\showDOI{\tempurl}


\bibitem[Krupalija et~al\mbox{.}(2023b)]%
        {krupalija2023usage}
\bibfield{author}{\bibinfo{person}{Ehlimana Krupalija}, \bibinfo{person}{Emir Cogo}, \bibinfo{person}{Damir Pozderac}, \bibinfo{person}{Aya Ali Al~Zayat}, {and} \bibinfo{person}{Ingmar Be{\v{s}}i{\'c}}.} \bibinfo{year}{2023}\natexlab{b}.
\newblock \showarticletitle{Usage of machine learning methods for cause-effect graph feasibility prediction}.
\newblock In \bibinfo{booktitle}{\emph{Machine Learning and Artificial Intelligence}}. \bibinfo{publisher}{IOS Press}, \bibinfo{pages}{126--131}.
\newblock


\bibitem[Krupalija et~al\mbox{.}(2024)]%
        {krupalija2024etf}
\bibfield{author}{\bibinfo{person}{Ehlimana Krupalija}, \bibinfo{person}{Emir Cogo}, \bibinfo{person}{Damir Pozderac}, \bibinfo{person}{Samir Omanovi{\'c}}, \bibinfo{person}{Almir Karabegovi{\'c}}, \bibinfo{person}{Razija~Tur{\v{c}}inhod{\v{z}}i{\'c} Mulahasanovi{\'c}}, {and} \bibinfo{person}{Ingmar Be{\v{s}}i{\'c}}.} \bibinfo{year}{2024}\natexlab{}.
\newblock \showarticletitle{ETF-RI-CEG-Advanced: A graphical desktop tool for black-box testing by using cause--effect graphs}.
\newblock \bibinfo{journal}{\emph{SoftwareX}}  \bibinfo{volume}{25} (\bibinfo{year}{2024}), \bibinfo{pages}{101625}.
\newblock


\bibitem[Laplante and Kassab(2022)]%
        {laplante2022requirements[5]2}
\bibfield{author}{\bibinfo{person}{Phillip~A Laplante} {and} \bibinfo{person}{Mohamad Kassab}.} \bibinfo{year}{2022}\natexlab{}.
\newblock \bibinfo{booktitle}{\emph{Requirements engineering for software and systems}}.
\newblock \bibinfo{publisher}{Auerbach Publications}.
\newblock
\urldef\tempurl%
\url{https://doi.org/10.1201/9781003129509}
\showDOI{\tempurl}


\bibitem[Ma et~al\mbox{.}(2025)]%
        {10.1145/3731756}
\bibfield{author}{\bibinfo{person}{Qianou Ma}, \bibinfo{person}{Weirui Peng}, \bibinfo{person}{Chenyang Yang}, \bibinfo{person}{Hua Shen}, \bibinfo{person}{Kenneth Koedinger}, {and} \bibinfo{person}{Tongshuang Wu}.} \bibinfo{year}{2025}\natexlab{}.
\newblock \showarticletitle{What Should We Engineer in Prompts? Training Humans in Requirement-Driven LLM Use}.
\newblock \bibinfo{journal}{\emph{ACM Trans. Comput.-Hum. Interact.}} (\bibinfo{date}{April} \bibinfo{year}{2025}).
\newblock
\showISSN{1073-0516}
\urldef\tempurl%
\url{https://doi.org/10.1145/3731756}
\showDOI{\tempurl}
\newblock
\shownote{Just Accepted}.


\bibitem[Marques et~al\mbox{.}(2024)]%
        {marques2024using}
\bibfield{author}{\bibinfo{person}{Nuno Marques}, \bibinfo{person}{Rodrigo~Rocha Silva}, {and} \bibinfo{person}{Jorge Bernardino}.} \bibinfo{year}{2024}\natexlab{}.
\newblock \showarticletitle{Using ChatGPT in Software Requirements Engineering: A Comprehensive Review}.
\newblock \bibinfo{journal}{\emph{Future Internet}} \bibinfo{volume}{16}, \bibinfo{number}{6} (\bibinfo{year}{2024}), \bibinfo{pages}{180}.
\newblock


\bibitem[Mu et~al\mbox{.}(2024)]%
        {10.1145/3660810}
\bibfield{author}{\bibinfo{person}{Fangwen Mu}, \bibinfo{person}{Lin Shi}, \bibinfo{person}{Song Wang}, \bibinfo{person}{Zhuohao Yu}, \bibinfo{person}{Binquan Zhang}, \bibinfo{person}{ChenXue Wang}, \bibinfo{person}{Shichao Liu}, {and} \bibinfo{person}{Qing Wang}.} \bibinfo{year}{2024}\natexlab{}.
\newblock \showarticletitle{ClarifyGPT: A Framework for Enhancing LLM-Based Code Generation via Requirements Clarification}.
\newblock \bibinfo{journal}{\emph{Proc. ACM Softw. Eng.}} \bibinfo{volume}{1}, \bibinfo{number}{FSE}, Article \bibinfo{articleno}{103} (\bibinfo{date}{July} \bibinfo{year}{2024}), \bibinfo{numpages}{23}~pages.
\newblock
\urldef\tempurl%
\url{https://doi.org/10.1145/3660810}
\showDOI{\tempurl}


\bibitem[Nardi(1993)]%
        {nardi1993small}
\bibfield{author}{\bibinfo{person}{Bonnie~A Nardi}.} \bibinfo{year}{1993}\natexlab{}.
\newblock \bibinfo{booktitle}{\emph{A small matter of programming: perspectives on end user computing}}.
\newblock \bibinfo{publisher}{MIT press}.
\newblock


\bibitem[Nguyen et~al\mbox{.}(2024)]%
        {nguyen2024agilecoder}
\bibfield{author}{\bibinfo{person}{Minh~Huynh Nguyen}, \bibinfo{person}{Thang~Phan Chau}, \bibinfo{person}{Phong~X Nguyen}, {and} \bibinfo{person}{Nghi~DQ Bui}.} \bibinfo{year}{2024}\natexlab{}.
\newblock \showarticletitle{AgileCoder: Dynamic Collaborative Agents for Software Development based on Agile Methodology}.
\newblock \bibinfo{journal}{\emph{arXiv preprint arXiv:2406.11912}} (\bibinfo{year}{2024}).
\newblock


\bibitem[Northrop et~al\mbox{.}(2006)]%
        {northrop2006ultra}
\bibfield{author}{\bibinfo{person}{Linda Northrop}, \bibinfo{person}{Peter Feiler}, \bibinfo{person}{Richard~P Gabriel}, \bibinfo{person}{John Goodenough}, \bibinfo{person}{Rick Linger}, \bibinfo{person}{Tom Longstaff}, \bibinfo{person}{Rick Kazman}, \bibinfo{person}{Mark Klein}, \bibinfo{person}{Douglas Schmidt}, \bibinfo{person}{Kevin Sullivan}, {et~al\mbox{.}}} \bibinfo{year}{2006}\natexlab{}.
\newblock \showarticletitle{Ultra-large-scale systems: The software challenge of the future}.
\newblock  (\bibinfo{year}{2006}).
\newblock


\bibitem[Parsa(2023)]%
        {parsa2023acceptance}
\bibfield{author}{\bibinfo{person}{Saeed Parsa}.} \bibinfo{year}{2023}\natexlab{}.
\newblock \showarticletitle{Acceptance testing and behavior driven development (BDD)}.
\newblock In \bibinfo{booktitle}{\emph{Software Testing Automation: Testability Evaluation, Refactoring, Test Data Generation and Fault Localization}}. \bibinfo{publisher}{Springer}, \bibinfo{pages}{79--158}.
\newblock


\bibitem[Puspita et~al\mbox{.}(2024)]%
        {puspita2024analysis}
\bibfield{author}{\bibinfo{person}{Safriya~Murni Puspita}, \bibinfo{person}{Alvina~Waihda Ardhani}, \bibinfo{person}{Dea~Ayu Retnaningrum}, \bibinfo{person}{Afreza~Restu Firmansyah}, {and} \bibinfo{person}{Dwi Rolliawati}.} \bibinfo{year}{2024}\natexlab{}.
\newblock \showarticletitle{ANALYSIS OF SOFTWARE QUALITY USING THE FURPS+ MODEL}.
\newblock \bibinfo{journal}{\emph{JURTEKSI (Jurnal Teknologi dan Sistem Informasi)}} \bibinfo{volume}{11}, \bibinfo{number}{1} (\bibinfo{year}{2024}), \bibinfo{pages}{131--138}.
\newblock


\bibitem[Qian et~al\mbox{.}(2023)]%
        {qian2023experiential}
\bibfield{author}{\bibinfo{person}{Chen Qian}, \bibinfo{person}{Yufan Dang}, \bibinfo{person}{Jiahao Li}, \bibinfo{person}{Wei Liu}, \bibinfo{person}{Zihao Xie}, \bibinfo{person}{Yifei Wang}, \bibinfo{person}{Weize Chen}, \bibinfo{person}{Cheng Yang}, \bibinfo{person}{Xin Cong}, \bibinfo{person}{Xiaoyin Che}, {et~al\mbox{.}}} \bibinfo{year}{2023}\natexlab{}.
\newblock \showarticletitle{Experiential co-learning of software-developing agents}.
\newblock \bibinfo{journal}{\emph{arXiv preprint arXiv:2312.17025}} (\bibinfo{year}{2023}).
\newblock


\bibitem[Qian et~al\mbox{.}(2024a)]%
        {qian2024chatdev}
\bibfield{author}{\bibinfo{person}{Chen Qian}, \bibinfo{person}{Wei Liu}, \bibinfo{person}{Hongzhang Liu}, \bibinfo{person}{Nuo Chen}, \bibinfo{person}{Yufan Dang}, \bibinfo{person}{Jiahao Li}, \bibinfo{person}{Cheng Yang}, \bibinfo{person}{Weize Chen}, \bibinfo{person}{Yusheng Su}, \bibinfo{person}{Xin Cong}, {et~al\mbox{.}}} \bibinfo{year}{2024}\natexlab{a}.
\newblock \showarticletitle{Chatdev: Communicative agents for software development}. In \bibinfo{booktitle}{\emph{Proceedings of the 62nd Annual Meeting of the Association for Computational Linguistics (Volume 1: Long Papers)}}. \bibinfo{pages}{15174--15186}.
\newblock


\bibitem[Qian et~al\mbox{.}(2024b)]%
        {qian-etal-2024-chatdev}
\bibfield{author}{\bibinfo{person}{Chen Qian}, \bibinfo{person}{Wei Liu}, \bibinfo{person}{Hongzhang Liu}, \bibinfo{person}{Nuo Chen}, \bibinfo{person}{Yufan Dang}, \bibinfo{person}{Jiahao Li}, \bibinfo{person}{Cheng Yang}, \bibinfo{person}{Weize Chen}, \bibinfo{person}{Yusheng Su}, \bibinfo{person}{Xin Cong}, \bibinfo{person}{Juyuan Xu}, \bibinfo{person}{Dahai Li}, \bibinfo{person}{Zhiyuan Liu}, {and} \bibinfo{person}{Maosong Sun}.} \bibinfo{year}{2024}\natexlab{b}.
\newblock \showarticletitle{{C}hat{D}ev: Communicative Agents for Software Development}. In \bibinfo{booktitle}{\emph{Proceedings of the 62nd Annual Meeting of the Association for Computational Linguistics (Volume 1: Long Papers)}}, \bibfield{editor}{\bibinfo{person}{Lun-Wei Ku}, \bibinfo{person}{Andre Martins}, {and} \bibinfo{person}{Vivek Srikumar}} (Eds.). \bibinfo{publisher}{Association for Computational Linguistics}, \bibinfo{address}{Bangkok, Thailand}, \bibinfo{pages}{15174--15186}.
\newblock
\urldef\tempurl%
\url{https://doi.org/10.18653/v1/2024.acl-long.810}
\showDOI{\tempurl}


\bibitem[Qian et~al\mbox{.}(2024c)]%
        {qian2024scaling}
\bibfield{author}{\bibinfo{person}{Chen Qian}, \bibinfo{person}{Zihao Xie}, \bibinfo{person}{Yifei Wang}, \bibinfo{person}{Wei Liu}, \bibinfo{person}{Yufan Dang}, \bibinfo{person}{Zhuoyun Du}, \bibinfo{person}{Weize Chen}, \bibinfo{person}{Cheng Yang}, \bibinfo{person}{Zhiyuan Liu}, {and} \bibinfo{person}{Maosong Sun}.} \bibinfo{year}{2024}\natexlab{c}.
\newblock \showarticletitle{Scaling large-language-model-based multi-agent collaboration}.
\newblock \bibinfo{journal}{\emph{arXiv preprint arXiv:2406.07155}} (\bibinfo{year}{2024}).
\newblock


\bibitem[Robinson et~al\mbox{.}(2024)]%
        {robinson2024requirements}
\bibfield{author}{\bibinfo{person}{Diana Robinson}, \bibinfo{person}{Christian Cabrera}, \bibinfo{person}{Andrew~D. Gordon}, \bibinfo{person}{Neil~D. Lawrence}, {and} \bibinfo{person}{Lars Mennen}.} \bibinfo{year}{2024}\natexlab{}.
\newblock \showarticletitle{Requirements are All You Need: The Final Frontier for End-User Software Engineering}.
\newblock \bibinfo{journal}{\emph{ACM Trans. Softw. Eng. Methodol.}} (\bibinfo{date}{Dec.} \bibinfo{year}{2024}).
\newblock
\showISSN{1049-331X}
\urldef\tempurl%
\url{https://doi.org/10.1145/3708524}
\showDOI{\tempurl}
\newblock
\shownote{Just Accepted}.


\bibitem[SANTOS et~al\mbox{.}({[n.\,d.]})]%
        {santos4933832perception}
\bibfield{author}{\bibinfo{person}{SHEXMO SANTOS}, \bibinfo{person}{Michel dos Santos~Soares}, {and} \bibinfo{person}{Fabio Gomes~Rocha}.} \bibinfo{year}{[n.\,d.]}\natexlab{}.
\newblock \showarticletitle{Perception of Professionals Regarding the Adoption of Behavior-Driven Development (Bdd): A Descriptive and Statistical Study Through a Survey}.
\newblock \bibinfo{journal}{\emph{Available at SSRN 4933832}} (\bibinfo{year}{[n.\,d.]}).
\newblock


\bibitem[Shannon(1948)]%
        {shannon1948mathematical}
\bibfield{author}{\bibinfo{person}{Claude~E Shannon}.} \bibinfo{year}{1948}\natexlab{}.
\newblock \showarticletitle{A mathematical theory of communication}.
\newblock \bibinfo{journal}{\emph{The Bell system technical journal}} \bibinfo{volume}{27}, \bibinfo{number}{3} (\bibinfo{year}{1948}), \bibinfo{pages}{379--423}.
\newblock


\bibitem[Van~Lamsweerde(2000)]%
        {van2000requirements}
\bibfield{author}{\bibinfo{person}{Axel Van~Lamsweerde}.} \bibinfo{year}{2000}\natexlab{}.
\newblock \showarticletitle{Requirements engineering in the year 00: A research perspective}. In \bibinfo{booktitle}{\emph{Proceedings of the 22nd international conference on Software engineering}}. \bibinfo{pages}{5--19}.
\newblock


\bibitem[Vogelsang et~al\mbox{.}(2025)]%
        {vogelsang2025impact}
\bibfield{author}{\bibinfo{person}{Andreas Vogelsang}, \bibinfo{person}{Alexander Korn}, \bibinfo{person}{Giovanna Broccia}, \bibinfo{person}{Alessio Ferrari}, \bibinfo{person}{Jannik Fischbach}, {and} \bibinfo{person}{Chetan Arora}.} \bibinfo{year}{2025}\natexlab{}.
\newblock \showarticletitle{On the Impact of Requirements Smells in Prompts: The Case of Automated Traceability}.
\newblock \bibinfo{journal}{\emph{arXiv preprint arXiv:2501.04810}} (\bibinfo{year}{2025}).
\newblock


\bibitem[Wei et~al\mbox{.}(2022)]%
        {wei2022chain}
\bibfield{author}{\bibinfo{person}{Jason Wei}, \bibinfo{person}{Xuezhi Wang}, \bibinfo{person}{Dale Schuurmans}, \bibinfo{person}{Maarten Bosma}, \bibinfo{person}{Fei Xia}, \bibinfo{person}{Ed Chi}, \bibinfo{person}{Quoc~V Le}, \bibinfo{person}{Denny Zhou}, {et~al\mbox{.}}} \bibinfo{year}{2022}\natexlab{}.
\newblock \showarticletitle{Chain-of-thought prompting elicits reasoning in large language models}.
\newblock \bibinfo{journal}{\emph{Advances in neural information processing systems}}  \bibinfo{volume}{35} (\bibinfo{year}{2022}), \bibinfo{pages}{24824--24837}.
\newblock


\bibitem[White et~al\mbox{.}(2024)]%
        {white2024chatgpt}
\bibfield{author}{\bibinfo{person}{Jules White}, \bibinfo{person}{Sam Hays}, \bibinfo{person}{Quchen Fu}, \bibinfo{person}{Jesse Spencer-Smith}, {and} \bibinfo{person}{Douglas~C Schmidt}.} \bibinfo{year}{2024}\natexlab{}.
\newblock \showarticletitle{Chatgpt prompt patterns for improving code quality, refactoring, requirements elicitation, and software design}.
\newblock In \bibinfo{booktitle}{\emph{Generative AI for Effective Software Development}}. \bibinfo{publisher}{Springer}, \bibinfo{pages}{71--108}.
\newblock


\bibitem[Wiegers and Beatty(2013)]%
        {wiegers2013software}
\bibfield{author}{\bibinfo{person}{Karl~E Wiegers} {and} \bibinfo{person}{Joy Beatty}.} \bibinfo{year}{2013}\natexlab{}.
\newblock \bibinfo{booktitle}{\emph{Software requirements}}.
\newblock \bibinfo{publisher}{Pearson Education}.
\newblock


\bibitem[Wu et~al\mbox{.}(2025)]%
        {wu2025clarifycoder}
\bibfield{author}{\bibinfo{person}{Jie~JW Wu}, \bibinfo{person}{Manav Chaudhary}, \bibinfo{person}{Davit Abrahamyan}, \bibinfo{person}{Arhaan Khaku}, \bibinfo{person}{Anjiang Wei}, {and} \bibinfo{person}{Fatemeh~H Fard}.} \bibinfo{year}{2025}\natexlab{}.
\newblock \showarticletitle{ClarifyCoder: Clarification-Aware Fine-Tuning for Programmatic Problem Solving}.
\newblock \bibinfo{journal}{\emph{arXiv preprint arXiv:2504.16331}} (\bibinfo{year}{2025}).
\newblock


\bibitem[Wu et~al\mbox{.}(2023)]%
        {wu2023autogen}
\bibfield{author}{\bibinfo{person}{Qingyun Wu}, \bibinfo{person}{Gagan Bansal}, \bibinfo{person}{Jieyu Zhang}, \bibinfo{person}{Yiran Wu}, \bibinfo{person}{Beibin Li}, \bibinfo{person}{Erkang Zhu}, \bibinfo{person}{Li Jiang}, \bibinfo{person}{Xiaoyun Zhang}, \bibinfo{person}{Shaokun Zhang}, \bibinfo{person}{Jiale Liu}, {et~al\mbox{.}}} \bibinfo{year}{2023}\natexlab{}.
\newblock \showarticletitle{Autogen: Enabling next-gen llm applications via multi-agent conversation}.
\newblock \bibinfo{journal}{\emph{arXiv preprint arXiv:2308.08155}} (\bibinfo{year}{2023}).
\newblock


\bibitem[Xing et~al\mbox{.}(2025)]%
        {xing2025prompt}
\bibfield{author}{\bibinfo{person}{Zhenchang Xing}, \bibinfo{person}{Yang Liu}, \bibinfo{person}{Zhuo Cheng}, \bibinfo{person}{Qing Huang}, \bibinfo{person}{Dehai Zhao}, \bibinfo{person}{Daniel SUN}, {and} \bibinfo{person}{Chenhua Liu}.} \bibinfo{year}{2025}\natexlab{}.
\newblock \showarticletitle{When Prompt Engineering Meets Software Engineering: CNL-P as Natural and Robust" APIs''for Human-AI Interaction}. In \bibinfo{booktitle}{\emph{The Thirteenth International Conference on Learning Representations}}.
\newblock


\bibitem[Zhang et~al\mbox{.}(2025)]%
        {zhang2024empowering}
\bibfield{author}{\bibinfo{person}{Sai Zhang}, \bibinfo{person}{Zhenchang Xing}, \bibinfo{person}{Ronghui Guo}, \bibinfo{person}{Fangzhou Xu}, \bibinfo{person}{Lei Chen}, \bibinfo{person}{Zhaoyuan Zhang}, \bibinfo{person}{Xiaowang Zhang}, \bibinfo{person}{Zhiyong Feng}, {and} \bibinfo{person}{Zhiqiang Zhuang}.} \bibinfo{year}{2025}\natexlab{}.
\newblock \showarticletitle{Empowering Agile-Based Generative Software Development through Human-AI Teamwork}.
\newblock \bibinfo{journal}{\emph{ACM Trans. Softw. Eng. Methodol.}} \bibinfo{volume}{34}, \bibinfo{number}{6}, Article \bibinfo{articleno}{156} (\bibinfo{date}{July} \bibinfo{year}{2025}), \bibinfo{numpages}{46}~pages.
\newblock
\showISSN{1049-331X}
\urldef\tempurl%
\url{https://doi.org/10.1145/3702987}
\showDOI{\tempurl}


\end{thebibliography}

\end{document}